\documentclass[a4paper,12pt]{article}

\usepackage{array}
\usepackage{tabularx}
\usepackage{makecell}
\usepackage{tikz}
\usepackage{tikz-cd}
\usetikzlibrary{decorations.markings}
\usepackage{amsmath,amssymb,amsfonts,epsfig,cite,setspace,bigstrut,framed}
\usepackage[all]{xy}
\usepackage{color}
\usepackage{pifont}
\usepackage{xcolor}
\usepackage{tikz}
\usepackage{tikz-cd}
\usepackage[export]{adjustbox}
\usetikzlibrary{shapes.geometric}
\usepackage[colorlinks=true,urlcolor=blue,anchorcolor=blue,citecolor=blue,filecolor=blue,linkcolor=blue,menucolor=blue,linktocpage=true,pdfproducer=medialab,pdfa=true]{hyperref}	 
\usetikzlibrary{positioning}
\usepackage[perpage]{footmisc}
\usepackage[compat=1.1.0]{tikz-feynman}
\usepackage{colortbl}

\usepackage[numbers,sort&compress]{natbib} 

\usepackage{todonotes}
\usepackage{draft,hyperref,tikz,cancel,subfig,float,commutative-diagrams,multirow}
\usetikzlibrary{snakes}
\usetikzlibrary{shapes.misc}
\usetikzlibrary{cd}
\usepackage{dsfont}

\definecolor{dgreen}{rgb}{0, 0.55, 0}
\definecolor{llightyellow}{rgb}{1.0, 0.95, 0.7}
\definecolor{llightblue}{rgb}{0.7, 0.9, 1.0}
\definecolor{llightpink}{rgb}{1.0, 0.85, 0.95}
\definecolor{llightgreen}{rgb}{0.7, 1.0, 0.4}
\colorlet{lightyellow}{llightyellow!50!white}
\colorlet{lightblue}{llightblue!50!white}
\colorlet{lightgreen}{llightgreen!50!white}
\colorlet{lightpink}{llightpink!50!white}

\usepackage{mathtools}

\newcommand{\dsl}{\pa \kern-0.5em /}

\newcommand{\pa}{\partial}

\newcommand{\bit}{\begin{itemize}}
\newcommand{\eit}{\end{itemize}}

\def\bZ{\mathbb Z}

\newcommand{\ba}{\begin{array}}
\newcommand{\ea}{\end{array}}
\def\bZ{\mathbb Z}

\makeatletter \@addtoreset{equation}{section} \makeatother

\newcommand{\comment}[1]{}

\newcommand{\bi}{\begin{itemize}}
\newcommand{\ei}{\end{itemize}}
\newcommand{\beq}{\begin{equation}}
\newcommand{\eeq}{\end{equation}}

\newcommand{\orcid}[1]{\href{https://orcid.org/#1}{\includegraphics[width=8pt]{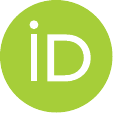}}}
\usepackage{bm}

\newcommand{\Scal}{\mathcal{S}}

\definecolor{nblue}{rgb}{0.2,0.2,0.7}
\definecolor{ngreen}{rgb}{0.2,0.6,0.2}
\definecolor{nred}{rgb}{0.7,0.2,0.2}
\definecolor{nblack}{rgb}{0,0,0}

\usepackage{titlesec}
\titleformat{\section}
  {\large\bfseries} 
  {\thesection}{0.5em}{}
\titleformat{\subsection}
  {\bfseries}
  {\thesubsection}{0.5em}{}

\begin{document}

\begin{titlepage}
\begin{center}

\title{\fontsize{15pt}{15pt}\selectfont \bfseries Subsystem Symmetry-Protected Topological Phases from Subsystem SymTFT of 2-Foliated Exotic Tensor Gauge Theory}

\vskip0.5cm 
{Qiang Jia$^{a,*}$\orcid{0009-0002-1890-8431} and Zhian Jia$^{b,c,\dagger}$\orcid{0000-0001-8588-173X}} 
\vskip.5cm
 {\small{\textit{$^a$Department of Physics, Korea Advanced Institute of Science \& Technology, Daejeon 34141, Korea}}}
 
 \vskip.2cm
 {\small{\textit{$^b$Centre for Quantum Technologies, National University of Singapore, Singapore 117543, Singapore \\}}}
 {\small{\textit{$^c$Department of Physics, National University of Singapore, Singapore 117543, Singapore}}}
  \vskip.2cm
  
  {\small{\textit{$^*$Email: qjia1993@kaist.ac.kr}}; \small{\textit{$^{\dagger}$Email: giannjia@foxmail.com}}}
 
\end{center}

\vskip.5cm
\baselineskip 16pt

\begin{abstract}
Symmetry topological field theory (SymTFT), or topological holography, posits a correspondence between symmetries in a $d$-dimensional theory and topological order in a $(d+1)$-dimensional theory. In this work, we extend this framework to subsystem symmetries and develop subsystem SymTFT as a systematic tool to characterize and classify subsystem symmetry-protected topological (SSPT) phases. For $(2+1)$D gapped phases, we introduce a  2-foliated $(3+1)$D exotic tensor gauge theory (which is equivalent to 2-foliated $(3+1)$D BF theory via exotic duality) as the subsystem SymTFT and systematically analyze its topological boundary conditions and linearly rigid subsystem symmetries. Taking subsystem symmetry groups $G = \mathbb{Z}_N$ and $G=\mathbb{Z}_N \times \mathbb{Z}_M$ as examples, we demonstrate how to recover the classification scheme $\mathcal{C}[G] = H^{2}(G^{\times 2}, U(1)) / \left( H^2(G, U(1)) \right)^3$, which was previously derived by examining topological invariant under linear subsystem-symmetric local unitary transformations in the lattice Hamiltonian formalism. To illustrate the correspondence between field-theoretic and lattice descriptions, we further analyze $\mathbb{Z}_2 \times \mathbb{Z}_2$ and $\mathbb{Z}_N \times \mathbb{Z}_M$ cluster state models as concrete examples.

\end{abstract}

\vfill

\end{titlepage}

\hrule
\tableofcontents
\vspace{2em}
\hrule

\section{Introduction}
\label{sec:intro}

The symmetry-protected topological (SPT) phases \cite{Haldane1983nonlinear, haldane1983continuum, Affleck1987AKLT, Gu2009SPT, Chen2011classification, Pollmann2012symmetry, Chen20112dSPT, Pollmann2012detection, Schuch2011classifyingSPT, Chen2012SPTboson, Else2014SPT} are some of the most fundamental phases of quantum matter and have garnered significant attention in the past several decades. The ground state of an SPT phase is short-range entangled, and it is characterized by an invertible topological field \cite{Wen2017zoo, freed2021reflection}. This implies that, on any closed manifold, SPT phases have a unique ground state. Besides its fundamental importance, it also has applications in measurement-based quantum computation \cite{Raussendorf2001}, where quantum states that have SPT order can be used as resource states for implementing quantum computation via performing only single-qubit measurements.
The classification of both bosonic and fermionic SPT phases has been extensively studied from various perspectives \cite{Wen2013SPT, Chen2013SPTclassification, Vishwanath2013physics, Wang2013SPT, Burnell2014spt, kapustin2014SPT, kapustin2014bosonicSPT, freed2021reflection, yonekura2019cobordism, kitaev2009periodic, ryu2010topological, Gu2014SPTfermion, kong2020classification}.

Recent studies have shown that the so-called Symmetry Topological Field Theory (SymTFT) is a unifying framework to study various properties of the symmetry, such as gauging and their anomalies ~\cite{Witten:1998wy, Kong2020algebraic, gaiotto2021orbifold, Apruzzi:2021nmk, Kaidi:2022cpf, vanBeest:2022fss, Freed:2022qnc, Apruzzi:2022dlm, Kaidi:2023maf, Baume:2023kkf, Cvetic:2024dzu, Jia:2025jmn}. It is also useful to understand and classify SPT phases, as demonstrated in, e.g., \cite{Kong2020algebraic, Bhardwaj:2023idu, Bhardwaj:2023fca, SchaferNameki2024ICTP, huang2023topologicalholo, bhardwaj2024lattice, Freed:2022qnc, gaiotto2021orbifold, bhardwaj2023generalizedcharge, apruzzi2023symmetry, Bhardwaj:2023fca, Zhang2024anomaly, Ji2020categoricalsym, bhardwaj2024clubsandwich, bhardwaj2024hassediagramsgaplessspt, luo2023lecture, shao2024whats, Bhardwaj:2024qiv, Bhardwaj:2025piv}. The central idea is to separate the symmetry information and dynamics of the $d$-dimensional system $\mathfrak{T}_{\mathcal{S}}$ into two distinct boundaries of a topological quantum field theory (TQFT) $\mathcal{Z}(\mathcal{S})$ in $(d+1)$-dimension and then implement compactification to obtain the original system\footnote{In this work, we use $d$ to denote the spatial dimension and $D$ for the spacetime dimension.}. 
The SymTFT framework can be understood via a sandwich construction: the theory $\mathfrak{T}_{\mathcal{S}}$ lives on $\mathcal{M}_d$ is expanded into $\mathcal{Z}(\mathcal{S})$ on $[0,1] \times \mathcal{M}_d$, and the symmetry $\mathcal{S}$ lives on the topological boundary $\mathcal{B}_{\text{top}}$ at $\{0\} \times \mathcal{M}_d$. On the other hand, the physical boundary $\mathcal{B}_{\text{phys}}$ at $\{1\}\times \mathcal{M}_d$ encodes the dynamics and non-topological information of $\mathfrak{T}_{\mathcal{S}}$. To study the $d$-dimensional gapped phase of symmetry $\mathcal{S}$, we need to choose the physical boundary $\mathcal{B}_{\text{phys}}$ to be also topological, so that the SymTFT construction gives a $d$-dimensional TQFT. By changing the physical boundaries, we will obtain various gapped phases, including spontaneous symmetry-breaking (SSB) phases and SPT phases.

On the other hand, the notion of global symmetry, which is traditionally mathematically modeled by a group, has been generalized by increasing the codimension of the manifold supporting the symmetry and by relaxing the conditions of invertibility and unitarity. The manifold that supports a group symmetry can be extended to a higher codimension, giving rise to higher-form symmetry \cite{gaiotto2015generalized, kapustin2017higher, gomes2023introduction, Bhardwaj2024lecture}. The group algebra can be generalized to fusion categories and higher fusion categories, leading to non-invertible symmetries \cite{cordova2022snowmass, brennan2023introduction, mcgreevy2023generalized, luo2023lecture, shao2024whats, SchaferNameki2024ICTP, Bhardwaj2024lecture}. It can also be extended to Hopf and weak Hopf algebras, as well as other quantum algebras that characterize the various symmetries a system may exhibit \cite{Buerschaper2013a, meusburger2017kitaev, chen2021ribbon, jia2023boundary, Jia2023weak, jia2024generalized, jia2024weakTube, Choi2024duality, jia2024weakhopfnoninvertible, jia2024quantumcluster, meng2024noninvertiblespt,jia2025weakhopftubealgebra,jia2025quantumclusterstatespin,inamura2023fermionization,inamura2022lattice}.
The above generalizations typically still require the symmetry operator to be topological. However, this constraint can also be relaxed. If the symmetry operator has restricted mobility, it leads to the notion of subsystem symmetry \cite{Paramekanti2002subsys, You2018SSPT, Devakul2018classifcation, Pretko2018fracton, Yan2019hyperbolic, Shirley2019foliated, Gromov2019fracton, Devakul2019fractal, You2020fractonicBF, Devakul2020planar, Ibieta-Jimenez2020fracton, Tantivasadakarn2020SSPT,Seiberg2020subsym, Tantivasadakarn2020JW,Seiberg2021fraton, Distler2022subsym, Yamaguchi2022gapless, Stahl2022multipole, Gorantla2022global, Katsura2022subsym, Burnell2022anomaly, Cao2022boson, Cao2023subsystem, Yamaguchi2023sl2, Cao2024SymTFT, Pace2025gauging,Ohmori2023Foliated,Spieler2023foliated}.

Subsystem symmetries can act on various submanifolds (leaves of a foliation), which may exhibit intricate geometries such as lines, planes, or even fractals.
Subsystem symmetries have attracted considerable attention, particularly due to their connections with fracton phases, foliated quantum field theory, higher-order topological phases, and measurement-based quantum computation. Fracton models can be realized by gauging subsystem symmetries \cite{Shirley2019foliated,Ibieta-Jimenez2020fracton,Devakul2019fractal,Devakul2020planar}. The corresponding quasiparticle excitations typically exhibit restricted mobility, the system possesses extensive ground state degeneracy, and the entanglement entropy usually features a large subleading correction.
Since subsystem symmetries are supported on foliated submanifolds, the corresponding effective field theory becomes a foliated quantum field theory \cite{You2020fractonicBF,Seiberg2021fraton,Distler2022subsym,Yamaguchi2022gapless,Burnell2022anomaly,Ohmori2023Foliated,Spieler2023foliated,Cao2023subsystem,Cao2024SymTFT}. For higher-order topological phases, protected edge states are not confined to the boundaries of the system but are instead located on higher-dimensional boundaries. These higher-order edge modes can also be protected by subsystem symmetries \cite{May-Mann2022higherorder,Zhang2023classification}.
Certain subsystem symmetry-protected quantum phases have been shown to possess computational universality for measurement-based quantum computation. Moreover, by utilizing quantum cellular automata, one can engineer more exotic types of subsystem symmetries—such as fractal subsystem symmetries—and investigate their rich physical consequences \cite{stephen2019subsystem,Zhang2024subsymSPT}.
It is natural to generalize familiar concepts associated with global symmetries to the realm of subsystem symmetries, including selection rules \cite{Gorantla2021lowenergy}, spontaneous symmetry breaking \cite{Qi2021fracton,Distler2022subsym,Rayhaun2023subsysm}, and anomaly inflow \cite{Burnell2022anomaly}, duality \cite{Cao2022boson,Cao2023subsystem,Cao2024SymTFT,Ohmori2023Foliated}, among others.

In this work, we will primarily focus on linearly rigid subsystem symmetries.
For such symmetries, there are two distinct types of subsystem symmetry-protected topological (SSPT) phases. The first type, known as \emph{weak SSPT}, can be constructed by stacking lower-dimensional SPT phases. The second type, referred to as \emph{strong SSPT}, cannot be realized through stacking~\cite{You2018SSPT}.
The classification of SSPT phases is significantly more challenging, and currently, only partial results are available.
In Ref.~\cite{Devakul2018classifcation}, a classification of SSPT phases is provided. By introducing linear subsystem-symmetric local unitary transformations, it has been shown that $2d$ (hereinafter, we use $d$ and $D$ to denote the spatial and spacetime dimensions, respectively) strong SSPT phases are classified by
\begin{equation}\label{eq:ClassGgeneral}
    \mathcal{C}[G] = {H^{2}(G^{\times 2}, U(1))}/{\left(H^2(G, U(1))\right)^3},
\end{equation}
where $G$ is the subsystem symmetry group, and $H^2(\bullet, U(1))$ denotes the second cohomology group of the corresponding symmetry group.

Conventional SPT phases can be systematically understood and classified within the framework of SymTFT. This naturally raises the question of whether such a framework can be extended to incorporate subsystem symmetries.  In this direction, Ref.~\cite{Cao2024SymTFT} proposed a 2-foliated BF theory \cite{Spieler2023foliated,Gorantla2021fcc} as a candidate for subsystem SymTFTs, providing detailed analyses of $\bZ_2$ and $\bZ_N$ cases, including subsystem Kramers-Wannier(KW)/Jordan-Wigner(JW) dualities and emergent $SL(2,\bZ_2)$ symmetries. Building on this perspective, we aim to explore how subsystem SymTFTs can be employed to characterize and classify SSPT phases.
By carefully analyzing the topological boundary conditions of the 2-foliated BF theory, we find that SSPT phases naturally fit within the framework of subsystem SymTFT. The topological invariants derived from the field-theoretic approach show excellent agreement with those obtained from the lattice formalism \cite{Devakul2018classifcation}. This consistency provides strong evidence that the SymTFT framework effectively captures the essential features of subsystem symmetries.

This paper is organized as follows. In Section~\ref{sec:2dSPT}, we review $(1+1)$D SPT phases and their formulation within the SymTFT framework. Section~\ref{sec:subSymTFT} introduces subsystem SymTFTs based on foliated exotic tensor gauge theory.  The construction also applies to foliated BF theory. However, since foliated BF theory is equivalent to exotic tensor gauge theory via duality, and the subsystem symmetry is more manifest in the latter, we focus on exotic tensor gauge theory.
We review its canonical quantization, topological boundary conditions, and the action of the $SL(2,\mathbb{Z}_N)$ symmetry, highlighting their roles in the context of subsystem SymTFTs.
In Section~\ref{sec:SymTFTclass}, we present a classification scheme for SSPT phases using this framework. We illustrate the subsystem SymTFT picture of subsystem SSB and subsystem SPT phases through explicit examples with $\bZ_N$ and $\bZ_N \times \bZ_M$ symmetry. 
Section~\ref{sec:cluster} connects the field-theoretic constructions to lattice realizations, focusing on models derived from cluster states. 
We conclude with a summary of our results and a discussion of open questions and future research directions.

\section{Review of (1+1)D SPT Phases and SymTFT}
\label{sec:2dSPT}

In this section, we will briefly review the (1+1)D SPT phase, its edge modes, anomaly inflow, and its SymTFT construction with an \emph{abelian} global symmetry group $G$. The reader is referred to, e.g., Refs.~\cite{Chen2011classification,Chen2013SPTclassification,Pollmann2012symmetry,Schuch2011classifyingSPT,Else2014SPT,Devakul2018classifcation,Kong2020algebraic,SchaferNameki2024ICTP,Bhardwaj2024lecture} for more details.

\subsection{Edge modes and (1+1)D SPT phases}

A non-trivial SPT phase in (1+1)D can be identified by its non-trivial edge modes where the symmetry group is realized projectively \cite{Chen2011classification,Pollmann2012symmetry,Schuch2011classifyingSPT,Else2014SPT}. Suppose we have an infinite 1-dimensional lattice and bosonic degrees of freedom live at the site $i$ with $i \in \mathbb{Z}$. Each site transforms under an on-site $G$-symmetry characterized by a linear unitary representation $u_i(g)$ with $g\in G$. The symmetry operator of the global symmetry $G$ is constructed as
    \begin{equation}
        U(g) = \bigotimes_{i=-\infty}^{\infty} u_i(g)\,,\quad (g\in G)
    \end{equation}
which also form a linear unitary representation.

Let $|\Psi_{\rm GS}\rangle$ denote the unique ground state of a gapped, symmetric, local Hamiltonian defined on a circle $\mathbb{S}^1$ (i.e., in the absence of a boundary). This ground state is invariant under the global symmetry $G$ 
\begin{equation}
    U(g)\, |\Psi_{\rm GS}\rangle = |\Psi_{\rm GS}\rangle\,.
\end{equation}
Now consider a truncated symmetry operator supported on an open interval $[x_0, x_1) \subset \mathbb{S}^1$, defined as
\begin{equation}
    U_{[x_0,x_1)}(g) \equiv \bigotimes_{i = x_0}^{x_1 - 1} u_i(g),
\end{equation}
which acts nontrivially only on sites $i \in [x_0, x_1)$. We assume that the endpoints $x_0$ and $x_1$ are well separated, such that the interval length is much greater than any intrinsic correlation length of the system.
Under this assumption, the operator $U_{[x_0,x_1)}(g)$ commutes with all local terms of the Hamiltonian, except those near the endpoints $x_0$ and $x_1$. As a result, applying $U_{[x_0,x_1)}(g)$ to the ground state generally creates a pair of localized excitations near the two edges. 
We may effectively write
\begin{equation}\label{eq:Vdecompose}
    U_{[x_0,x_1)}(g) = V_{x_0}^L(g)\otimes V_{x_1}^R(g),
\end{equation}
where we have redefined $V_x^{\dagger}$ as $V_x$ for notational convenience. This expression reveals the phenomenon of \emph{symmetry fractionalization}: the global symmetry operator decomposes into localized operators associated with the edges.

Since $U_{[x_0,x_1)}(g)$ is a representation of $G$, we have $U(g_1g_2)=U(g_1)U(g_2)$, then we see that edge operators at two edges form projective representations of $G$
    \begin{equation}
        V^L_{x_0}(g_1) V^L_{x_0}(g_2) = \omega^L (g_1,g_2) V_x(g_1 g_2),\quad
         V^R_{x_1}(g_1) V^R_{x_1}(g_2) = \omega^R (g_1,g_2) V_x(g_1 g_2)\,,
    \end{equation}
with $\omega^{L,R}(g_1,g_2)$ the $U(1)$-valued anomalous phase. 
The left and right anomalous phases must cancel, $\omega^L(g_1, g_2)\, \omega^R(g_1, g_2) = 1$, so we may focus on just one of them, denoted as $\omega(g_1, g_2)$.
The associativity of the symmetry implies
\begin{equation}\label{review-projective-phase}
    \omega(g_1, g_2)\, \omega(g_1 g_2, g_3) = \omega(g_1, g_2 g_3)\, \omega(g_2, g_3),
\end{equation}
which is the 2-cocycle condition. The unit property of the group implies the normalization condition $\omega(1, g) = \omega(g, 1) = 1$.

It is important to note that this fractionalization is only well-defined up to a phase ambiguity. Specifically, under the transformation
\begin{equation}
    V^L_{x_0}(g) \mapsto \alpha^L(g)\, V^L_{x_0}(g)\,, \quad V^R_{x_1}(g) \mapsto \alpha^R(g)\, V^R_{x_1}(g)\,, \quad (\alpha^L(g)\,, \alpha^R(g) \in \mathrm{U}(1))
\end{equation}
with the constraint $\alpha^L(g)\, \alpha^R(g) = 1$, the operator $U_{[x_0, x_1)}(g)$ remains invariant. Hence, the fractionalized symmetry operators are determined only up to $\mathrm{U}(1)$ phase factors.
As a result, the projective phase $\omega(g_1, g_2)$ is defined only up to an equivalence relation:
\begin{equation}
    \omega(g_1, g_2) \sim \omega(g_1, g_2) \frac{\alpha(g_1 g_2)}{\alpha(g_1)\, \alpha(g_2)},
\end{equation}
where $\alpha(g)$ is an arbitrary $\mathrm{U}(1)$-valued function. The equivalence class of $\omega$ is an element of the second group cohomology $H^2(G, \mathrm{U}(1))$, which classifies the $(1+1)$D bosonic $G$-anomalies.

For abelian groups, the projective representation also implies (suppressing the superscripts $L, R$ for simplicity)
\begin{equation}\label{review-edge-algebra}
    V_x(g_1)\, V_x(g_2) = \phi(g_1, g_2)\, V_x(g_2)\, V_x(g_1),
\end{equation}
where the $\mathrm{U}(1)$ phase $\phi(g_1, g_2)$ is defined by
\begin{equation}
    \phi(g_1, g_2) = \frac{\omega(g_1, g_2)}{\omega(g_2, g_1)}.
\end{equation}
Using the cocycle condition in Eq.~\eqref{review-projective-phase}, one can verify that $\phi$ satisfies the relation
\begin{equation}
    \phi(g_1, g_2)\, \phi(g_1, g_3) = \phi(g_1, g_2 g_3),
\end{equation}
which shows that $\phi$ is multiplicative in the second argument. Moreover, using the identity $\phi(g, h) = \phi(h, g)^{-1}$, we find that $\phi$ is also multiplicative in the first argument. Therefore, $\phi$ defines a $\mathrm{U}(1)$-valued bilinear form on $G$.

From the perspective of quantum field theory, the partition function for an SPT associated to $\omega\in H^2(G,U(1))$ on the spacetime manifold $M$ is
\begin{equation}
    Z[M,\mathcal{A}]=\exp(2\pi i \int_M \mathcal{A}^* (\omega))
\end{equation}
where $\mathcal{A}: M\to BG$ is a  background $G$ gauge field and $\mathcal{A}^*$ is pullback.
The connection between a $(1+1)$D SPT phase and its $(0+1)$D boundary degrees of freedom is manifested through the mechanism of anomaly inflow.

The edge operators $V_x(g)$ can be viewed as $(0+1)$D symmetry generators localized at the boundaries, and they implement the projective group multiplication as~\cite{web:Yuji}
    \begin{equation}
    \begin{gathered}
        \begin{tikzpicture}
            \draw[line,thick] (0,0)--(0,3);
            \filldraw[black] (0,1) circle (1.5pt) node[anchor=east]{$V_x(g_2)$};
            \filldraw[black] (0,2) circle (1.5pt) node[anchor=east]{$V_x(g_1)$};            
        \end{tikzpicture}
    \end{gathered} \quad = \quad \omega(g_1,g_2)\quad \times \quad
    \begin{gathered}
        \begin{tikzpicture}
            \draw[line,thick] (0,0)--(0,3);
            \filldraw[black] (0,1.5) circle (1.5pt) node[anchor=east]{$V_x(g_1 g_2)$};
        \end{tikzpicture}
    \end{gathered}
\end{equation}
The insertions of $V_x(g)$ operator can be equivalently understood as turning on a background field of $G$-symmetry, such that any other operators charged under $G$ will transform according to $g$ when they pass through $V_x(g)$. We will continue to use this viewpoint in the following and think of a $G$-background as an insertion of symmetry operators.

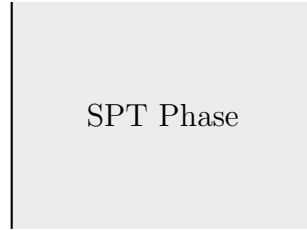
\begin{figure}[!h]
    \begin{equation}
    \begin{gathered}
        \begin{tikzpicture}
            \fill[lightgray!30!white] (0,0) rectangle (4,3);
            \draw[line,thick] (0,0)--(0,3);
            \node at (2,1.5) {SPT Phase};
        \end{tikzpicture}
    \end{gathered}\nonumber
\end{equation}
\caption{The $(1+1)$D SPT phase coupled to the $1$D boundary.}
\label{fig:2D-SPT}
\end{figure}

Suppose we couple a (1+1)D bulk on the right such that the (0+1)D system lives on the boundary. The (0+1)D $G$-background generated by $V_x(g)$ should be extended to the bulk, represented as line operators that generate the (1+1)D $G$-symmetry and end at $V_x(g)$ on the boundary. Conversely, it is equivalent to say a symmetry action along the half-space generates the boundary mode $V_x$. When we fuse the $V_x$ operators on the boundary, we also need to fuse the line operators, and it will introduce a 3-way junction in the bulk, where $g_1$ and $g_2$ fuse to a single symmetry operator $g_1 g_2$
    \begin{equation}
    \begin{gathered}
        \begin{tikzpicture}
                    \fill[lightgray!30!white] (0,0) rectangle (3,3);
            \draw[line,thick] (0,0)--(0,3);
            \filldraw[black] (0,1) circle (1.5pt) node[anchor=east]{$V_x(g_2)$};
            \filldraw[black] (0,2) circle (1.5pt) node[anchor=east]{$V_x(g_1)$};      
            \begin{scope}[line,thick,decoration={markings,mark=at position 0.75 with {\arrow{>}}}] 
            \draw [postaction={decorate}] (0,1)--(2,1);
            \draw [postaction={decorate}] (0,2)--(2,2);            
            \end{scope}
            \node at (2.25,1) {$g_2$};
            \node at (2.25,2) {$g_1$};
        \end{tikzpicture}
    \end{gathered} \quad \stackrel{?}{=} \quad 
    \begin{gathered}
        \begin{tikzpicture}
                     \fill[lightgray!30!white] (0,0) rectangle (3,3);
            \draw[line,thick] (0,0)--(0,3);
            \filldraw[black] (0,1.5) circle (1.5pt) node[anchor=east]{$V_x(g_1 g_2)$};
            \filldraw[black] (1,1.5) circle (1.5pt);
            \begin{scope}[line,thick,decoration={markings,mark=at position 0.75 with {\arrow{>}}}] 
            \draw [postaction={decorate}](0,1.5)--(1,1.5);
            \draw [postaction={decorate}] (1,1.5)--(2,2);   
            \draw [postaction={decorate}] (1,1.5)--(2,1);
            \end{scope}
            \node at (0.5,1.85) {$g_1 g_2$};
            \node at (2.25,1) {$g_2$};
            \node at (2.25,2) {$g_1$};
        \end{tikzpicture}
    \end{gathered}
\end{equation}
Recall that the fusion of two $V_x(g)$ operators will introduce an anomalous phase factor $\omega(g_1,g_2)$. However, we can attach an additional phase factor $\omega^*(g_1,g_2)$ to the junction point that cancels the phase factor on the boundary, then the whole system is anomaly-free. In other words, the anomaly at the boundary is canceled by the bulk inflow. To be concrete, let us assign the following phase factors to the junctions of symmetry generators in the bulk
    \begin{equation}
        \begin{gathered}
        \begin{tikzpicture}
                 \fill[lightgray!30!white] (0,.5) rectangle (2.5,2.5);
        \begin{scope}[line,thick,decoration={markings,mark=at position 0.75 with {\arrow{>}}}] 
            \draw[postaction={decorate}] (0,1.5)--(1,1.5);
            \draw[postaction={decorate}] (1,1.5)--(2,2);
            \draw[postaction={decorate}] (1,1.5)--(2,1);
        \end{scope}
            \filldraw[black] (1,1.5) circle (1.5pt);
            \node at (0.5,1.85) {$g_1 g_2$};
            \node at (2.25,1) {$g_2$};
            \node at (2.25,2) {$g_1$};
        \end{tikzpicture}        
        \end{gathered}: \omega^*(g_1,g_2)\,, \qquad
        \begin{gathered}
        \begin{tikzpicture}
         \fill[lightgray!30!white] (-0.6,.5) rectangle (2,2.5);
        \begin{scope}[line,thick,decoration={markings,mark=at position 0.5 with {\arrow{>}}}] 
            \draw[postaction={decorate}] (1,1.5)--(2,1.5);
            \draw[postaction={decorate}] (0,2)--(1,1.5);
            \draw[postaction={decorate}] (0,1)--(1,1.5);
        \end{scope}
            \filldraw[black] (1,1.5) circle (1.5pt);
            \node at (1.5,1.85) {$g_1 g_2$};
            \node at (-0.25,1) {$g_2$};
            \node at (-0.25,2) {$g_1$};
        \end{tikzpicture}        
        \end{gathered} : \omega(g_1,g_2)\,,
    \end{equation}
where each line operator flows from left to right. 
Then the bulk theory with such a phase factor imposed on the junctions is a (1+1)D SPT phase shown in figure~\ref{fig:2D-SPT}.

To illustrate this, let us consider the partition functions of the $(1+1)$D SPT phase on the tours with a background field represented by a network of symmetry operators. Consider the torus with periodic coordinates $(x,y)\sim(x+2\pi,y)\sim(x,y+2\pi)$,
we will denote the 1-cycles along $x$ and $y$ directions separately as $\Gamma_1$ and $\Gamma_2$. Turning on the backgrounds of $G$-symmetry on the torus is equivalent to assigning the holonomies along the two 1-cycles
    \begin{equation}
        \textrm{Hol}(\Gamma_1) = g\,, \quad \textrm{Hol}(\Gamma_2) = h\,, \quad
    \end{equation}
with $g,h\in G$ and $gh=hg$. Since we assume $G$ is abelian, the constraint $gh=hg$ is automatically satisfied. We can represent the holonomies using the following networks
    \begin{equation}
        \begin{gathered}
        \begin{tikzpicture}
            \draw[very thick] (0,0) rectangle (2,2);
        \begin{scope}[line,thick,decoration={markings,mark=at position 0.5 with {\arrow{<}}}] 
            \draw[postaction={decorate}] (1,2)--(0.8,1.2);
            \draw[postaction={decorate}] (0,1)--(0.8,1.2);
            \draw[postaction={decorate}] (0.8,1.2)--(1.2,0.8);
            \draw[postaction={decorate}] (1.2,0.8)--(1,0);
            \draw[postaction={decorate}] (1.2,0.8)--(2,1); 
        \end{scope}
            \node at (1,-0.4) {$x$};
            \node at (-0.4,1) {$y$};           
            \node at (0.3,1.4) {$h$};
            \node at (1.2,1.6) {$g$};
            \node at (0.75,0.75) {$gh$};
            \node at (1.7,0.7) {$h$};
            \node at (1.3,0.3) {$g$};
        \end{tikzpicture}
        \end{gathered}
    \end{equation}
We will write down the torus partition function $Z(g,h)$ as a function of the two holonomies, and we assume the partition function of the trivial sector $g,h=1$ is one 
    \begin{equation}
        Z(1,1) = 1\,,
    \end{equation}
so that there exists a unique ground state. The partition function with a general background is read from the junctions of the line operators as
    \begin{equation}
        Z(g,h) = \omega(g,h) \omega^*(h,g) = \frac{\omega(g,h)}{\omega(h,g)} = \phi(g,h)\,.
    \end{equation}
To get some intuition of the partition function, if we set $h=1$ we have
    \begin{equation}
        Z(g,1) = 1\,,
    \end{equation}
where we use $\phi(1,g)=\phi(g,1)=1$. Since $g$ is the symmetry defect inserted along the time direction, it indicates a unique ground state $|\psi_g\rangle$ in the twist (defect) Hilbert space $\mathcal{H}_g$. Moreover, with $h$ inserted as a symmetry operator along the spatial direction, the partition function is understood as
    \begin{equation}\label{Partition-function-SPT}
        Z(g,h) = \langle \psi_g| U(h) |\psi_g\rangle = \phi(g,h)\,,
    \end{equation}
where we act the symmetry generator $U(h)$ on the ground state $|\psi_g\rangle$. Therefore, $\phi(g,h)$ is the phase generated by the symmetry transformation, and it gives the $G$-charge carried by the ground state $|\psi_g\rangle$. We will see how that is interpreted in the SymTFT picture in the following section.

\subsection{Review of the (2+1)D SymTFT}

\begin{figure}[!h]
    \begin{equation}
            \begin{gathered}
        \begin{tikzpicture}
            \draw[thick] (0,0)--(0,3);
            \draw[thick] (0,3)--(2,3.5);
            \draw[thick] (2,3.5)--(2,0.5);
            \draw[thick] (2,0.5)--(0,0);
            \filldraw[red] (1,1.75) circle (1.5pt); 
            \node at (1.35,1.5) {$\psi$};
            \draw[thick,blue] (0,2)--(2,2.5);
            \node at (1.5,2.75) {$U(g)$};
            \node at (1,-0.5) {$\mathcal{M}_2$};
            \node at (4,1.75) {$\Leftrightarrow$};
        \end{tikzpicture}\\
        \Updownarrow\qquad \qquad \qquad \\
        \begin{tikzpicture}
            \draw[thick] (0,0)--(0,3);
            \draw[thick] (0,3)--(2,3.5);
            \draw[thick] (2,3.5)--(2,0.5);
            \draw[thick] (2,0.5)--(0,0);
            \filldraw[red] (1,1.75) circle (1.5pt); 
            \node at (1.35,1.5) {$\rho_g \psi$};
            \draw[thick,blue] (0,0.75)--(2,1.25);
            \node at (1.5,0.75) {$U(g)$};
            \node at (1,-0.5) {$\mathcal{M}_2$};
            \node at (4,1.75) {$\Leftrightarrow$};
        \end{tikzpicture}
    \end{gathered} \qquad \qquad
            \begin{gathered}
        \begin{tikzpicture}
            \draw[thick] (0,0)--(0,3);
            \draw[thick] (0,3)--(2,3.5);
            \draw[thick] (2,3.5)--(2,0.5);
            \draw[thick] (2,0.5)--(0,0);
            \draw[thick] (4,0)--(4,3);
            \draw[thick] (4,3)--(6,3.5);
            \draw[thick] (6,3.5)--(6,0.5);
            \draw[thick] (6,0.5)--(4,0);
            \draw[thick] (0,0)--(4,0);
            \draw[thick] (0,3)--(4,3);
            \draw[thick] (2,3.5)--(6,3.5);
            \draw[thick] (2,0.5)--(4,0.5);
            \draw[thick,dashed] (4,0.5)--(6,0.5);
            \draw[thick,red] (1,1.75)--(4,1.75);
            \draw[thick,red,dashed] (4,1.75)--(5,1.75);
            \filldraw[red] (1,1.75) circle (1.5pt);
            \filldraw[red] (5,1.75) circle (1.5pt); 
            \node at (3,2) {$W$};
            \draw[thick,blue] (0,2)--(2,2.5);
            \node at (1.5,2.75) {$U(g)$};
            \node at (5,1.25) {$\widetilde{\psi}$};
            \node at (1,-0.5) {$\mathcal{B}_{\textrm{top}}$};
            \node at (1.35,1.5) {$v$};
            \node at (5,-0.5) {$\mathcal{B}_{\textrm{phys}}$};
        \end{tikzpicture}\\
        \Updownarrow\\
        \begin{tikzpicture}
            \draw[thick] (0,0)--(0,3);
            \draw[thick] (0,3)--(2,3.5);
            \draw[thick] (2,3.5)--(2,0.5);
            \draw[thick] (2,0.5)--(0,0);
            \draw[thick] (4,0)--(4,3);
            \draw[thick] (4,3)--(6,3.5);
            \draw[thick] (6,3.5)--(6,0.5);
            \draw[thick] (6,0.5)--(4,0);
            \draw[thick] (0,0)--(4,0);
            \draw[thick] (0,3)--(4,3);
            \draw[thick] (2,3.5)--(6,3.5);
            \draw[thick] (2,0.5)--(4,0.5);
            \draw[thick,dashed] (4,0.5)--(6,0.5);
            \draw[thick,red] (1,1.75)--(4,1.75);
            \draw[thick,red,dashed] (4,1.75)--(5,1.75);
            \filldraw[red] (1,1.75) circle (1.5pt);
            \filldraw[red] (5,1.75) circle (1.5pt); 
            \node at (3,2) {$W$};
            \draw[thick,blue] (0,0.75)--(2,1.25);
            \node at (1.5,0.75) {$U(g)$};
            \node at (5,1.25) {$\widetilde{\psi}$};
            \node at (1.35,1.5) {$\rho_g v$};
            \node at (1,-0.5) {$\mathcal{B}_{\textrm{top}}$};
            \node at (5,-0.5) {$\mathcal{B}_{\textrm{phys}}$};
        \end{tikzpicture}
    \end{gathered}\nonumber 
    \end{equation}
    \caption{The SymTFT description of the (1+1)D theory on $\mathcal{M}_2$ with symmetry $G$.}
    \label{Fig-SymTFT}
\end{figure}
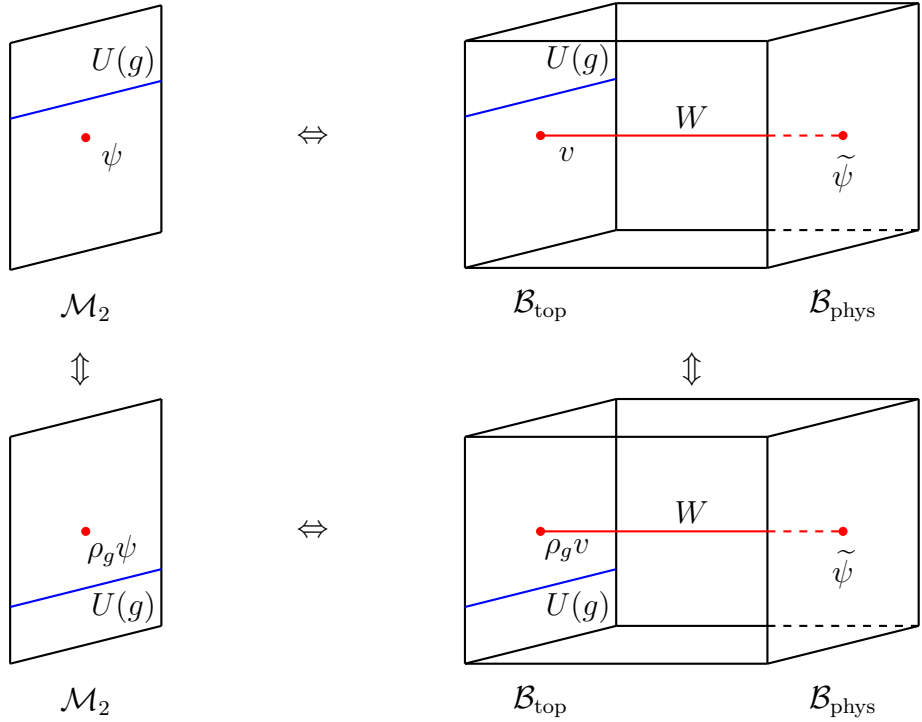

In this section, we will give a brief review of (2+1)D SymTFT with $G$-symmetry. The essence of SymTFT is to expand a $(1+1)$D theory $\mathfrak{T}_G$ with symmetry $G$ on $\mathcal{M}_2$ to a (2+1)D TQFT living on $[0,1] \times \mathcal{M}_2$ with two boundaries $\{0\} \times \mathcal{M}_2$ and $\{1\} \times \mathcal{M}_2$, as depicted in Figure~\ref{Fig-SymTFT}. The information on symmetry and dynamics are separately stored in the topological boundary $\mathcal{B}_{\textrm{top}}=\{0\} \times \mathcal{M}_2$ and the physical boundary $\mathcal{B}_{\textrm{phys}}=\{1\} \times \mathcal{M}_{2}$. The symmetry generator $U(g)$ lives on the topological boundary $\mathcal{B}_{\textrm{top}}$, and a local operator $\psi$ is represented as a bulk line operator $W$ (anyon string) stretching between a local operator $\widetilde{\psi}$ on $\mathcal{B}_{\textrm{phys}}$ and a topological endpoint $v$ at $\mathcal{B}_{\textrm{top}}$ 
which carries the vector space of a given representation $\rho$ of $G$
    \begin{equation}
        \psi = v W \widetilde{\psi}\,.
    \end{equation}  
The symmetry transformation on $\psi$ is
    \begin{equation}
        U(g) \psi U^{\dagger}(g) = \rho(g)\psi \quad \Leftrightarrow \quad U(g) \psi = \rho(g) \psi U(g)\,,
    \end{equation}
and it can be understood as passing the $U(g)$ operator through $\psi$ as shown on the LHS in Figure~\ref{Fig-SymTFT}. In the SymTFT picture, the symmetry action is performed on $\mathcal{B}_{\textrm{top}}$, and the endpoint $v$ will transform accordingly as $\rho(g) v$.

Given a topological boundary $\mathcal{B}_{\textrm{top}}$, not every line operator in the bulk can end on $\mathcal{B}_{\textrm{top}}$, and some of them will transit to symmetry generators living on the boundary. We will define the topological boundary $\mathcal{B}_{\textrm{top}}$ via the collection of line operators in the bulk that can end at the boundary simultaneously and consistently. We can introduce a composite line operator $\mathcal{L}$ and write
    \begin{equation}
        \mathcal{L} = \bigoplus_{\alpha} n_{\alpha} W_{\alpha}\,,\quad (n_{\alpha} \in \mathbb{Z}_{\geq 0})
    \end{equation}
with $n_{\alpha}$ non-negative integers equaling the dimension of the endpoint $v_{\alpha}$. In particular, if $n_{\alpha}=0$, the corresponding line operator $W_{\alpha}$ cannot end on $\mathcal{B}_{\textrm{top}}$. Mathematically, $\mathcal{L}$ is the algebraic object of a Lagrangian algebra (see e.g., \cite{Kong2014}) of the SymTFT, and the topological boundary $\mathcal{B}_{\textrm{top}}$ is one-to-one corresponding to the choice of Lagrangian algebras of the SymTFT. On the other hand, the physical boundary is determined by the detailed dynamics of the theory.

From now on, we will assume the symmetry $G$ is abelian. To illustrate the basic idea of SymTFT, we consider a $(1+1)$D theory $\mathfrak{T}_{\mathbb{Z}_N}$ with the anomaly-free $\mathbb{Z}_N$ symmetry. The corresponding SymTFT $\mathcal{Z}(\mathbb{Z}_N)$ is the $(2+1)$D BF theory with level $N$
    \begin{equation}
        S_{BF}[\hat{A},A] = \frac{N}{2\pi} \int \hat{A} \wedge d A\,,
        \end{equation}
where $\hat{A},A$ are $U(1)$ 1-form gauge fields. It is a $\mathbb{Z}_N$ gauge theory and is the low-energy description of the toric code for $N=2$ in the condensed matter literature. The gauge invariant operators are Wilson loops defined as
    \begin{equation}
        W[\Gamma] = \exp \left(i \oint_{\Gamma} A \right)\,,\quad \hat{W}[\Gamma] = \exp \left(i \oint_{\Gamma} \hat{A} \right)\,,
    \end{equation}
with $\Gamma \in H_1(T^2,\mathbb{Z})$. 
The holonomies of $A$ and $\hat{A}$ are quantized as,
\begin{equation}
    \frac{N}{2\pi} \oint_{\Gamma} A =0,1,\cdots,N-1\,,\quad \frac{N}{2\pi} \oint_{\Gamma} \hat{A} = 0,1,\cdots,N-1\,,
\end{equation}
and one has $W^N[\Gamma] = \hat{W}^N[\Gamma] = 1$. The two kinds of Wilson loops have a non-trivial linking rule given by
    \begin{equation}
        \langle W[\Gamma] \hat{W}[\Gamma'] \rangle = \omega^{-\textrm{Link}(\Gamma,\Gamma')}\,,
    \end{equation}
where $\omega=e^{\frac{2\pi i}{N}}$ is the $N$-th root of unity and the $\operatorname{Link}(\Gamma,\Gamma')$ is the linking number between $\Gamma$ and $\Gamma'$. One way to see that is to fix a gauge $A_0=\hat{A}_0=0$ and consider the canonical quantization
    \begin{equation}
        \left[A_i(x,y) , \hat{A}_j(x',y') \right] =  \frac{2\pi i}{N} \epsilon_{ij} \delta(x-x',y-y')\,,
    \end{equation}
where $A$ and $\hat{A}$ are conjugated with each other like position and momentum. 
For simplicity, we place the BF theory on a spatial torus $T^2$, and the operators satisfy the commutation relation
\begin{equation}
    W[\Gamma_i] \hat{W}[\Gamma_j] = \omega^{-\int \gamma_i \wedge \gamma_j} \hat{W}[\Gamma_j] W[\Gamma_i]\,,
\end{equation}
where $\gamma \in H^1(T^2,\mathbb{Z})$ is the Poincare dual of the 1-cycle $\Gamma$ defined as $\oint_{\Gamma} \alpha = \int \gamma \wedge \alpha$ for any $1$-form $\alpha$.

There exist two kinds of bosonic topological boundary $\mathcal{B}_{\textrm{top}}$.
\begin{itemize}
    \item The Dirichlet boundary $\mathcal{B}_{\textrm{Dir}}$ for $A$ where $W$ can end on the boundary and the Lagrangian algebra is $\mathcal{L}_{\textrm{Dir}}= \bigoplus_{i=0}^{N-1} W^i$. On the other hand, $\hat{W}$ will survive and serve as the $\mathbb{Z}_N$ generator at the boundary. 
    \begin{equation}
            \begin{gathered}
        \begin{tikzpicture}
            \draw[line,thick] (0,0)--(0,3);
            \draw[line,thick] (0,3)--(2,3.5);
            \draw[line,thick] (2,3.5)--(2,0.5);
            \draw[line,thick] (2,0.5)--(0,0);
            \draw[line,thick,red] (1,1.75)--(4,1.75);
            \filldraw[red] (1,1.75) circle (1.5pt);  
            \draw[thick,blue] (0,2)--(2,2.5);
            \node at (3,2) {$W$};
            \node at (1.5,2.75) {$\hat{W}$};
        \end{tikzpicture}
    \end{gathered}\nonumber
    \end{equation}
On the torus, we can consider a Dirichlet boundary state $|a\rangle$ which satisfies
            \begin{equation}\label{Z2-Dirichlet-Boundary-State}
                \left\{ \begin{array}{l}
                    W[\Gamma] |a\rangle = \omega^{\int \gamma \wedge a} |a\rangle\,,\\
                    \hat{W}[\Gamma] |a\rangle = |a - \gamma\rangle\,,
                \end{array}\right.
            \end{equation}
where $2\pi a =NA  = a_x dx + a_y dy \in H^2(T^2,\mathbb{Z}_2)$ whose components $(a_x,a_y)$ are the holonomies of $A$ along $\Gamma_1$ and $\Gamma_2$.

    \item The Neumann boundary $\mathcal{B}_{\textrm{Neu}}$ for $A$ where $\hat{W}$ can end at the boundary and the Lagrangian algebra is $\mathcal{L}_{\textrm{Neu}}= \bigoplus_{i=0}^{N-1} \hat{W}^i$. Similarly, $W$ will become the $\mathbb{Z}_N$ generators at the boundary. 
    \begin{equation}
            \begin{gathered}
        \begin{tikzpicture}
            \draw[line,thick] (0,0)--(0,3);
            \draw[line,thick] (0,3)--(2,3.5);
            \draw[line,thick] (2,3.5)--(2,0.5);
            \draw[line,thick] (2,0.5)--(0,0);
            \draw[line,thick,blue] (1,1.75)--(4,1.75);
            \filldraw[blue] (1,1.75) circle (1.5pt);  
            \draw[thick,red] (0,2)--(2,2.5);
            \node at (3,2.05) {$\hat{W}$};
            \node at (1.5,2.75) {$W$};
        \end{tikzpicture}
    \end{gathered}    \nonumber
    \end{equation}
The Neumann boundary state $|\hat{a}\rangle$ on the torus satisfies
            \begin{equation}\label{Z2-Neumann-Boundary-State}
                \left\{ \begin{array}{l}
                    \hat{W}[\Gamma] |\hat{a}\rangle = \omega^{\int \gamma \wedge \hat{a}} |\hat{a}\rangle\,,\\
                    W[\Gamma] |\hat{a}\rangle = |\hat{a} - \gamma\rangle\,,
                \end{array}\right.
            \end{equation}
where $2\pi \hat{a} =N \hat{A}  = \hat{a}_x dx + \hat{a}_y dy \in H^2(T^2,\mathbb{Z}_2)$ whose components $(\hat{a}_x,\hat{a}_y)$ are the holonomies of $\hat{A}$ along $\Gamma_1$ and $\Gamma_2$. It is the discrete Fourier transformation of the Dirichlet boundary state
    \begin{equation}
        |\hat{a}\rangle = \frac{1}{N} \sum_{a\in H^1(T^2,\mathbb{Z}_N)} \omega^{\int a \wedge \hat{a}}|a\rangle\,.
    \end{equation}
    \end{itemize}

On the other hand, the physical boundary  $\mathcal{B}_{\textrm{phys}}$ is characterized by the state vector $|\chi\rangle$ on the torus which depends on the partition function $Z_{\mathfrak{T}_{\mathbb{Z}_N}}[a]$ of theory $\mathfrak{T}_{\mathbb{Z}_N}$ as
    \begin{equation}
        |\chi\rangle = \sum_{a \in H^1(T^2,\mathbb{Z}_N)} Z_{\mathfrak{T}_{\mathbb{Z}_N}}[a] |a\rangle.
    \end{equation}
Choosing different topological boundaries, the path integral of the BF theory on the slab gives,
    \begin{equation}
        Z[a] = \langle a | e^{iHt} |\chi\rangle = \langle a  |\chi\rangle,\quad Z[\hat{a}] = \langle \hat{a} | e^{iHt} |\chi \rangle = \langle \hat{a} | \chi \rangle\,,
    \end{equation}
where the Hamiltonian of the topological theory is zero. Here $Z[a] = Z_{\mathfrak{T}_{\mathbb{Z}_N}}[a] $ agrees with the torus partition function of $\mathfrak{T}_{\mathbb{Z}_N}$ and
    \begin{equation}
        Z[\hat{a}] = \frac{1}{N} \sum_{a} \omega^{\int \hat{a} \wedge a} Z_{\mathfrak{T}_{\mathbb{Z}_N}}[a] \equiv Z_{\mathfrak{T}_{\mathbb{Z}_N}/\mathbb{Z}_N}[\hat{a}],
    \end{equation}
which is the partition function of the orbifold theory $\mathfrak{T}_{\mathbb{Z}_N}/\mathbb{Z}_N$, the Kramers-Wannier duaity of $\mathfrak{T}_{\mathbb{Z}_N}$. In other words, the $\mathbb{Z}_N$ gauging of $\mathfrak{T}_{\mathbb{Z}_N}$ can be viewed from the SymTFT as switching the topological boundary $\mathcal{B}_{\textrm{Dir}}$ to $\mathcal{B}_{\textrm{Neu}}$.

When $N=2$, there also exists a fermionic topological boundary $\mathcal{B}_{\textrm{Fer}}$ such that the combination $W\hat{W}$ can end at the boundary and the fermionic Lagrangian algebra  is $\mathcal{L}_{f} = 1 \oplus W \hat{W}$ \cite{Wan2017fermionic,bhardwaj2024fermionic}
    \begin{equation}
            \begin{gathered}
        \begin{tikzpicture}
            \draw[line,thick] (0,0)--(0,3);
            \draw[line,thick] (0,3)--(2,3.5);
            \draw[line,thick] (2,3.5)--(2,0.5);
            \draw[line,thick] (2,0.5)--(0,0);
            \draw[line,thick,purple] (1,1.75)--(4,1.75);
            \filldraw[purple] (1,1.75) circle (1.5pt);  
            \draw[thick,blue] (0,2)--(2,2.5);
            \node at (3,2) {$W\hat{W}$};
            \node at (1.4,2.75) {$W/\hat{W}$};
        \end{tikzpicture}
    \end{gathered}    \nonumber
    \end{equation}
and the symmetry generator $(-1)^F$ at the boundary can be either $W$ or $\hat{W}$. The topological boundary states are denoted as $|s\rangle$, where $s= s_x dx + s_y dy \in H^1(T^2,\mathbb{Z}_2)$ and the $\mathbb{Z}_2$ components $(s_x,s_y)$ can be understood as the spin structure on torus ($s=0$ is chosen to be anti-periodic)
        \begin{equation}\label{review-fermionic-boundary-state}
            \left\{ \begin{array}{l}
                W_F[\Gamma] |s\rangle = (-1)^{\textrm{Arf}(s+\gamma) - \textrm{Arf}(s)} |s\rangle\,,\\
                \hat{W}[\Gamma] |s\rangle = |s + \gamma\rangle\,,
              \end{array}\right.
        \end{equation}
where we choose $\hat{W}$ as the generator $(-1)^F$, and we will consider the other choice in the following section when we discuss the fermionic SPT phase. Here $\textrm{Arf}(s) \equiv s_x s_y$ is the Arf-invariant. The topological boundary state $|s\rangle$ can also be expressed as,
    \begin{equation}
        |s\rangle = \frac{1}{2} \sum_{a \in H^1(T^2,\mathbb{Z}_2)} (-1)^{\textrm{Arf}(s+a)}|a\rangle\,,
    \end{equation}
and the transition amplitude $\langle s | \chi \rangle$ is
\begin{equation}
    Z_{F}[s] = \langle s | \chi \rangle = \frac{1}{2} \sum_{a\in H^1(T^2,\mathbb{Z}_2)} (-1)^{\textrm{Arf}(s+a)} Z_{\mathfrak{T}_{\mathbb{Z}_2}}[a]\,,
\end{equation}
which gives the partition function of the fermionic theory after the Jordan-Wigner transformation.

\subsection{Gapped phase from SymTFT viewpoint}

We can also interpret the (1+1)D gapped phases in the SymTFT picture, and the strategy is as follows: Given any $(1+1)$D symmetry $G$, we choose a topological boundary $\mathcal{B}_{\textrm{top}}$ which supports the $G$-symmetry and is represented by a Lagrangian algebra $\mathcal{L}_{\textrm{top}}$. For the physical boundary $\mathcal{B}_{\textrm{phys}}$, we will also set that to be topological and is determined by another Lagrangian algebra $\mathcal{L}_{\textrm{phys}}$. After we shrink the interval, we have different (1+1)D topological field theories depending on the choice of $\mathcal{B}_{\textrm{phys}}$ and $\mathcal{L}_{\textrm{phys}}$. They are the candidates of the $(1+1)$D gapped phases.

\begin{figure}[!h]
    \begin{equation}
            \begin{gathered}
        \begin{tikzpicture}
            \draw[line,thick] (0,0)--(0,3);
            \draw[line,thick] (0,3)--(2,3.5);
            \draw[line,thick] (2,3.5)--(2,0.5);
            \draw[line,thick] (2,0.5)--(0,0);
            \draw[line,thick] (4,0)--(4,3);
            \draw[line,thick] (4,3)--(6,3.5);
            \draw[line,thick] (6,3.5)--(6,0.5);
            \draw[line,thick] (6,0.5)--(4,0);
            \draw[thick] (0,0)--(4,0);
            \draw[thick] (0,3)--(4,3);
            \draw[thick] (2,3.5)--(6,3.5);
            \draw[thick] (2,0.5)--(4,0.5);
            \draw[thick,dashed] (4,0.5)--(6,0.5);
            \begin{scope}[line,thick,decoration={markings,mark=at position 0.45 with {\arrow{<}}}] 
            \draw[red,postaction={decorate}] (1,1.75)--(4,1.75);
            \draw[blue,postaction={decorate}] (2,1.75+0.25)--(1,1.75);
            \draw[line,thick,red,dashed] (4,1.75)--(5,1.75);
        \end{scope}
            \filldraw[red] (1,1.75) circle (1.5pt);
            \filldraw[red] (5,1.75) circle (1.5pt); 
            \node at (3,2) {$W$};
            \node at (1.5,1.75+0.75) {$U(g)$};
            \node at (1,1.25) {$v_g$};
            \node at (1,-0.5) {$\mathcal{B}_{\textrm{top}}$};
            \node at (5,-0.5) {$\mathcal{B}_{\textrm{phys}}$};
            \node at (7,1.5) {$\Leftrightarrow$};
        \end{tikzpicture}\\ \Updownarrow \\
        \begin{tikzpicture}
            \draw[line,thick] (0,0)--(0,3);
            \draw[line,thick] (0,3)--(2,3.5);
            \draw[line,thick] (2,3.5)--(2,0.5);
            \draw[line,thick] (2,0.5)--(0,0);
            \draw[line,thick] (4,0)--(4,3);
            \draw[line,thick] (4,3)--(6,3.5);
            \draw[line,thick] (6,3.5)--(6,0.5);
            \draw[line,thick] (6,0.5)--(4,0);
            \draw[thick] (0,0)--(4,0);
            \draw[thick] (0,3)--(4,3);
            \draw[thick] (2,3.5)--(6,3.5);
            \draw[thick] (2,0.5)--(4,0.5);
            \draw[thick,dashed] (4,0.5)--(6,0.5);
            \begin{scope}[line,thick,decoration={markings,mark=at position 0.45 with {\arrow{<}}}] 
            \draw[red,postaction={decorate}] (1,1.75)--(4,1.75);
            \draw[blue,postaction={decorate}] (1,3.25)--(1,1.75);
            \draw[line,thick,red,dashed] (4,1.75)--(5,1.75);
        \end{scope}
            \filldraw[red] (1,1.75) circle (1.5pt);
            \filldraw[red] (5,1.75) circle (1.5pt); 
            \node at (3,2) {$W$};
            \node at (1.5,1.75+0.75) {$U(g)$};
            \node at (1,1.25) {$v_g$};
            \node at (1,-0.5) {$\mathcal{B}_{\textrm{top}}$};
            \node at (5,-0.5) {$\mathcal{B}_{\textrm{phys}}$};
            \node at (7,1.5) {$\Leftrightarrow$};
        \end{tikzpicture}
    \end{gathered} \qquad 
            \begin{gathered}
        \begin{tikzpicture}
            \draw[line,thick] (0,0)--(0,3);
            \draw[line,thick] (0,3)--(2,3.5);
            \draw[line,thick] (2,3.5)--(2,0.5);
            \draw[line,thick] (2,0.5)--(0,0);
            \begin{scope}[line,thick,decoration={markings,mark=at position 0.45 with {\arrow{<}}}] 
            \draw[blue,postaction={decorate}] (2,1.75+0.25)--(1,1.75);
            \end{scope}
            \filldraw[red] (1,1.75) circle (1.5pt); 
            \node at (1.5,1.75+0.75) {$U(g)$};
            \node at (1,1.25) {$\psi_g$};
            \node at (1,-0.5) {$\mathcal{M}_2$};
        \end{tikzpicture} \\ \Updownarrow \\
        \begin{tikzpicture}
            \draw[line,thick] (0,0)--(0,3);
            \draw[line,thick] (0,3)--(2,3.5);
            \draw[line,thick] (2,3.5)--(2,0.5);
            \draw[line,thick] (2,0.5)--(0,0);
            \begin{scope}[line,thick,decoration={markings,mark=at position 0.45 with {\arrow{<}}}] 
            \draw[blue,postaction={decorate}] (1,3.25)--(1,1.75);
            \end{scope}
            \filldraw[red] (1,1.75) circle (1.5pt); 
            \node at (1.5,1.75+0.75) {$U(g)$};
            \node at (1,1.25) {$\psi_g$};
            \node at (1,-0.5) {$\mathcal{M}_2$};
        \end{tikzpicture}
    \end{gathered} \nonumber
    \end{equation}
    \caption{The twist operator in the SymTFT picture.}
    \label{SymTFT-Twist-Operator}
    \end{figure}
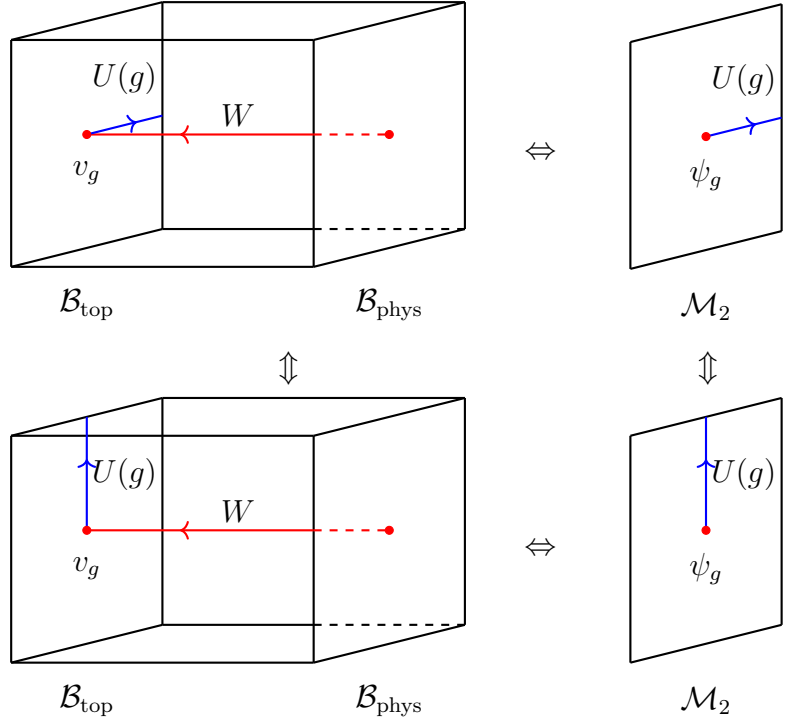

To extract the information of the gapped phases, we need to examine the line operators $W \in \mathcal{L}_{\textrm{phys}}$, which can end on $\mathcal{B}_{\textrm{phys}}$ and provide physical degrees of freedom. If $W \notin \mathcal{L}_{\textrm{top}}$, it cannot end on $\mathcal{B}_{\textrm{top}}$ and must transit to a symmetry generator $U(g)$ on the topological boundary. After we shrink the interval, it gives a (1+1)D twist (disorder) operator $\psi_g$ attached to the symmetry generator $U(g)$ in the TQFT, as shown in Figure~\ref{SymTFT-Twist-Operator}.

Let the vertical direction in Figure~\ref{SymTFT-Twist-Operator} be the time direction. We can choose the direction of the symmetry operator $U(g)$ along either spatial or time direction. For the former choice, we have a symmetry generator $U(g)$ truncated by a local operator $\psi_g$ acting on the Hilbert space $\mathcal{H}$. On the other hand, if we choose $U(g)$ lying along the time direction as a defect line, then $\psi_g$ is an operator which maps $\mathcal{H}$ to the defect Hilbert space $\mathcal{H}_g$. 

We adopt the first interpretation that the twist operator is a truncated symmetry generator $U(g)$ acting only on half of the space. At the beginning of this section, we consider acting symmetry $U(g)$ along an interval $[x_0,x_1)$ on the lattice. Suppose we have an infinite 1-dimensional lattice and send $x_1 \rightarrow +\infty$, it matches the twist operator we consider in Figure~\ref{SymTFT-Twist-Operator} and the edge mode $V_x(g)$ is carried by $\psi_g$.

\begin{figure}[!h]
    \begin{equation}
            \begin{gathered}
        \begin{tikzpicture}
            \draw[line,thick] (0,0)--(0,3);
            \draw[line,thick] (0,3)--(2,3.5);
            \draw[line,thick] (2,3.5)--(2,0.5);
            \draw[line,thick] (2,0.5)--(0,0);
            \draw[line,thick] (4,0)--(4,3);
            \draw[line,thick] (4,3)--(6,3.5);
            \draw[line,thick] (6,3.5)--(6,0.5);
            \draw[line,thick] (6,0.5)--(4,0);
            \draw[thick] (0,0)--(4,0);
            \draw[thick] (0,3)--(4,3);
            \draw[thick] (2,3.5)--(6,3.5);
            \draw[thick] (2,0.5)--(4,0.5);
            \draw[thick,dashed] (4,0.5)--(6,0.5);
            \begin{scope}[line,thick,decoration={markings,mark=at position 0.45 with {\arrow{<}}}] 
            \draw[red,postaction={decorate}] (1,1.75)--(4,1.75);
            \draw[thick,postaction={decorate},blue] (2,2.5)--(0,2);
            \draw[blue,postaction={decorate}] (2,1.75+0.25)--(1,1.75);
            \draw[line,thick,red,dashed] (4,1.75)--(5,1.75);
        \end{scope}
            \filldraw[red] (1,1.75) circle (1.5pt);
            \filldraw[red] (5,1.75) circle (1.5pt); 
            \node at (3,2) {$W$};
            \node at (1,1.75+0.9) {$U(h)$};
            \node at (0.6,1.5) {$v_g$};
            \node at (1,-0.5) {$\mathcal{B}_{\textrm{top}}$};
            \node at (5,-0.5) {$\mathcal{B}_{\textrm{phys}}$};
            \node at (8,1.75) {$\Leftrightarrow$};
        \end{tikzpicture}\\ \Updownarrow\qquad \qquad \\
        \begin{tikzpicture}
            \draw[line,thick] (0,0)--(0,3);
            \draw[line,thick] (0,3)--(2,3.5);
            \draw[line,thick] (2,3.5)--(2,0.5);
            \draw[line,thick] (2,0.5)--(0,0);
            \draw[line,thick] (4,0)--(4,3);
            \draw[line,thick] (4,3)--(6,3.5);
            \draw[line,thick] (6,3.5)--(6,0.5);
            \draw[line,thick] (6,0.5)--(4,0);
            \draw[thick] (0,0)--(4,0);
            \draw[thick] (0,3)--(4,3);
            \draw[thick] (2,3.5)--(6,3.5);
            \draw[thick] (2,0.5)--(4,0.5);
            \draw[thick,dashed] (4,0.5)--(6,0.5);
            \begin{scope}[line,thick,decoration={markings,mark=at position 0.45 with {\arrow{<}}}] 
            \draw[red,postaction={decorate}] (1,1.75)--(4,1.75);
            \draw[thick,blue,postaction={decorate}] (2,1.5)--(0,1);
            \draw[blue,postaction={decorate}] (2,1.75+0.25)--(1,1.75);
            \draw[line,thick,red,dashed] (4,1.75)--(5,1.75);
        \end{scope}
            \filldraw[red] (1,1.75) circle (1.5pt);
            \filldraw[red] (5,1.75) circle (1.5pt); 
            \node at (3,2) {$W$};
            \node at (1,0.75) {$U(h)$};
            \node at (1,2.25) {$\phi(g,h)v_g$};
            \node at (1,-0.5) {$\mathcal{B}_{\textrm{top}}$};
            \node at (5,-0.5) {$\mathcal{B}_{\textrm{phys}}$};
            \node at (8,1.75) {$\Leftrightarrow$};
        \end{tikzpicture}
    \end{gathered} \qquad \qquad
            \begin{gathered}
        \begin{tikzpicture}
            \draw[line,thick] (0,0)--(0,3);
            \draw[line,thick] (0,3)--(2,3.5);
            \draw[line,thick] (2,3.5)--(2,0.5);
            \draw[line,thick] (2,0.5)--(0,0);
            \begin{scope}[line,thick,decoration={markings,mark=at position 0.45 with {\arrow{<}}}] 
            \draw[blue,postaction={decorate}] (2,1.75+0.25)--(1,1.75);
            \draw[thick,postaction={decorate},blue] (2,2.5)--(0,2);
            \end{scope}
            \filldraw[red] (1,1.75) circle (1.5pt); 
            \node at (1,1.75+0.9) {$U(h)$};
            \node at (0.6,1.5) {$\psi_g$};
            \node at (1,-0.5) {$\mathcal{M}_2$};
        \end{tikzpicture}\\ \Updownarrow\\
        \begin{tikzpicture}
            \draw[line,thick] (0,0)--(0,3);
            \draw[line,thick] (0,3)--(2,3.5);
            \draw[line,thick] (2,3.5)--(2,0.5);
            \draw[line,thick] (2,0.5)--(0,0);
            \begin{scope}[line,thick,decoration={markings,mark=at position 0.45 with {\arrow{<}}}] 
            \draw[blue,postaction={decorate}] (2,1.75+0.25)--(1,1.75);
            \draw[thick,blue,postaction={decorate}] (2,1.5)--(0,1);
            \end{scope}
            \filldraw[red] (1,1.75) circle (1.5pt); 
            \node at (1,0.75) {$U(h)$};
            \node at (1,2.25) {$\phi(g,h)\psi_g$};
            \node at (1,-0.5) {$\mathcal{M}_2$};
        \end{tikzpicture}
    \end{gathered} \nonumber
    \end{equation}
    \caption{The action of the symmetry on the twist operator in SymTFT picture.}
    \label{SymTFT-Twist-Operator-symmetry-action}
    \end{figure}
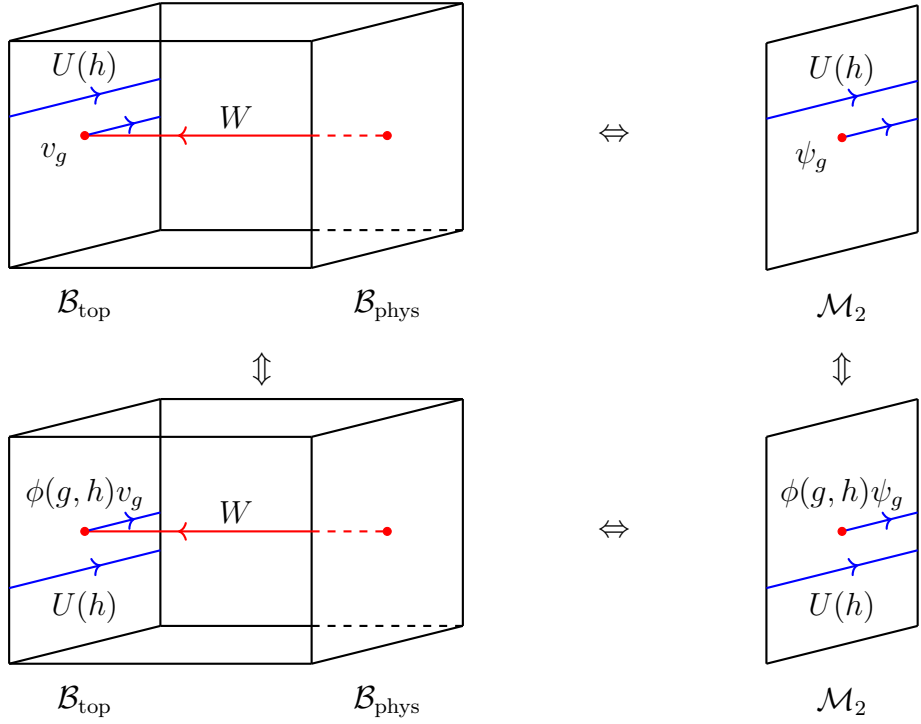

Moreover, we can consider the action of symmetry generator $U(h)$ on the twist operator $\psi_g$, and it is represented as passing the $U(h)$ operator through the endpoint $v_g$ in the SymTFT picture as shown in Figure~\ref{SymTFT-Twist-Operator-symmetry-action}.

Since we assume $G$ is abelian, the symmetry operator $U(h)$ commutes the tail $U(g)$ attached to $v_g$ on $\mathcal{B}_{\textrm{top}}$. However, $U(h)$ will also link with the bulk line operator $W$ as it passes through $v_g$. The linking can generate a phase factor $\phi(g,h)$, which gives the $G$-charge of the operator $\psi_g$. Moreover, we can apply operator-state correspondence in (1+1)D TQFT and map the operator $\psi_g$ to the ground state $|\psi_g\rangle$ in the twist Hilbert space $\mathcal{H}_g$. If the gapped phase is an SPT phase so that the ground state $|\psi_g\rangle$ is unique, we recover the phase factor introduced in \eqref{Partition-function-SPT}. We will continue to use operator-state correspondence in the following when we discuss the (1+1)D examples.

Suppose $\mathcal{M}_2$ is a compact torus, we can assign the physical boundary $\mathcal{B}_{\textrm{phys}}$ a state vector $|\textrm{Gapped}\rangle$ such that the torus partition function is recovered by
    \begin{equation}
        Z[A] = \langle A |\textrm{Gapped}\rangle \,,
    \end{equation}
where $A$ is the background of the $G$-symmetry on $\mathcal{B}_{\textrm{top}}$.

In the following, we will work out two examples to illustrate the construction in SymTFT. 

\subsubsection{Examples : (1+1)D $\mathbb{Z}_2$ symmetry}
As discussed above, there are three kinds of topological boundaries given by
\begin{itemize}
    \item The Dirichlet boundary $\mathcal{B}_{\textrm{Dir}}$ where $W$ can end, and the topological boundary state on the torus is $|a\rangle$.
    \item The Neumann boundary $\mathcal{B}_{\textrm{Neu}}$ where $\hat{W}$ can end and the topological boundary state on torus is $|\hat{a}\rangle$. It is related to $|a\rangle$ according to
        \begin{equation}
            |\hat{a}\rangle = \frac{1}{2} \sum_{a\in H^1(T^2,\mathbb{Z}_2)} (-1)^{\int a \wedge \tilde{a}} |a\rangle \nonumber\,.
        \end{equation}
    \item The fermionic boundary $\mathcal{B}_{\textrm{Fer}}$ where the combination $W\hat{W}$ can end, and the topological boundary state $|s\rangle$ is defined as
        \begin{equation}
            |s\rangle = \frac{1}{2} \sum_{a\in H^1(T^2,\mathbb{Z}_2)} (-1)^{\textrm{Arf}(s+a)} |a\rangle \nonumber\,.
        \end{equation}
\end{itemize}
We will first consider the bosonic phases and choose the topological boundary to be the Dirichlet boundary $\mathcal{B}_{\textrm{Dir}}$.

\subsubsection*{SSB phase}
If the physical boundary $\mathcal{B}_{\textrm{phys}}$ is also of the Dirichlet type, then all line operators starting from $\mathcal{B}_{\textrm{phys}}$ can end on the topological boundary. That means there exist two states in the trivial sector and none in the twist sector. For the state corresponding to $W$ ending at the boundary, the $\mathbb{Z}_2$ charge can be read from the linking between $\hat{W}$ with $W$ and is simply one. So we have two ground states with $\mathbb{Z}_2$ charges $0,1$, which implies an SSB phase of $\mathbb{Z}_2$. The physical boundary state on the torus is chosen as $|\textrm{SSB}\rangle = 2 |0\rangle$, where $|0\rangle$ is the vacuum of the Dirichlet boundary state introduced in \eqref{Z2-Dirichlet-Boundary-State}, and the normalization factor $2$ equals the order of the group. The partition function is
        \begin{equation}
            Z_{\textrm{SSB}}[a] = \langle a | \textrm{SSB} \rangle = 2 \delta_{a_x,0}\delta_{a_y,0}\,,
        \end{equation}
where the delta functions are defined mod $\mathbb{Z}_2$.

\subsubsection*{Trivial phase}
If the physical boundary $\mathcal{B}_{\textrm{phys}}$ is of the Neumann type, then the $\hat{W}$ operator starting from $\mathcal{B}_{\textrm{phys}}$ will transit to the symmetry generator of $\mathbb{Z}_2$ on $\mathcal{B}_{\textrm{top}}$ and give a twist operator. Therefore, we have a single ground state in the twist sector. Moreover, since $\hat{W}$ links with itself trivially, the ground state in the twist sector is neutral under the $\mathbb{Z}_2$ symmetry, and we have a trivial phase. The physical boundary state is chosen as $|\textrm{Tri}\rangle = 2|\hat{0}\rangle$ where $|\hat{0}\rangle$ is the vacuum of Neumann boundary state introduced in \eqref{Z2-Neumann-Boundary-State}. The partition function is simply
    \begin{equation}
        Z_{\textrm{Tri}}[a] = \langle a | \textrm{Tri} \rangle = 1\,.
    \end{equation}

\subsubsection*{(Fermionic) SPT phase}
There is no non-trivial (1+1)D bosonic SPT phases for $\mathbb{Z}_N$ which is implied from the classification $H^2(\mathbb{Z}_N,U(1)) = \bZ_1$. Nevertheless, when $N=2$ we can still have a non-trivial fermionic SPT phase. We will also review the fermionic SPT phase from the SymTFT picture for completeness. To do that, we need to set the topological boundary state to the fermionic boundary $\mathcal{B}_{\textrm{Fer}}$, which supports the fermionic parity $(-1)^F$ as a symmetry. However, the choice of $(-1)^F$ on the topological boundary is not unique, and we have two possibilities
    \begin{equation}
        (-1)^F = \hat{W}\,,\quad (-1)^F = W\,,
    \end{equation}
and they will lead to two different fermionic phases, as we will see soon.

Let us set the physical boundary $\mathcal{B}_{\textrm{phys}}$ to be the Neumann type such that $\hat{W}$-operator can end on it. On the other hand, it cannot end at the topological boundary and will transit to $(-1)^F$ regardless of the choice we made above. Therefore, we have a single ground state in the $(-1)^F$-twist sector. However, depending on the choice of $(-1)^F$, the statistics of the state will be different. If we choose $(-1)^F=\hat{W}$ as we did in \eqref{review-fermionic-boundary-state}, then since $\hat{W}$ links trivially with itself, the twist sector ground state does not carry $(-1)^F$ charge and is bosonic. Thus, we get the fermionic trivial phase whose partition function is
    \begin{equation}
        Z_{F,\textrm{Tri}}[s] = 2 \langle s | \textrm{F,Tri}\rangle = 1\,,\quad |\textrm{F,Tri}\rangle = 2|\hat{0}\rangle\,.
    \end{equation}
On the other hand, if we choose $(-1)^F = W$ instead, then $W$ will link $\hat{W}$ non-trivially so that the twist sector ground state carries charge one under $(-1)^F$ and is fermionic. We therefore obtain a non-trivial fermionic SPT phase. The partition function can be read directly, which is given by the Arf-invariant $(-1)^{\textrm{Arf}(s)}$. We can also obtain the phase from the topological boundary state $|s\rangle$. Recall that $|s\rangle$ is defined where $\hat{W}$ is considered as the fermionic parity and we have
    \begin{equation}
        |s\rangle = \hat{W}[s] |s=0\rangle = (\hat{W}[\Gamma_1])^{s_y} (\hat{W}[\Gamma_2])^{s_x} |s=0\rangle\,,
    \end{equation}
where the generic states $|s\rangle$ are raised using $\hat{W}$. If we switch to $(-1)^F=W$, we need to replace $\hat{W}$ to $W$ and we have
    \begin{equation}
        |s\rangle \rightarrow  (W[\Gamma_1])^{s_y} (W[\Gamma_2])^{s_x} |s=0\rangle = (-1)^{\textrm{Arf}(s)} |s\rangle\,,
    \end{equation}
where we use \eqref{review-fermionic-boundary-state}. Therefore, changing $(-1)^F$ from $\hat{W}$ to $W$ is equivalent to stacking the fermionic SPT phase on the system. Following the same method, we can obtain the partition function as
    \begin{equation}
        Z_{F,\textrm{SPT}}[s] = (-1)^{\textrm{Arf}(s)}\,.
    \end{equation}
    
Alternatively, we can keep the fermionic parity $(-1)^F = \hat{W}$ and switch the dynamical boundary to the Dirichlet boundary, and the fermionic SPT phase can also be obtained as $Z_{\textrm{F,SPT}}[s] = \langle s|\textrm{F,SPT}\rangle$ with $|\textrm{F,SPT}\rangle = 2|0\rangle$.

\subsubsection{Examples : (1+1)D $\mathbb{Z}_2\times \mathbb{Z}_2$ symmetry}
Let us turn to another example and consider the bosonic gapped phase of $(1+1)$D $\mathbb{Z}_2\times \mathbb{Z}_2$ symmetry. The SymTFT is simply two copies of level-2 BF theories, and the action is
    \begin{equation}
        S_{\mathbb{Z}_2\times \mathbb{Z}_2} = \frac{2}{2\pi} \int \hat{A} \wedge dA+ \frac{2}{2\pi} \int \hat{B} \wedge dB\,,
    \end{equation}
where $A,\hat{A}$ are 1-form gauge fields for first $\bZ_2$  and   $B,\hat{B}$ are 1-form gauge fields for second $\bZ_2$.
Let's denote the line operators of the first $\mathbb{Z}_2$ as $W_A\,,\hat{W}_{A}$ and the second $\mathbb{Z}_2$ as $W_B\,, \hat{W}_B$. There are six bosonic topological boundaries and the corresponding torus boundary states are
    \begin{itemize}
        \item Dirichlet-Dirichlet : $|a,b\rangle \equiv |a\rangle \otimes |b\rangle$ and $W_A, W_B$ can end at the boundary.
        \item Neumann-Dirichlet : $|\hat{a},b\rangle \equiv |\hat{a}\rangle \otimes |b\rangle$ and $\hat{W}_A, W_B$ can end at the boundary.
        \item Dirichlet-Neumann : $|a,\hat{b}\rangle \equiv |a\rangle\otimes |\hat{b}\rangle$ and $W_A, \hat{W}_B$ can end at the boundary.
        \item Neumann-Neumann : $|\hat{a},\hat{b}\rangle \equiv |\hat{a}\rangle\otimes|\hat{b}\rangle$ and $\hat{W}_A, \hat{W}_B$ can end at the boundary.
        \item Mixed boundary $|a',b'\rangle_1$ defined as
            \begin{equation}
                |a',b'\rangle_1 = \frac{1}{2} \sum_{a\in H^1(T^2,\mathbb{Z}_2)} (-1)^{\int a\wedge b'}|a+a',a\rangle\,,
            \end{equation}
        and $W_A W_B, \hat{W}_A\hat{W}_B$ can end at the boundary.
        \item Mixed boundary $|a',b'\rangle_2$ defined as
            \begin{equation}
                |a',b'\rangle_2 = \frac{1}{2^2} \sum_{a,b\in H^1(T^2,\mathbb{Z}_2)} (-1)^{\int a \wedge b}|a+a',b+b'\rangle\,, 
            \end{equation}
        and $W_A \hat{W}_B, \hat{W}_A W_B$ can end at the boundary.
    \end{itemize}
In the above, all states are labeled by two holonomy variables $(a_x,a_y)$ and $(b_x,b_y)$ for each $\mathbb{Z}_2$. We will choose the Dirichlet-Dirichlet boundary as the topological boundary, which supports the $\mathbb{Z}_2\times\mathbb{Z}_2$ symmetry generated by $\hat{W}_A$ and $\hat{W}_B$.
\subsubsection*{SSB phase of the whole $\mathbb{Z}_2\times \mathbb{Z}_2$}
    If we choose the physical boundary $\mathcal{B}_{\textrm{phys}}$ also to be the Dirichlet-Dirichlet boundary, then we have the SSB phase, and the whole $\mathbb{Z}_2\times \mathbb{Z}_2$ symmetry is broken. The partition function is
    \begin{equation}
        Z_{\textrm{(SSB,SSB)}}[a,b] = \langle a,b|\textrm{SSB, SSB}\rangle = 4\delta_{a_x,0}\delta_{a_y,0}\delta_{b_x,0}\delta_{b_y,0}\,,\quad |\textrm{SSB, SSB}\rangle=2^2 |0,0\rangle\,.
    \end{equation}
\subsubsection*{SSB phase of the first $\mathbb{Z}_2$}
    If we choose the physical boundary $\mathcal{B}_{\textrm{phys}}$ to be the Dirichlet-Neumann boundary, then only the first $\mathbb{Z}_2$ is broken. The partition function is
    \begin{equation}
        Z_{\textrm{(SSB,Tri)}}[a,b] = \langle a,b|\textrm{SSB, Tri}\rangle = 2 \delta_{a_x,0} \delta_{a_y,0}\,,\quad |\textrm{SSB, Tri}\rangle=2^2|0,\hat{0}\rangle\,.
    \end{equation}
\subsubsection*{SSB phase of the second $\mathbb{Z}_2$}
    If we choose the physical boundary $\mathcal{B}_{\textrm{phys}}$ to be the Neumann-Dirichlet boundary, then only the second $\mathbb{Z}_2$ is broken. The partition function is
    \begin{equation}
        Z_{\textrm{(Tri,SSB)}}[a,b] = \langle a,b|\textrm{Tri, SSB}\rangle = 2 \delta_{b_x,0} \delta_{b_y,0}\,,\quad |\textrm{Tri, SSB}\rangle = 2^2|\hat{0},0\rangle\,.
    \end{equation}
\subsubsection*{Trivial phase}
If we choose the physical boundary $\mathcal{B}_{\textrm{phys}}$ to be the Neumann-Neumann boundary, then we get the trivial phase. The partition function is
    \begin{equation}
        Z_{\textrm{(Tri,Tri)}}[a,b] = \langle a,b |\textrm{Tri, Tri}\rangle = 1   \,,\quad |\textrm{Tri, Tri}\rangle=2^2|\hat{0},\hat{0}\rangle\,.
        \end{equation}
\subsubsection*{Another $\mathbb{Z}_2$ SSB phase}  
If we choose the physical boundary $\mathcal{B}_{\textrm{phys}}$ to be the first mixed boundary, we will have two states in the trivial sector and another two states in the twist sector of the diagonal $\mathbb{Z}_2$. They are separately given by $1,W_AW_B$ and $\hat{W}_A \hat{W}_B, W_A W_B \hat{W}_A \hat{W}_B$. Therefore the $\mathbb{Z}_2 \times \mathbb{Z}_2$ is broken to the diagonal part $\mathbb{Z}_2 \times \mathbb{Z}_2 \rightarrow \mathbb{Z}_{2,\textrm{Diag}}$, and the partition is
    \begin{equation}
        Z_{\textrm{DiagSSB}}[a,b] = \langle a,b|\textrm{DiagSSB}\rangle  = 2 \delta_{a_1,b_1} \delta_{a_2,b_2}\,, \quad |\textrm{DiagSSB}\rangle = 2^2|0,0\rangle_1\,.
    \end{equation}
\subsubsection*{SPT phase corresponding to generator of $H^2(\mathbb{Z}_2\times \mathbb{Z}_2,U(1))=\mathbb{Z}_2$}
Finally, if we choose the physical boundary $\mathcal{B}_{\textrm{phys}}$ to be the second mixed boundary, then for each twist sector of $\mathbb{Z}_2 \times \mathbb{Z}_2$ we will have a single ground state. Moreover, since the $\hat{W}_A$ and $\hat{W}_B$ operators stretching from $\mathcal{B}_{\textrm{phys}}$ are also dressed by $W_B$ and $W_A$, they link non-trivially with the symmetry generators on $\mathcal{B}_{\textrm{top}}$ given by $\hat{W}_A$ and $\hat{W}_B$. Therefore we will get a non-trivial SPT phase where the twist sector for each $\mathbb{Z}_2$ will carry the charge of the other $\mathbb{Z}_2$. The partition function is
    \begin{equation}
        Z_{\textrm{SPT}}[a,b] =  \langle a,b|\textrm{SPT}\rangle = (-1)^{\int a \wedge b}\,,\quad |\textrm{SPT}\rangle = 2^2|0,0\rangle_2\,.
    \end{equation}

\section{SymTFT for (2+1)D Subsystem Symmetry }
\label{sec:subSymTFT}

In this section, we will review the SymTFT construction for (2+1)D theory $\mathfrak{T}_{\textrm{sub}}$ with a $\mathbb{Z}_N$ subsystem symmetry living on $\mathcal{M}_3$ with coordinates $(x,y,z)$, where $z$ is the time direction.  
The 2-foliation lies along the $x$ and $y$ directions of the spacetime manifold, and the corresponding SymTFT is given by a 2-foliated BF theory~\cite{Spieler2023foliated,Ohmori2023Foliated,Gorantla2021fcc,Cao2024SymTFT}:
\begin{equation}\label{eq:foliatedBF}
    S_{\textrm{foliated BF}} = \frac{N}{2\pi}\int b \wedge dc + \sum_{k=1,2} d B^k \wedge C^k \wedge d x^k + \sum_{k=1,2} b \wedge C^k \wedge d x^k\,,
\end{equation}
which is dual to the $(3+1)$D 2-foliated exotic tensor gauge theory~\cite{Spieler2023foliated,Gorantla2021fcc} via the exotic-foliated duality~\cite{Spieler2023foliated,Ohmori2023Foliated} (see also \cite[Appendix B]{Cao2024SymTFT} for further details about the equivalence of the two theories). We will work with the dual theory, in which the subsystem symmetry is more manifest. The relevant topological boundaries used to construct SSPT phases will be summarized in the next section. For further details, we refer the reader to~\cite{Cao2024SymTFT}.

In terms of exotic tensor gauge theory, the action of the SymTFT for a $(2+1)$D $\mathbb{Z}_N$ subsystem symmetry is
\begin{equation}\label{eq:exotic}
    S_{\textrm{sub},\mathbb{Z_N}} = \frac{N}{2\pi} \int d^4x\left[A^{\tau} (\partial_z \hat{A}^{xy} - \partial_x \partial_y \hat{A}^z) - A^z (\partial_{\tau} \hat{A}^{xy} - \partial_x \partial_y \hat{A}^{\tau})- A^{xy} (\partial_{\tau} \hat{A}^z - \partial_z \hat{A}^{\tau})  \right]\,,
\end{equation}
with coordinates $(x,y,z,\tau)$. Here $A = (A^{\tau}, A^z, A^{xy})$ and $\hat{A} = (\hat{A}^{\tau},\hat{A}^z,\hat{A}^{xy} )$ are electric and magnetic gauge fields with the following gauge transformations
\begin{align}
    \begin{split}
         &A^{\tau} \sim A^{\tau} + \partial_{\tau} \lambda\,,\quad A^{z} \sim A^{z} + \partial_{z} \lambda\,,\quad A^{xy}\sim A^{xy} + \partial_{x}\partial_y \lambda\,,\\
         &\hat{A}^{\tau} \sim \hat{A}^{\tau} + \partial_{\tau} \hat{\lambda}\,,\quad \hat{A}^z \sim \hat{A}^z + \partial_z \hat{\lambda}\,,\quad \hat{A}^{xy} \sim \hat{A}^{xy} + \partial_x \partial_y \hat{\lambda}\,,
    \end{split}
\end{align}
where $\lambda,\hat{\lambda}$ are gauge parameters. 
The equations of motion for gauge fields $A$ and $\hat{A}$ are
 \begin{align}\label{Foliated-Theory-EOM}
 \begin{split}
     &\partial_z A^{\tau} - \partial_{\tau} A^z=0\,,\quad \partial_{\tau} A^{xy} - \partial_x \partial_y A^{\tau}=0\,,\quad \partial_{z} A^{xy} - \partial_x \partial_y A^{z}=0\,,\\
      &\partial_z \hat{A}^{\tau} - \partial_{\tau} \hat{A}^z=0\,,\quad \partial_{\tau} \hat{A}^{xy} - \partial_x \partial_y \hat{A}^{\tau}=0\,,\quad \partial_{z} \hat{A}^{xy} - \partial_x \partial_y \hat{A}^{z}=0\,.
 \end{split}
 \end{align}
In the exotic theory~\eqref{eq:exotic}, there exists a $SL(2,\mathbb{Z}_N)$ symmetry
\begin{align}\label{eq:sym}
    \begin{split}
        &S:\quad A\rightarrow \hat{A}\,,\quad \hat{A}\rightarrow -A\,,\\
        &  T:\quad A\rightarrow A\,,\quad \hat{A} \rightarrow \hat{A} + A\,,
    \end{split}
\end{align}
which should be modified if we put the theory on the lattice, as we will discuss soon. 

The gauge-invariant operators in this theory exhibit restricted mobility. In particular, there exist electric and magnetic line operators that are topological along the $z$–$\tau$ plane but cannot move freely in the $x$ or $y$ directions
\begin{align}\label{Foliated-Theory-Line-Operators}
    \begin{split}
        W(C_{z,\tau}(x,y)) &= \exp\left(i \oint_{C_{z,\tau}(x,y)} A^{\tau} \, d\tau + A^z \, dz\right), \\
        \hat{W}(C_{z,\tau}(x,y)) &= \exp\left(i \oint_{C_{z,\tau}(x,y)} \hat{A}^{\tau} \, d\tau + \hat{A}^z \, dz\right),
    \end{split}
\end{align}
where $C_{z,\tau}(x,y)$ is a closed curve in the $z$–$\tau$ plane, localized at the spatial position $(x, y)$ in the ambient $(3+1)$D spacetime.
The exotic theory also admits gauge-invariant strip operators that extend along the $x$ or $y$ directions:
\begin{align}\label{Foliated-Theory-Strip-Operators}
    \begin{split}
        W(x_1,x_2,C_{y,z,\tau}(x)) &= \exp\left(i\int_{x_1}^{x_2} dx \oint_{C_{y,z,\tau}(x)} A^{xy} \, dy + \partial_x A^{z} \, dz + \partial_x A^{\tau} \, d\tau \right), \\
        W(y_1,y_2,C_{x,z,\tau}(y)) &= \exp\left(i\int_{y_1}^{y_2} dy \oint_{C_{x,z,\tau}(y)} A^{xy} \, dx + \partial_y A^{z} \, dz + \partial_y A^{\tau} \, d\tau \right),
    \end{split}
\end{align}
where $A$ is the electric gauge field. Magnetic counterparts $\hat{W}$ can be defined analogously using the dual gauge field $\hat{A}$. Here, $C_{x,z,\tau}(y)$ denotes a curve in the $x$–$z$–$\tau$ plane at fixed $y$, while $C_{y,z,\tau}(x)$ lies in the $y$–$z$–$\tau$ plane at fixed $x$. These curves can be smoothly deformed within their respective planes but cannot move freely along the transverse $y$ or $x$ directions. This restricted mobility follows directly from the equations of motion~\eqref{Foliated-Theory-EOM} for the gauge fields $A$ and $\hat{A}$.

\begin{figure}[!h]
    \begin{equation}
            \begin{gathered}
        \begin{tikzpicture}
            \draw[->,thick] (0,0)--(0,1);
            \draw[->,thick] (0,0)--(0.7071,0.7071);
            \draw[->,thick] (0,0)--(1,0);
            \node at (0,1.3) {$z$ or $\tau$};
            \node at (1.3,0) {$x$};
            \node at (0.9,0.9) {$y$};
            \node at (0,-3) {};
        \end{tikzpicture}
    \end{gathered}\qquad
            \begin{gathered}
        \begin{tikzpicture}
            \draw[line,thick] (0,0)--(4,0);
            \draw[line,thick] (0,0)--(2,2);
            \draw[line,thick] (2,2)--(6,2);
            \draw[line,thick] (6,2)--(4,0);
            \begin{scope}[thick,decoration={markings,mark=at position 0.5 with {\arrow{>}}}] 
            \draw[line,red,postaction={decorate}] (2,3)--(2,0.5);
            \draw[line,red,postaction={decorate}] (3,1.5)--(3,4);
            \end{scope}      
            \node (v1) at (2,0.5) {};
            \node (v2) at (3,1.5) {};
            \node (v4) at (5.5,1.5) {};
            \node (v3) at (4.5,0.5) {};        \draw[opacity=0.25,fill=red] (v1.center)--(v2.center)--(v4.center)--(v3.center)--(v1.center); 
        \end{tikzpicture}
    \end{gathered} \qquad                      \begin{gathered}
        \begin{tikzpicture}
            \draw[line,thick] (0,0)--(4,0);
            \draw[line,thick] (0,0)--(2,2);
            \draw[line,thick] (2,2)--(6,2);
            \draw[line,thick] (6,2)--(4,0);
            \begin{scope}[thick,decoration={markings,mark=at position 0.5 with {\arrow{>}}}] 
            \draw[line,red,postaction={decorate}] (1.5,4)--(1.5,0.5);
            \draw[line,red,postaction={decorate}] (3.5,0.5)--(3.5,4);
            \end{scope}      
            \node (v1) at (1.5,0.5) {};
            \node (v2) at (3.5,0.5) {};
            \node (v4) at (5,2) {};
            \node (v3) at (3,2) {};        \draw[opacity=0.25,fill=red] (v1.center)--(v2.center)--(v4.center)--(v3.center)--(v1.center); 
        \end{tikzpicture}
    \end{gathered}\nonumber
    \end{equation}
    \caption{A pair of line operators can transit to a strip operator.}
    \label{Fig-Line-Strip-Transit}
\end{figure}
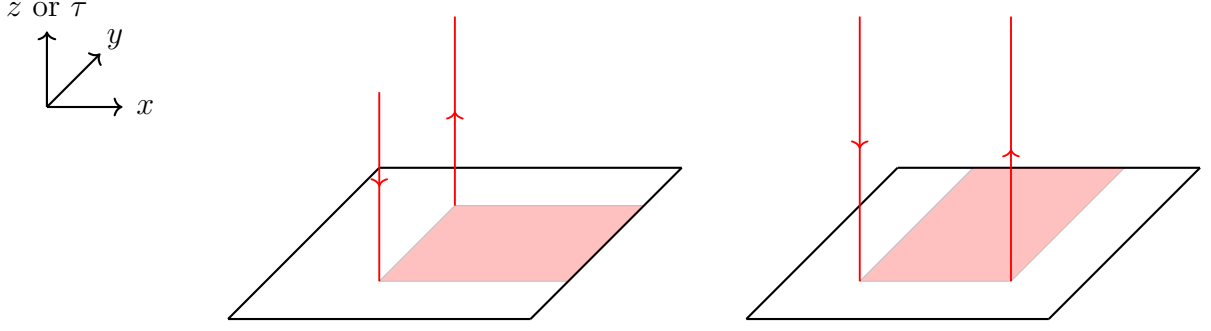

From the expression of the strip operator, we see that if we have a pair of $W$ (or $\hat{W}$) line operators with opposite orientations located at $(x,y)$ and $(x,y')$, with the same $x$-coordinates but different $y$-coordinates, they can be bent into a strip operator spanned between $(y,y')$ and extended along $x$-direction. Similarly, a pair of line operators with opposite orientations located at $(x,y)$ and $(x',y)$ can be bent into a strip operator spanned between $(x,x')$ and extended along $y$-direction. See Figure~\ref{Fig-Line-Strip-Transit} for an illustration.

\subsection{Canonical Quantization}
To get some feeling of the SymTFT, we can quantize the exotic theory~\eqref{eq:exotic} by picking $\tau$ as the time direction with the Coulomb gauge $A^{\tau} = \hat{A}^{\tau}=0$. In the Coulomb gauge, the action~\eqref{eq:exotic} takes the form
\begin{equation}\label{Foliated-Theory-Action-Coulomb-Gauge}
    S_{\textrm{exotic}} = \frac{N}{2\pi} \int \left[- A^{xy} (\partial_{\tau} \hat{A}^{z}) - A^z (\partial_{\tau} \hat{A}^{xy}) \right]\,,
\end{equation}
with the canonical commutation relations between the conjugate fields $A$ and $\hat{A}$
\begin{align}
    \left[ A^{xy} (x,y,z),\, \hat{A}^z (x',y',z') \right] &= \frac{2\pi i}{N} \, \delta^3(x - x', y - y', z - z')\,, \\
    \left[ A^z(x,y,z),\, \hat{A}^{xy} (x',y',z') \right] &= \frac{2\pi i}{N} \, \delta^3(x - x', y - y', z - z')\,.
\end{align}
The Gauss law constraints
\begin{equation}\label{Foliated-Theory-Gauss-Law}
    \partial_x \partial_y \hat{A}^{z} - \partial_z \hat{A}^{xy} = 0\,, \quad
    \partial_x \partial_y A^z - \partial_z A^{xy} = 0\,,
\end{equation}
impose a flatness condition on the gauge fields.

We will put the exotic tensor theory on a manifold $\mathcal{M}_3 = \mathbb{R}^2 \times \mathbb{R}^1$, where $(x,y)$ parameterize the spatial $\mathbb{R}^2$ plane and $z$ is the coordinate of the boundary time direction $\mathbb{R}^1$. The gauge invariant operators \eqref{Foliated-Theory-Line-Operators},\eqref{Foliated-Theory-Strip-Operators} restricting to $\mathcal{M}_3$ gives the electric line/strip operators
\begin{align}\label{Foliated-Theory-Operators-electric}
    \begin{split}
        &W_z(x,y)=\exp\left(i\int dz A^z\right)\,,\\
        &W(x_1,x_2)=\exp\left(i\int_{x_1}^{x_2}dx\int dy A^{xy}\right)\,,\\
        &W(y_1,y_2)=\exp\left(i\int_{y_1}^{y_2}dy\int dx A^{xy}\right)\,,
    \end{split}
\end{align}
and the magnetic line/strip operators
\begin{align}\label{Foliated-Theory-Operators-magnetic}
    \begin{split}
        &\hat{W}_z(x,y)=\exp\left(i\int dz \hat{A}^{z}\right)\,,\\
        &\hat{W}(x_1,x_2)=\exp\left(i\int_{x_1}^{x_2}dx\int dy \hat{A}^{xy}\right)\,,\\
        &\hat{W}(y_1,y_2)=\exp\left(i\int_{y_1}^{y_2}dy\int dx\hat{A}^{xy}\right)\,.
    \end{split}
\end{align}
They are $\mathbb Z_N$-valued operators 
\begin{equation}
    W^N = \hat{W}^N = 1\,,
\end{equation} 
with the following commutation relations 
\begin{align}
    \begin{split}
    W(x_1,x_2) \hat{W}_z (x,y) &=\omega \hat{W}_z (x,y) W(x_1,x_2)\,,\quad (x_1<x<x_2),\\
    W(y_1,y_2) \hat{W}_z (x,y) &= \omega \hat{W}_z (x,y) W(y_1,y_2)\,,\quad (y_1 < y <y_2),     
    \end{split}
\end{align}
and
\begin{align}
    \begin{split}
    \hat{W}(x_1,x_2) W_z (x,y) &= \omega^{-1} W_z (x,y) \hat{W}(x_1,x_2)\,,\quad ( x_1<x<x_2),\\
    \hat{W}(y_1,y_2) W_z (x,y) &= \omega^{-1} W_z (x,y) \hat{W}(y_1,y_2)\,,\quad  ( y_1<y<y_2),        
    \end{split}
\end{align}
where the phase factor $\omega = \exp (2\pi i/N)$ is the $N$-th root of unity.

Using the Gauss law constraints, the holonomy of the electric gauge field $A^z$ can be decomposed as
\begin{equation}\label{eq:split}
    \int dz\, A^z = \mathcal{A}^y(x) + \mathcal{A}^x(y)\,,
\end{equation}
where $\mathcal{A}^y(x)$ and $\mathcal{A}^x(y)$ are operators depending only on $x$ and $y$, respectively. This decomposition of the holonomy~\eqref{eq:split} leads to a factorization of the $z$-directional line operator:
\begin{equation}
    W_z(x,y) = W_{z,y}(x)\, W_{z,x}(y)\,,
\end{equation}
where $W_{z,y}(x)$ and $W_{z,x}(y)$ are line operators along the $z$ direction with restricted mobility along the $y$ and $x$ directions, respectively.
These operators satisfy the following commutation relations
\begin{align}\label{Foliated-Theory-Operators-Algebra-1}
    \begin{split}
    W(x_1, x_2)\, \hat{W}_{z,y}(x) &= \omega\, \hat{W}_{z,y}(x)\, W(x_1, x_2)\,, \quad (x_1 < x < x_2), \\
    W(y_1, y_2)\, \hat{W}_{z,x}(y) &= \omega\, \hat{W}_{z,x}(y)\, W(y_1, y_2)\,, \quad (y_1 < y < y_2),
    \end{split}
\end{align}
and similarly
\begin{align}\label{Foliated-Theory-Operators-Algebra-2}
    \begin{split}
    \hat{W}(x_1, x_2)\, W_{z,y}(x) &= \omega^{-1}\, W_{z,y}(x)\, \hat{W}(x_1, x_2)\,, \quad (x_1 < x < x_2), \\
    \hat{W}(y_1, y_2)\, W_{z,x}(y) &= \omega^{-1}\, W_{z,x}(y)\, \hat{W}(y_1, y_2)\,, \quad (y_1 < y < y_2).
    \end{split}
\end{align}

In a real physical system, the theory is defined on a discrete lattice. Therefore, we must also discretize the $\mathbb{R}^2$ plane in the SymTFT formulation
\begin{equation*}
    \begin{tikzpicture}
        \draw[thick, dotted, red] (0.5,0) -- (0.5, 3.5);
        \draw[thick, dotted, red] (1.5,0) -- (1.5, 3.5);
        \draw[thick, dotted, red] (2.5,0) -- (2.5, 3.5);
        \draw[thick] (1,0) -- (1, 3.5);
        \draw[thick] (2,0) -- (2, 3.5);
        \draw[thick] (3,0) -- (3, 3.5);
        \draw[thick, dotted, red] (0,0.5) -- (3.5, 0.5);
        \draw[thick, dotted, red] (0,1.5) -- (3.5, 1.5);
        \draw[thick, dotted, red] (0,2.5) -- (3.5, 2.5);
        \draw[thick] (0,1) -- (3.5, 1);
        \draw[thick] (0,2) -- (3.5, 2);
        \draw[thick] (0,3) -- (3.5, 3);
        \node[below] at (1,0) {$\scriptstyle i-1$};
        \node[below] at (2,0) {$\scriptstyle i$};
        \node[below] at (3,0) {$\scriptstyle i+1$};
        \node[left] at (0,1) {$\scriptstyle j-1$};
        \node[left] at (0,2) {$\scriptstyle j$};
        \node[left] at (0,3) {$\scriptstyle j+1$};
        \node[right] at (3.5,0.5) {\color{red}$\scriptstyle j-1-\frac{1}{2}$};
        \node[right] at (3.5,1.5) {\color{red}$\scriptstyle j-\frac{1}{2}$};
        \node[right] at (3.5,2.5) {\color{red}$\scriptstyle j+\frac{1}{2}$};
        \node[above] at (0.5,3.5) {\color{red}$\scriptstyle i-1-\frac{1}{2}$};
        \node[above] at (1.5,3.5) {\color{red}$\scriptstyle i-\frac{1}{2}$};
        \node[above] at (2.5,3.5) {\color{red}$\scriptstyle i+\frac{1}{2}$};
    \end{tikzpicture}
\end{equation*}
where the direct lattice is shown in solid black, while the dual lattice is indicated by dotted red lines.
Let us label the spatial lattice by $(x_i,y_j)$ with $i,j\in \bZ$, the discrete version of the algebras between $W$ and $\hat{W}$ is
\begin{align}\label{Foliated-Theory-Operators-Albetra-Discrete-1}
    \begin{split}
    W(x_{i},x_{i+1}) \hat{W}_{z,y}(x_{i+\frac{1}{2}}) &= \omega \hat{W}_{z,y}(x_{i+\frac{1}{2}}) W(x_{i},x_{i+1})\,,\\
    W(y_{j},y_{j+1}) \hat{W}_{z,x}(y_{j+\frac{1}{2}}) &= \omega \hat{W}_{z,x}(y_{j+\frac{1}{2}}) W(y_{j},y_{j+1})\,,
    \end{split}
\end{align}
and
\begin{align}\label{Foliated-Theory-Operators-Albetra-Discrete-2}
    \begin{split}
    \hat{W}(x_{i-\frac{1}{2}},x_{i+\frac{1}{2}}) W_{z,y}(x_i) &= \omega^{-1} W_{z,y}(x_i) \hat{W}(x_{i-\frac{1}{2}},x_{i+\frac{1}{2}})\,,\\
    \hat{W}(y_{j-\frac{1}{2}},y_{j+\frac{1}{2}}) W_{z,x}(y_j) &= \omega^{-1} W_{z,x}(y_j) \hat{W}(y_{j-\frac{1}{2}},y_{j+\frac{1}{2}})\,, 
    \end{split}
\end{align}
where $\hat{W}$ lives on the dual lattice labeled by $(x_{i+\frac{1}{2}},y_{j+\frac{1}{2}})$. As a last comment, the algebras also hold if we switch the role between $z$ and $\tau$ in \eqref{Foliated-Theory-Operators-Albetra-Discrete-1} and $\eqref{Foliated-Theory-Operators-Albetra-Discrete-2}$ since $z$ and $\tau$ are democratic in the SymTFT action.

In the following, we will also consider compactifying $\mathcal{M}_3$ to $T^3$ to discuss the partition functions. Denote the periods of the spatial lattice as $x_i \sim x_{i +L_x}$ and $y_j \sim y_{j+L_y}$, we have in total $2(L_x+L_y)$ electric operators: line operators $W_{z,y}(x_i),W_{z,x}(y_j)$ and strip operators $W(x_i,x_{i+1}),W(y_j,y_{j+1})$ with $i=1,\cdots, L_x$ and $j=1,\cdots, L_y$. However, when the $z$-direction is also compactified, the decomposition of $W_z(x_i,y_j) = W_{z,y}(x_i) W_{z,x}(y_j)$ is not unique because of the redundancy
\begin{equation}\label{eq:gaugere}
    \mathcal{A}^y(x_i) \rightarrow \mathcal{A}^y(x_i) + \frac{2\pi}{N}\,,\quad \mathcal{A}^x(y_j) \rightarrow \mathcal{A}^x(y_j) -\frac{2\pi}{N}\,,\quad \forall (x_i,y_j)\,,
\end{equation}
which leaves $\oint dz A^z$ invariant. On the other hand, the strip operators $W(x_i,x_{i'})$ and $W(y_i,y_{j'})$ also satisfy the constraint
\begin{equation}\label{eq:gaugecon}
    W(x_i,x_{i'}+L_x) = W(y_j,y_{j'}+L_y)\,,
\end{equation}
since both of them are the strip operator over the whole discrete torus. As a result, we need to impose the ``gauge" redundancy
\begin{equation}\label{Foliated-Theory-Operators-electric-Gauge-Redundancy-electric}
    (W_{z,y}(x_i),W_{z,x}(y_j)) \sim (\omega W_{z,y}(x_i),\omega^{-1} W_{z,x}(y_j))\,,
\end{equation}
and the constraint
\begin{equation}\label{Foliated-Theory-Operators-electric-Constraints-electric}
    \prod_{i=1}^{L_x} W(x_i,x_{i+1})  = \prod_{j=1}^{L_y} W(y_j,y_{j+1})\,,
\end{equation}
leaving only $2(L_x+L_y-1)$ operators independent. Similarly, there are $2(L_x+L_y-1)$ independent magnetic $\hat{W}$ operators on the dual lattice.

\subsection{SymTFT construction and topological boundaries}

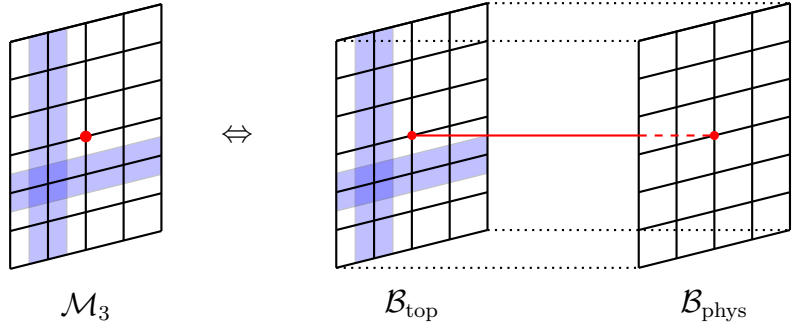
\begin{figure}[!h]
    \begin{equation}
            \begin{gathered}
        \begin{tikzpicture}
            \node (v1) at (0.25,0.0625) {};
            \node (v2) at (0.75,0.125+0.0625) {};
            \node (v4) at (0.25,0.0625+3) {};
            \node (v3) at (0.75,0.125+0.0625+3) {};        \draw[opacity=0.25,fill=blue] (v1.center)--(v2.center)--(v3.center)--(v4.center)--(v1.center); 

            \node (u1) at (0,0.75) {};
            \node (u2) at (2,0.75+0.5) {};
            \node (u4) at (0,1.25) {};
            \node (u3) at (2,1.25+0.5) {};        \draw[opacity=0.25,fill=blue] (u1.center)--(u2.center)--(u3.center)--(u4.center)--(u1.center); 
            \draw[thick] (0,0)--(0,3);
            \draw[thick] (0.5,0.125)--(0.5,3+0.125);
            \draw[thick] (1,0.25)--(1,3.25);
            \draw[thick] (1.5,0.375)--(1.5,3.375);
            \draw[thick] (2,3.5)--(2,0.5);
            \draw[thick] (0,3)--(2,3.5);
            \draw[thick] (2,0.5)--(0,0);
            \draw[thick] (0,0.5)--(2,1);
            \draw[thick] (0,1)--(2,1.5);
            \draw[thick] (0,1.5)--(2,2);
            \draw[thick] (0,2)--(2,2.5);
            \draw[thick] (0,2.5)--(2,3);
            \draw[thick] (0,3)--(2,3.5);
            \filldraw[red] (1,1.75) circle (2pt); 
            \node at (1,-0.5) {$\mathcal{M}_{3}$};           
            \node at (3,1.75) {$\Leftrightarrow$};
        \end{tikzpicture}
    \end{gathered} \qquad
            \begin{gathered}
        \begin{tikzpicture}
            \node (v1) at (0.25,0.0625) {};
            \node (v2) at (0.75,0.125+0.0625) {};
            \node (v4) at (0.25,0.0625+3) {};
            \node (v3) at (0.75,0.125+0.0625+3) {};        \draw[opacity=0.25,fill=blue] (v1.center)--(v2.center)--(v3.center)--(v4.center)--(v1.center); 

            \node (u1) at (0,0.75) {};
            \node (u2) at (2,0.75+0.5) {};
            \node (u4) at (0,1.25) {};
            \node (u3) at (2,1.25+0.5) {};        \draw[opacity=0.25,fill=blue] (u1.center)--(u2.center)--(u3.center)--(u4.center)--(u1.center); 
            \draw[thick] (0,0)--(0,3);
            \draw[thick] (0.5,0.125)--(0.5,3+0.125);
            \draw[thick] (1,0.25)--(1,3.25);
            \draw[thick] (1.5,0.375)--(1.5,3.375);
            \draw[thick] (2,3.5)--(2,0.5);
            \draw[thick] (0,3)--(2,3.5);
            \draw[thick] (2,0.5)--(0,0);
            \draw[thick] (0,0.5)--(2,1);
            \draw[thick] (0,1)--(2,1.5);
            \draw[thick] (0,1.5)--(2,2);
            \draw[thick] (0,2)--(2,2.5);
            \draw[thick] (0,2.5)--(2,3);
            \draw[thick] (0,3)--(2,3.5);
            
            \draw[thick] (4,0)--(4,3);
            \draw[thick] (4.5,0.125)--(4.5,3+0.125);
            \draw[thick] (5,0.25)--(5,3.25);
            \draw[thick] (5.5,0.375)--(5.5,3.375);
            \draw[thick] (6,3.5)--(6,0.5);
            \draw[thick] (4,3)--(6,3.5);
            \draw[thick] (6,0.5)--(4,0);
            \draw[thick] (4,0.5)--(6,1);
            \draw[thick] (4,1)--(6,1.5);
            \draw[thick] (4,1.5)--(6,2);
            \draw[thick] (4,2)--(6,2.5);
            \draw[thick] (4,2.5)--(6,3);
            \draw[thick] (4,3)--(6,3.5);
            
            \draw[thick,dotted] (0,0)--(4,0);
            \draw[thick,dotted] (0,3)--(4,3);
            \draw[thick,dotted] (2,3.5)--(6,3.5);
            \draw[thick,dotted] (2,0.5)--(6,0.5);
            \draw[thick,red] (1,1.75)--(4,1.75);
            \draw[thick,red,dashed] (4,1.75)--(5,1.75);
            \filldraw[red] (1,1.75) circle (1.5pt);
            \filldraw[red] (5,1.75) circle (1.5pt); 
            \node at (1,-0.5) {$\mathcal{B}_{\textrm{top}}$};           
            \node at (5,-0.5) {$\mathcal{B}_{\textrm{phys}}$};
        \end{tikzpicture}
    \end{gathered}\nonumber 
    \end{equation}
    \caption{An illustration for subsystem SymTFT construction.}
    \label{Fig-Subsystem-SymTFT}
\end{figure}

Suppose we have a (2+1)D theory $\mathfrak{T}_{\textrm{sub}}$ on $\mathcal{M}_3$ with coordinates $(x,y,z)$ that enjoys a $\mathbb{Z}_N$ subsystem symmetry along $x$ and $y$ direction, we can similarly expand it into a (3+1)D subsystem SymTFT formulated on $[0,1] \times \mathcal{M}_3$ with $\tau \in [0,1]$ the bulk direction, as illustrated in Figure~\ref{Fig-Subsystem-SymTFT}. The 2d lattice is $x$-$y$ plane, the bulk direction is $\tau$, and we omit the $z$ direction in the figure. The $\mathbb{Z}_N$ subsystem symmetry is encoded in the topological boundary $\mathcal{B}_{\textrm{top}}=\{0\}\times \mathcal{M}_3$, and generated by the strip operators in the figure. The dynamical details are stored in the physical boundary $\mathcal{B}_{\textrm{phys}}=\{1\}\times \mathcal{M}_3$. A fracton operator at $(x_i,y_j)$ is represented as a line operator stretching between two boundaries.

We will similarly define the topological boundary $\mathcal{B}_{\textrm{top}}$ via the collection of $W_{\tau}$ and $\hat{W}_{\tau}$ operators in the bulk that can simultaneously end on the boundaries. Similar to the 3D $\mathbb{Z}_N$ BF theory introduced in the previous section, there also exist two kinds of topological boundary states where all $W_{\tau}$-operators or $\hat{W}_{\tau}$-operators in the bulk can end on the topological boundary.

\subsubsection*{Dirichlet boundary}
The Dirichlet boundary $\mathcal{B}_{\textrm{top}}$ is defined such that all $W_{\tau}$ operators can end on the boundary and will be denoted as
    \begin{equation}
        \mathcal{L}_{\textrm{Dir}} = \bigoplus_{i,j} W_{\tau}(x_{i},y_{j})\,,
    \end{equation}
and the $T^3$ Dirichlet boundary states  are written as
\begin{equation}\label{Dirichlet-state}
\ket{\mathbf{w}}:=\ket{w_{z,x;j},w_{z,y;i},w_{x;j+\frac12},w_{y;i+\frac12}}\,,
\end{equation}
where all components of the quartet $(w_{z,x;j},w_{z,y;i},w_{x;j+\frac12},w_{y;i+\frac12})$ are $\mathbb{Z}_N$-valued integers. They are subject to the gauge redundancy
\begin{equation}
    (w_{z,x;j},w_{z,y;i}) \sim (w_{z,x;j}+1,w_{z,y;i}+1)\,,\quad \forall(i,j)\,,
\end{equation}
and the constraint
    \begin{equation}
        \prod_{j=1}^{L_y}(-1)^{w_{x;j+\frac12}}\prod_{i=1}^{L_x}(-1)^{w_{y;i+\frac{1}{2}}}=1\,.
    \end{equation}
The electric operators $W$ are diagonalized as
    \begin{equation}\label{Dirichlet-algebra-1}
        \left\{ \begin{array}{l}
            W_{z,x}(y_{j}) \ket{\mathbf{w}} = \omega^{w_{z,x;j}} \ket{\mathbf{w}}\\
            W_{z,y}(x_{i}) \ket{\mathbf{w}} = \omega^{w_{z,y;i}} \ket{\mathbf{w}}\\
            W(y_{j},y_{j+1}) \ket{\mathbf{w}} = \omega^{w_{x;j+\frac12}} \ket{\mathbf{w}}\\
            W(x_i,x_{i+1}) \ket{\mathbf{w}} = \omega^{w_{y;i+\frac12}} \ket{\mathbf{w}}
        \end{array}\right. .
    \end{equation}
And the magnetic operators $\hat{W}$ will shift the eigenvalues when acting on the state $\ket{\mathbf{w}}$
\begin{equation}\label{Dirichlet-algebra-2}
        \left\{ \begin{array}{l}
            \hat{W}(y_{j'-\frac{1}{2}},y_{j'+\frac{1}{2}})\ket{\mathbf{w}} = \ket{w_{z,x;j} + \delta_{j,j'},w_{z,y;i},w_{x;j+\frac12},w_{y;i+\frac12}}\\
            \hat{W}(x_{i'-\frac{1}{2}},x_{i'+\frac{1}{2}})\ket{\mathbf{w}} = \ket{w_{z,x;j},w_{z,y;i} + \delta_{i,i'},w_{x;j+\frac12},w_{y;i+\frac12}}\\            
            \hat{W}_{z,x}(y_{j'+\frac{1}{2}})\ket{\mathbf{w}} = \ket{w_{z,x;j},w_{z,y;i},w_{x;j+\frac12} + \delta_{j,j'},w_{y;i+\frac12}}\\
            \hat{W}_{z,y}(x_{i'+\frac{1}{2}})\ket{\mathbf{w}} = \ket{w_{z,x;j},w_{z,y;i},w_{x;j+\frac12},w_{y;i+\frac12} + \delta_{i,i'}}            
        \end{array} \right. \,.
\end{equation}
Therefore, they are the symmetry operators and defects of the $\mathbb{Z}_N$ subsystem symmetry on the Dirichlet boundary.

The quartet $(w_{z,x;j},w_{z,y;i},w_{x;j+\frac12},w_{y;i+\frac12})$ is understood as the holonomies of the flat $\mathbb{Z}_N$-subsystem symmetry background field $(A^z,A^{xy})$ along $T^3$, which also reflects the twist boundary conditions of fracton/dipole operators. 
\begin{itemize}
    \item A fracton operator $\mathcal{O}(x_i,y_j,z)$ is topological along the $z$-direction. It satisfies the boundary condition given by
        \begin{equation}
            \mathcal{O}(x_i,y_j,z+2\pi R_z) = \omega^{q w_z(x_i,y_j)} \mathcal{O}(x_i,y_j,z)\,,
        \end{equation}
    where $w_z(x_i,y_j) = w_{z,y}(x_i)+w_{z,x}(y_j)$ is the holonomy of $A^z$ on the site $(x_i,y_j)$ and $q$ is the subsystem $\mathbb{Z}_N$ charge of $\mathcal{O}(x_i,y_j,z_k)$.
    \item A dipole operator $\mathcal{D}_{y_j,y_{j+1}}(x_i,z) =\mathcal{O}(x_i,y_j,z) \mathcal{O}^{\dagger}(x_i,y_{j+1},z)$ which can be understood as a pair of fracton operators is topological along $x$-direction. The boundary condition is characterized by
    \begin{equation}
        \mathcal{D}_{y_j,y_{j+1}}(x_{i+L_x},z) = \omega^{q w_{x;j+\frac{1}{2}}} \mathcal{D}_{y_j,y_{j+1}}(x_{i},z)\,.
    \end{equation}
    Similarly, the dipole operator $\mathcal{D}_{x_i,x_{i+1}}(y_j,z) =\mathcal{O}(x_i,y_j,z_k) \mathcal{O}^{\dagger}(x_{i+1},y_{j},z_k)$ is topological along $y$-direction with the boundary condition
    \begin{equation}
        \mathcal{D}_{x_i,x_{i+1}}(y_{j+L_y},z) = \omega^{q w_{y;i+\frac{1}{2}}} \mathcal{D}_{x_i,x_{i+1}}(y_j,z)\,.
    \end{equation}
    Here $(w_{x,j+\frac{1}{2}},w_{y,i+\frac{1}{2}})$ are holonomies of $A^{xy}$ along $x$ and $y$ directions.
\end{itemize}

\subsubsection*{Neumann boundary}
The Neumann boundary is defined such that all $\hat{W}_{\tau}$ operators can end on the boundary. We denote this boundary operator algebra by
\begin{equation}
    \hat{\mathcal{L}}_{\textrm{Neu}} = \bigoplus_{i,j} \hat{W}_{\tau}(x_{i+\frac{1}{2}}, y_{j+\frac{1}{2}})\,.
\end{equation}
Alternatively, one can consider a dual basis of states
\begin{equation}\label{Neumann-state}
    \ket{\hat{\mathbf{w}}} := \ket{\hat{w}_{z,x; j+\frac{1}{2}},\, \hat{w}_{z,y; i+\frac{1}{2}},\, \hat{w}_{x; j},\, \hat{w}_{y; i}}\,,
\end{equation}
with gauge redundancies and constraints analogous to those appearing in the Dirichlet boundary condition.
Here $\hat{W}$ operators are diagonalized as
    \begin{equation}\label{Neumann-algebra-1}
        \left\{ \begin{array}{l}
            \hat{W}_{z,x}(y_{j+\frac{1}{2}}) \ket{\hat{\mathbf{w}}} = \omega^{\hat{w}_{z,x;j+\frac{1}{2}}} \ket{\hat{\mathbf{w}}}\\
            \hat{W}_{z,y}(x_{i+\frac{1}{2}}) \ket{\hat{\mathbf{w}}} = \omega^{\hat{w}_{z,y;i+\frac{1}{2}}} \ket{\hat{\mathbf{w}}}\\
            \hat{W}(y_{j-\frac{1}{2}},y_{j+\frac{1}{2}}) \ket{\hat{\mathbf{w}}} = \omega^{\hat{w}_{x;j}} \ket{\hat{\mathbf{w}}}\\
            \hat{W}(x_{i-\frac{1}{2}},x_{i+\frac{1}{2}}) \ket{\hat{\mathbf{w}}} = \omega^{\hat{w}_{y;i}} \ket{\hat{\mathbf{w}}}
        \end{array}\right. .
    \end{equation}
When acting on the state $\ket{\hat{\mathbf{w}}}$, the electric operators $W$ will shift the dual holonomies
    \begin{equation}\label{Neumann-algebra-2}
        \left\{ \begin{array}{l}
            W(y_{j'},y_{j'+1})\ket{\hat{\mathbf{w}}} = \ket{\hat{w}_{z,x;j+\frac{1}{2}} + \delta_{j,j'},\hat{w}_{z,y;i+\frac{1}{2}},\hat{w}_{x;j},\hat{w}_{y;i}}\\
            W(x_{i'},x_{i'+1})\ket{\hat{\mathbf{w}}} = \ket{\hat{w}_{z,x;j+\frac{1}{2}},\hat{w}_{z,y;i+\frac{1}{2}} + \delta_{i,i'},\hat{w}_{x;j},\hat{w}_{y;i}}\\            
            W_{z,x}(y_{j'})\ket{\hat{\mathbf{w}}} = \ket{\hat{w}_{z,x;j+\frac12}, \hat{w}_{z,y;i+\frac12},\hat{w}_{x;j} + \delta_{j,j'},\hat{w}_{y;i}}\\
            W_{z,y}(x_{i'})\ket{\hat{\mathbf{w}}} = \ket{\hat{w}_{z,x;j+\frac12}, \hat{w}_{z,y;i+\frac12},\hat{w}_{x;j},\hat{w}_{y;i}+ \delta_{i,i'}}            
        \end{array} \right. .
    \end{equation}
Therefore, the electric operators $W$ can be identified as the $\mathbb{Z}_N$ subsystem symmetry generators. 

Notice that the labels of the dual holonomies $\hat{\mathbf{w}}$ are different from $\mathbf{w}$ in the Dirichlet boundary, and they are defined on the dual lattice. The Neumann boundary states $|\hat{\mathbf{w}}\rangle$ are related to the Dirichlet boundary states $|\mathbf{w}\rangle$ via a discrete Fourier transformation,
\begin{equation}\label{Dirichlet-Neumann-relation}
    \ket{\hat{\mathbf{w}}}= \frac{1}{N^{(L_x+L_y-1)}}\sum_{\mathbf{w}\in M_{w}}\omega^{\sum_{i}(\hat{w}_{z,y;i+\frac12}w_{y;i+\frac12}-\hat{w}_{y;i}w_{z,y;i})+\sum_{j}(\hat{w}_{z,x;j+\frac12}w_{x;j+\frac12}-\hat{w}_{x;j}w_{z,x;j})}\ket{\mathbf{w}}\,,
\end{equation}
where we introduce $M_{w}$ as the set of $\mathbb{Z}_N$-valued vector $\mathbf{w}$ satisfying the gauge redundancy and constraint
\begin{equation}
    M_{w} = \left\{\mathbf{w} \Big{|} \prod_{j=1}^{L_y}\omega^{w_{x;j+\frac12}}=\prod_{i=1}^{L_x}\omega^{w_{y;i+\frac{1}{2}}}\, ;  (w_{z,x;j},w_{z,y;i}) \sim (w_{z,x;j}+1,w_{z,y;i}-1) \right\}\,.
\end{equation}
We will also define $M_{\hat{w}}$ as the set of $\mathbb{Z}_N$-valued vector $\hat{\mathbf{w}}$ satisfying similar gauge redundancy and constraint.

\subsubsection*{Subsystem  Kramers-Wannier transformation}

Based on the SymTFT picture, given any $(2+1)$D theory $\mathfrak{T}_{\textrm{sub}}$ with a  subsystem $\mathbb{Z}_N$ symmetry, we can write down the dynamical boundary state as
    \begin{equation}\label{Foliated-Theory-Dynamical-Boundary-State}
        |\chi\rangle = \sum_{\mathbf{w}\in M_{v}} Z_{\mathfrak{T}_{\textrm{sub}}}[\mathbf{w}] |\mathbf{w}\rangle\,,
    \end{equation}
where the coefficient is the partition function of $\mathfrak{T}_{\textrm{sub}}$ on $\mathcal{M}_3$ coupled with the subsystem $\mathbb{Z}_N$ symmetry background $\mathbf{w}$. Choosing $|\mathbf{w}\rangle$ as the topological boundary state one has
    \begin{equation}
        Z_{\mathfrak{T}_{\textrm{sub}}} = \langle \mathbf{w}| \chi \rangle, 
    \end{equation}
which projects back to the partition function of $\mathfrak{T}_{\textrm{sub}}$. Alternatively, choosing the dual boundary state $|\hat{\mathbf{w}}\rangle$ reproduces the partition function of the dual theory 
\begin{align}
        \begin{split}
        Z_{\hat{\mathfrak{T}}_{\textrm{sub}}}(\hat{\mathbf{w}}) =& \langle \hat{\mathbf{w}} |\chi\rangle\\ 
        =& \frac{1}{2^{(L_x+L_y-1)}} \sum_{\mathbf{w}\in M_{v}}(-1)^{\sum_{i}(\hat{w}_{z,y;i+\frac12}w_{y;i+\frac12}+\hat{w}_{y;i}w_{z,y;i})+\sum_{j}(\hat{w}_{z,x;j+\frac12}w_{x;j+\frac12}+\hat{w}_{x;j}w_{z,x;j})} Z_{\mathfrak{T}_{\textrm{sub}}}(\mathbf{w}) \,.        
        \end{split}
\end{align}
The change of boundary conditions recovers the subsystem Kramers-Wannier transformation between the boundary theories.

\subsection{Subsystem \texorpdfstring{$T\ $}\ transformation and other boundary conditions}

In the exotic theory~\eqref{eq:exotic}, we identify a 0-form $SL(2,\mathbb Z_N)$ symmetry
\begin{align}\label{eq:sym2}
    \begin{split}
        &S:\quad A\rightarrow \hat{A}\,,\quad \hat{A}\rightarrow -A\,,\\
        &  T:\quad A\rightarrow A\,,\quad \hat{A} \rightarrow \hat{A} + A\,.
    \end{split}
\end{align}
Naively, the $S$-transformation should swap between the operators $W$ with $\hat{W}$ and exchange the Dirichlet boundary with the Neumann boundary. On the other hand, the $T$-transformation is expected to map the operators $\hat{W}$ to $\hat{W}W$ and leave $W$ invariant. However, since $W$ and $\hat{W}$ live on lattices that are dual to each other, one needs to define the transformation carefully on the lattices.

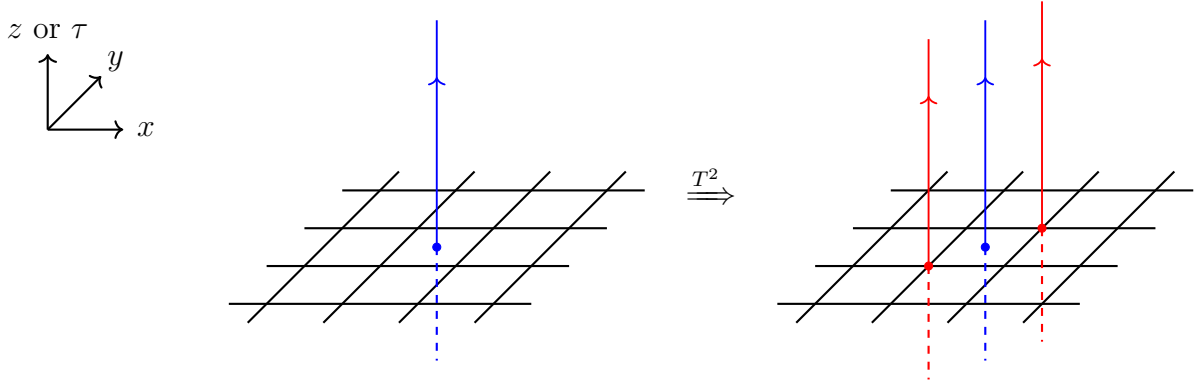
\begin{figure}
    \begin{equation}
            \begin{gathered}
        \begin{tikzpicture}
            \draw[->,thick] (0,0)--(0,1);
            \draw[->,thick] (0,0)--(0.7071,0.7071);
            \draw[->,thick] (0,0)--(1,0);
            \node at (0,1.3) {$z$ or $\tau$};
            \node at (1.3,0) {$x$};
            \node at (0.9,0.9) {$y$};
            \node at (0,-3) {};
        \end{tikzpicture}
    \end{gathered}\qquad
            \begin{gathered}
        \begin{tikzpicture}
            \draw[line,thick] (0,0)--(4,0);
            \draw[line,thick] (0.5,0.5)--(4.5,0.5);
            \draw[line,thick] (1,1)--(5,1);
            \draw[line,thick] (1.5,1.5)--(5.5,1.5);
            \draw[line,thick] (0.25,-0.25)--(0.25+2,-0.25+2);
            \draw[line,thick] (0.25+1,-0.25)--(0.25+3,-0.25+2);
            \draw[line,thick] (0.25+2,-0.25)--(0.25+4,-0.25+2);
            \draw[line,thick] (0.25+3,-0.25)--(0.25+5,-0.25+2);
            \filldraw[blue] (2.75,0.75) circle (1.5pt);  
            \begin{scope}[thick,decoration={markings,mark=at position 0.75 with {\arrow{>}}}] 
            \draw[line,blue,postaction={decorate}] (2.75,0.75)--(2.75,3.75);
            \end{scope}
            \draw[line,thick,dashed,blue] (2.75,0.75)--(2.75,-0.75);            
        \end{tikzpicture}
    \end{gathered} \quad \xRightarrow{T^2} \quad             \begin{gathered}
        \begin{tikzpicture}
            \draw[line,thick] (0,0)--(4,0);
            \draw[line,thick] (0.5,0.5)--(4.5,0.5);
            \draw[line,thick] (1,1)--(5,1);
            \draw[line,thick] (1.5,1.5)--(5.5,1.5);
            \draw[line,thick] (0.25,-0.25)--(0.25+2,-0.25+2);
            \draw[line,thick] (0.25+1,-0.25)--(0.25+3,-0.25+2);
            \draw[line,thick] (0.25+2,-0.25)--(0.25+4,-0.25+2);
            \draw[line,thick] (0.25+3,-0.25)--(0.25+5,-0.25+2);
            \filldraw[blue] (2.75,0.75) circle (1.5pt);  
            \filldraw[red] (3.5,1) circle (1.5pt);  
            \filldraw[red] (2,0.5) circle (1.5pt);  
            \begin{scope}[thick,decoration={markings,mark=at position 0.75 with {\arrow{>}}}] 
            \draw[line,blue,postaction={decorate}] (2.75,0.75)--(2.75,3.75);
             \draw[line,red,postaction={decorate}] (3.5,1)--(3.5,4);
             \draw[line,red,postaction={decorate}] (2,0.5)--(2,3.5);
             \end{scope}
            \draw[line,thick,dashed,blue] (2.75,0.75)--(2.75,-0.75);
             \draw[line,thick,dashed,red] (3.5,1)--(3.5,-0.5);
             \draw[line,thick,dashed,red] (2,0.5)--(2,-1);
        \end{tikzpicture}
    \end{gathered}\nonumber 
    \end{equation}
    \caption{The $T^2$-transformation dresses a pair of $W_z$ operators to $\hat{W}_z$.}
    \label{Fig-T-square-transformation}
\end{figure}

We will mainly focus on $T$-transformation, which generates a new set of operators and refers to the discussion of $S$-transformation in \cite{Cao2024SymTFT}. For the $T$-transformation, we need to dress every magnetic operator $\hat{W}$ with a nearby electric operator $W$. In the canoncial quantization picture, given $\hat{W}_{z,y}(x_{i+\frac{1}{2}})$ and $\hat{W}(x_{i-\frac{1}{2}},x_{i+\frac{1}{2}})$ operators which depends on $x$ only, one could consider the following transformation
\begin{equation}\label{SL2Z-Duality-T-Transformation-Fake}
    \hat{W}_{z,y}(x_{i+\frac{1}{2}}) \rightarrow  \hat{W}_{z,y}(x_{i+\frac{1}{2}}) W_{z,y}(x_{i+1})\,,\quad \hat{W}(x_{i-\frac{1}{2}},x_{i+\frac{1}{2}}) \rightarrow \hat{W}(x_{i-\frac{1}{2}},x_{i+\frac{1}{2}}) W (x_i,x_{i+1})\,,
\end{equation}
where the $W$-operators are on the right of $\hat{W}$-operators. However, the quantum algebras \eqref{Foliated-Theory-Operators-Albetra-Discrete-1} and~\eqref{Foliated-Theory-Operators-Albetra-Discrete-2} are not preserved, and the above transformation is not a good symmetry on the lattice. One can then try to modify the transformation \eqref{SL2Z-Duality-T-Transformation-Fake} in a way consistent with the algebra
\begin{equation}
    \hat{W}_{z,y}(x_{i+\frac{1}{2}}) \rightarrow W_{z,y}(x_i) \hat{W}_{z,y}(x_{i+\frac{1}{2}})\,,\quad \hat{W}(x_{i-\frac{1}{2}},x_{i+\frac{1}{2}}) \rightarrow \hat{W}(x_{i-\frac{1}{2}},x_{i+\frac{1}{2}}) W (x_i,x_{i+1})\,,
\end{equation}
where we move $W_{z,y}$ to the left of $\hat{W}_{z,y}$ and one can check the quantum algebras \eqref{Foliated-Theory-Operators-Albetra-Discrete-1} and~\eqref{Foliated-Theory-Operators-Albetra-Discrete-2} are preserved under the transformation.  But this is still inconsistent because it violates the fact that we can bend a pair of line operators into a strip operator as shown in Figure~\ref{Fig-Line-Strip-Transit}. To obtain a consistent transformation compatible with both quantum algebras and topological properties, we have to consider the $T^2$ transformation and dress the $W$ operators on both sides
\begin{equation}
    \begin{gathered}
    \hat{W}_{z,y}(x_{i+\frac{1}{2}}) \rightarrow W_{z,y}(x_i) \hat{W}_{z,y}(x_{i+\frac{1}{2}}) W_{z,y}(x_{i+1})\,,\\
    \hat{W}(x_{i-\frac{1}{2}},x_{i+\frac{1}{2}}) \rightarrow W (x_{i-1},x_{i}) \hat{W}(x_{i-\frac{1}{2}},x_{i+\frac{1}{2}}) W (x_i,x_{i+1})\,.
    \end{gathered}
\end{equation}
One can consider the similar transformation of $\hat{W}(y_{j-\frac{1}{2}},y_{j+\frac{1}{2}})$ and $\hat{W}_{x}(y_{j+\frac{1}{2}})$ independently. In this paper, we will mainly focus on the following transformation 
\begin{equation}\label{ZN-T-squre-transformation}
    T^2: \quad \left\{ \begin{array}{l}
        \hat{W}_{z,y}(x_{i+\frac{1}{2}}) \rightarrow W_{z,y}(x_i) \hat{W}_{z,y}(x_{i+\frac{1}{2}}) W_{z,y}(x_{i+1})\\
        \hat{W}_{z,x}(y_{j+\frac{1}{2}}) \rightarrow W_{z,x}(y_j) \hat{W}_{z,x}(y_{j+\frac{1}{2}}) W_{z,x}(y_{j+1})\\
        \hat{W}(x_{i-\frac{1}{2}},x_{i+\frac{1}{2}}) \rightarrow W (x_{i-1},x_{i}) \hat{W}(x_{i-\frac{1}{2}},x_{i+\frac{1}{2}}) W (x_i,x_{i+1})\\
        \hat{W}(y_{j-\frac{1}{2}},y_{j+\frac{1}{2}}) \rightarrow W (y_{j-1},y_{j}) \hat{W}(y_{j-\frac{1}{2}},y_{j+\frac{1}{2}}) W (y_j,y_{j+1})\\
    \end{array} \right. 
\end{equation}
where all $\hat{W}$ operators are sandwiched by a pair of $W$ operators in a symmetric way. Recall that
\begin{equation}
    W_z(x_i,y_j) = W_{z,y}(x_i) W_{z,x}(y_j)\,,\quad \hat{W}_{z}(x_{i+\frac{1}{2}},y_{j+\frac{1}{2}}) = \hat{W}_{z,y}(x_{i+\frac{1}{2}}) \hat{W}_{z,x}(y_{j+\frac{1}{2}})\,,
\end{equation}
and therefore the $T^2$ transformation maps the line operator $\hat{W}_z(x_{i+\frac{1}{2}},x_{j+\frac{1}{2}})$ according to
\begin{equation}
    \begin{split}
    T^2: \quad \hat{W}_z(x_{i+\frac{1}{2}},y_{j+\frac{1}{2}}) \rightarrow W_z(x_i,y_j)\hat{W}_z(x_{i+\frac{1}{2}},y_{j+\frac{1}{2}})W_z(x_{i+1},y_{j+1})\,,
    \end{split}
\end{equation}
where we dress two $W_z$ operators at $(x_i,y_j)$ and $(x_{i+1},y_{j+1})$\footnote{It is equivalent to dressing the $W_z$ line operators at $(x_{i+1},y_j)$ and $(x_{i},y_{j+1})$. To illustrate that, begin with the line operators $W_{z}(x_i,y_j)$ and $W_{z}(x_{i+1},y_{j+1})$, and we add a pair of $W_{z}(x_{i+1},y_{j})W^{-1}_{z}(x_{i+1},y_{j})$ at $(x_{i+1},y_j)$ and another pair of $W_{z}(x_{i},y_{j+1})W^{-1}_{z}(x_{i},y_{j+1})$ at $(x_{i},y_{j+1})$. Then we can move the combination $W_{z}(x_i,y_j) W^{-1}_{z}(x_i,y_{j+1})$ along the $x$-direction to annihilate with $W^{-1}_{z}(x_{i+1},y_{j})W_{z}(x_{i+1},y_{j+1})$. After that, only $W_{z}(x_{i},y_{j+1})$ and $W_{z}(x_{i+1},y_{j})$ survive.
} as shown in Figure~\ref{Fig-T-square-transformation}.
Moreover, since one can deform a $W_z(x_i,y_j)$ operator into $W_{\tau}(x_i,y_j)$ on the $(z,\tau)$ plane, one also expects the $\hat{W}_{\tau}$ operator is also dressed by two $W_{\tau}$ operators in the same way. By doing $T^2$-transformation, we will get $N-1$ different topological boundaries $\mathcal{L}_{\textrm{sub},k}$ such that $\hat{W}_{\tau}(x_{i+\frac{1}{2}},y_{j+\frac{1}{2}})$ operators dressed with $k$-pairs of $W_{\tau}(x_{i+1},y_{j+1})$ and $W_{\tau}(x_{i},y_{j})$ operators can end at the boundary. Therefore, we have the following algebra
    \begin{equation}\label{Subsystem-Lagrangian-Algebra-1}
        \hat{\mathcal{L}}_{\textrm{sub},k} = \bigoplus_{i,j} W^k_{\tau}(x_{i},y_{j})\hat{W}_{\tau}(x_{i+\frac{1}{2}},y_{j+\frac{1}{2}})W^k_{\tau}(x_{i+1},y_{j+1})\,.
    \end{equation}
Similarly, one can also consider another set of algebras where the roles of $\hat{W}$ and $W$ are exchanged
    \begin{equation}\label{Subsystem-Lagrangian-Algebra-2}
        \mathcal{L}_{\textrm{sub},k} = \bigoplus_{i,j}\hat{W}^k_{\tau}(x_{i-\frac{1}{2}},y_{j-\frac{1}{2}})W_{\tau}(x_{i},y_{j})\hat{W}^k_{\tau}(x_{i+\frac{1}{2}},y_{j+\frac{1}{2}})\,.
    \end{equation}
In particular, when $k=0$ they reduce to the $\hat{\mathcal{L}}_{\textrm{Neu}}$ and $\mathcal{L}_{\textrm{Dir}}$ algebra we introduced before.

        \begin{figure}
        \begin{equation}
          \begin{gathered}
        \begin{tikzpicture}
            \draw[line,thick] (0,0)--(4,0);
            \draw[line,thick] (0.5,0.5)--(4.5,0.5);
            \draw[line,thick] (1,1)--(5,1);
            \draw[line,thick] (1.5,1.5)--(5.5,1.5);
            \draw[line,thick] (0.25,-0.25)--(0.25+2,-0.25+2);
            \draw[line,thick] (0.25+1,-0.25)--(0.25+3,-0.25+2);
            \draw[line,thick] (0.25+2,-0.25)--(0.25+4,-0.25+2);
            \draw[line,thick] (0.25+3,-0.25)--(0.25+5,-0.25+2);
            \filldraw[blue] (2.75,0.75) circle (1.5pt);  
            \filldraw[red] (3.5,1) circle (1.5pt);  
            \filldraw[red] (2.5,1) circle (1.5pt);  
            \begin{scope}[thick,decoration={markings,mark=at position 0.75 with {\arrow{>}}}] 
            \draw[line,blue,postaction={decorate}] (2.75,0.75)--(2.75,3.75);
             \draw[line,red,postaction={decorate}] (3.5,1)--(3.5,4);
             \draw[line,red,postaction={decorate}] (2.5,1)--(2.5,4);
             \end{scope}
            \draw[line,thick,dashed,blue] (2.75,0.75)--(2.75,-0.75);
             \draw[line,thick,dashed,red] (3.5,1)--(3.5,-0.5);
             \draw[line,thick,dashed,red] (2.5,1)--(2.5,-0.5);
        \end{tikzpicture}
    \end{gathered} \qquad
          \begin{gathered}
        \begin{tikzpicture}
            \draw[line,thick] (0,0)--(4,0);
            \draw[line,thick] (0.5,0.5)--(4.5,0.5);
            \draw[line,thick] (1,1)--(5,1);
            \draw[line,thick] (1.5,1.5)--(5.5,1.5);
            \draw[line,thick] (0.25,-0.25)--(0.25+2,-0.25+2);
            \draw[line,thick] (0.25+1,-0.25)--(0.25+3,-0.25+2);
            \draw[line,thick] (0.25+2,-0.25)--(0.25+4,-0.25+2);
            \draw[line,thick] (0.25+3,-0.25)--(0.25+5,-0.25+2);
            \filldraw[blue] (2.75,0.75) circle (1.5pt);  
            \filldraw[red] (2.5,1) circle (1.5pt);  
            \filldraw[red] (2,0.5) circle (1.5pt);  
            \begin{scope}[thick,decoration={markings,mark=at position 0.75 with {\arrow{>}}}] 
            \draw[line,blue,postaction={decorate}] (2.75,0.75)--(2.75,3.75);
             \draw[line,red,postaction={decorate}] (2.5,1)--(2.5,4);
             \draw[line,red,postaction={decorate}] (2,0.5)--(2,3.5);
             \end{scope}
            \draw[line,thick,dashed,blue] (2.75,0.75)--(2.75,-0.75);
             \draw[line,thick,dashed,red] (2.5,1)--(2.5,-0.5);
             \draw[line,thick,dashed,red] (2,0.5)--(2,-1);
        \end{tikzpicture}
    \end{gathered} \nonumber
    \end{equation}
    \caption{Another two kinds of $T^2$-transformation which are related to the Jordan-Wigner transformations. Here, each red line is a $N/2$ stack of $W_z$ operators. }
    \label{Fig-Fermionic-Dressing}
    \end{figure}
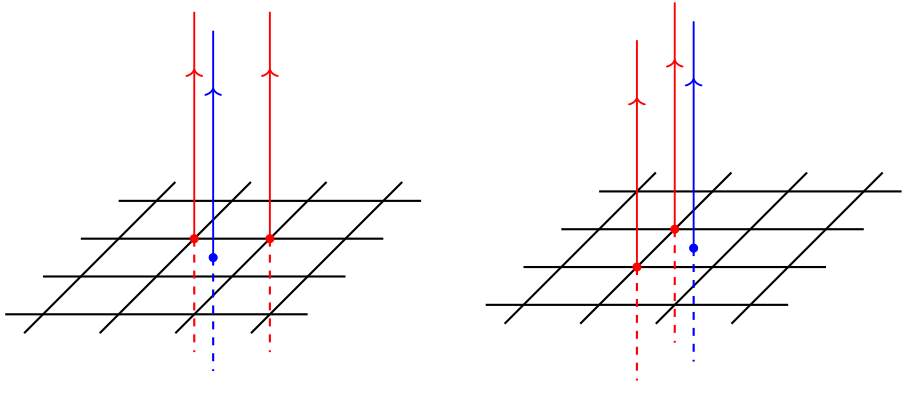

Before ending this section, we point out that there exist two other kinds of boundary conditions when $N$ is even, as shown in Figure~\ref{Fig-Fermionic-Dressing}. They are described by
\begin{equation}
    T_{F,x}^2: \quad \left\{ \begin{array}{l}
        \hat{W}_{z,y}(x_{i+\frac{1}{2}}) \rightarrow \left[W_{z,y}(x_i)\right]^{\frac{N}{2}} \hat{W}_{z,y}(x_{i+\frac{1}{2}}) \left[W_{z,y}(x_{i+1})\right]^{\frac{N}{2}}\\
        \hat{W}_{z,x}(y_{j+\frac{1}{2}}) \rightarrow \hat{W}_{z,x}(y_{j+\frac{1}{2}})\\
        \hat{W}(x_{i-\frac{1}{2}},x_{i+\frac{1}{2}}) \rightarrow \left[W (x_{i-1},x_{i})\right]^{\frac{N}{2}} \hat{W}(x_{i-\frac{1}{2}},x_{i+\frac{1}{2}}) \left[W (x_i,x_{i+1})\right]^{\frac{N}{2}}\\
        \hat{W}(y_{j-\frac{1}{2}},y_{j+\frac{1}{2}}) \rightarrow \hat{W}(y_{j-\frac{1}{2}},y_{j+\frac{1}{2}})\\
    \end{array} \right. 
\end{equation}
and also
\begin{equation}
    T_{F,y}^2: \quad \left\{ \begin{array}{l}
        \hat{W}_{z,y}(x_{i+\frac{1}{2}}) \rightarrow \hat{W}_{z,y}(x_{i+\frac{1}{2}})\\
        \hat{W}_{z,x}(y_{j+\frac{1}{2}}) \rightarrow \left[W_{z,x}(y_j)\right]^{\frac{N}{2}} \hat{W}_{z,x}(y_{j+\frac{1}{2}}) \left[W_{z,x}(y_{j+1})\right]^{\frac{N}{2}}\\
        \hat{W}(x_{i-\frac{1}{2}},x_{i+\frac{1}{2}}) \rightarrow \hat{W}(x_{i-\frac{1}{2}},x_{i+\frac{1}{2}})\\
        \hat{W}(y_{j-\frac{1}{2}},y_{j+\frac{1}{2}}) \rightarrow \left[W (y_{j-1},y_{j})\right]^{\frac{N}{2}} \hat{W}(y_{j-\frac{1}{2}},y_{j+\frac{1}{2}}) \left[W (y_j,y_{j+1})\right]^{\frac{N}{2}}\\
    \end{array} \right. 
\end{equation}
and both of them are compatible with the quantum algebras and topological properties. They are characterized by the following dressing of $W_z$ operators
\begin{equation}
    \begin{split}
    T_{F,x}^2:& \quad \hat{W}_z(x_{i+\frac{1}{2}},y_{j+\frac{1}{2}}) \rightarrow \left[W_z(x_i,y_{j+1})\right]^{\frac{N}{2}}\hat{W}_z(x_{i+\frac{1}{2}},y_{j+\frac{1}{2}})\left[W_z(x_{i+1},y_{j+1})\right]^{\frac{N}{2}}\,,\\
    T_{F,y}^2:& \quad \hat{W}_z(x_{i+\frac{1}{2}},y_{j+\frac{1}{2}}) \rightarrow \left[W_z(x_{i+1},y_j)\right]^{\frac{N}{2}}\hat{W}_z(x_{i+\frac{1}{2}},y_{j+\frac{1}{2}})\left[W_z(x_{i+1},y_{j+1})\right]^{\frac{N}{2}}\,,
    \end{split}
\end{equation}
which are drawn in the figure. As stated in \cite{Cao2024SymTFT}, they define the topological boundary states that are related to the Jordan-Wigner transformation for the $\mathbb{Z}_2$ subgroup of the $\mathbb{Z}_N$ subsystem symmetry. Since our focus is primarily on bosonic SSPT phases, we will not consider such topological boundaries in this paper, leaving their exploration to future work.

\section{Subsystem Gapped Phases and SymTFT}
\label{sec:SymTFTclass}

In this section, we will study the gapped phase of (2+1)D theory $\mathfrak{T}_{\textrm{sub}}$ with an abelian subsystem $G$-symmetry using the SymTFT method. We will consider $G=\mathbb{Z}_N$ first and then move to the more general case $G=\mathbb{Z}_N \times \mathbb{Z}_M$. The cases for general abelian group $G=\bZ_{N_1}\times \cdots \times \bZ_{N_k}$ can be generalized straightforwardly.

\subsection{$G=\mathbb{Z}_N$ subsystem symmetry}\label{sec-ZN-SSPT}
\label{sec:ZNSymTFT}
As discussed in the previous section, there are two sets of topological boundaries labeled by $\mathcal{L}_{\textrm{sub},k}$ and $\hat{\mathcal{L}}_{\textrm{sub},k}$ with $k=0,\cdots,N-1.$ We will choose the topological boundary $\mathcal{B}_{\textrm{top}}$ to be the Dirichlet boundary $\mathcal{L}_{\textrm{sub},0}= \mathcal{L}_{\textrm{Dir}}$ where all $W_{\tau}$ operators can end on the boundary, and the $\mathbb{Z}_N$ subsystem symmetry on $\mathcal{B}_{\textrm{top}}$ is generated by $\hat{W}$ operators. The topological boundary state on the torus is $|\mathbf{w}\rangle$, which satisfies the algebras \eqref{Dirichlet-algebra-1} and \eqref{Dirichlet-algebra-2}.

\subsubsection*{SSB phase}
If the physical boundary is also of Dirichlet type $\mathcal{L}_{\textrm{phys}} = \mathcal{L}_{\textrm{Dir}}$, we will get the SSB phase where all $W_{\tau}$ operators starting from the physical boundary can end on the topological boundary. Let us count the number of such line operators stretching between two boundaries. Since $W_{\tau}$ is topological along the $z$-direction, we only need to focus on the $(x,y)$-plane. Recall that we can decompose $W_{\tau}(x,y)$ into a pair of $W_{\tau,x}(y)$ and $W_{\tau,y}(x)$ and they are separately mobile along $x$ and $y$ directions, and we can move them to the boundary of the lattice. Therefore, we have in total $N^{L_x+L_y}$ numbers of combination by counting all $W_{\tau,x}(y)$ and $W_{\tau,y}(x)$ operators. Moreover, since $W_{\tau,x}$ and $W_{\tau,y}$ satisfy the gauge redundancy $W_{\tau,x} \sim \omega W_{\tau,x}\,,W_{\tau,y} \sim \omega^{-1} W_{\tau,y}\,$, we need to choose a gauge invariant configuration.  That reduces the number of combinations by a factor of $N$ so that there are $N^{L_x+L_y-1}$ different combinations, which equals the number of vacua in the SSB phase. The partition function on a discrete $T^3$ is
    \begin{equation}
        Z_{\textrm{sub,SSB}} [\mathbf{w}] =  \langle \mathbf{w} |\textrm{SSB}\rangle = N^{L_x+L_y-1} \delta_{\mathbf{w},0} \,,\quad  |\textrm{SSB}\rangle = N^{L_x+L_y-1}|\mathbf{0}\rangle\,,
    \end{equation}
where the delta symbol is defined as $\delta_{\mathbf{w},0}=\delta_{w_{z,x;j},0} \delta_{w_{z,y;i},0} \delta_{w_{x,j+\frac{1}{2}},0} \delta_{w_{y,i+\frac{1}{2}},0}$. We have in total $N^{L_x+L_y-1}$ number of vacua, and they carry different charges under the subsystem $\mathbb{Z}_N$ symmetry.
    
\subsubsection*{Trivial phase}
Consider the physical boundary to be the Neumann boundary $\mathcal{L}_{\textrm{phys}} = \hat{\mathcal{L}}_{\textrm{Neu}}$, we will get the trivial phase where only the identity operator starting from the physical boundary can end on the topological boundary, which indicates a unique vacuum. Other $\hat{W}_{\tau}$ operators can transit to the symmetry defect $\hat{W}_z(x_i,y_j)$ along the $z$-direction. After we shrink the interval, they are the operators that create different twist sectors, and one also expects that there exists a unique neutral ground state in each twist sector. Indeed, the partition function on the discrete $T^3$ is trivial
    \begin{equation}
        Z_{\textrm{sub,Trivial}}[\mathbf{w}] = \langle \mathbf{w} |\textrm{Tri}\rangle = 1\,, \quad |\textrm{Tri}\rangle = N^{L_x+L_y-1}|\hat{\mathbf{0}}\rangle\,,
    \end{equation}
and is independent of the holonomies $\mathbf{w}$, where $|\hat{\mathbf{0}}\rangle$ is the Neumann vacuum introduced in \eqref{Neumann-state}, which is related to the Dirichlet boundary state $|\mathbf{w}\rangle$ via a discrete Fourier transformation \eqref{Dirichlet-Neumann-relation} 
\begin{equation}
    \ket{\hat{\mathbf{0}}}= \frac{1}{N^{(L_x+L_y-1)}}\sum_{\mathbf{w}\in M_{v}}\ket{\mathbf{w}}\,.
\end{equation}

\subsubsection*{SSPT phase}
We then consider the physical boundary characterized by other $\mathcal{L}_{\textrm{phys}}=\hat{\mathcal{L}}_{\textrm{sub},k}$ introduced in \eqref{Subsystem-Lagrangian-Algebra-1}, where $\hat{W}_{\tau}$ with $k$-pair of $W_{\tau}$ operators can end at the physical boundary. Similar to the previous case, only the identity operator starting from the physical boundary can end on the topological boundary, which implies a unique ground state. Other $\hat{W}_{\tau}$ operators will transit to the symmetry defect along the $z$-direction that creates different twist sectors after we shrink the interval. However, since $\hat{W}_{\tau}$ operators are also decorated with a pair of $W_{\tau}$ operators, which can end on the topological boundary, so that the corresponding twist operators will carry non-trivial subsystem charges. Depending on the choices of $k$, we can obtain $N-1$ non-trivial SSPT phases.

The partition functions are
    \begin{equation}\label{SSPT-partition-function}
        Z_{\textrm{sub,SSPT};k}[\mathbf{w}] = \langle \mathbf{w} |\textrm{SSPT},k\rangle\,,\quad |\textrm{SSPT},k\rangle = N^{L_x+L_y-1}|\hat{\mathbf{0}}\rangle_{\hat{\mathcal{L}}_k}\,,
    \end{equation}
where $|\mathbf{0}\rangle_{\hat{\mathcal{L}}_k}$ is the vacuum of the topological boundary state that diagonalizes the operators
    \begin{equation}
        \left[W_{z,y}(x_i)\right]^k \hat{W}_{z,y}(x_{i+\frac{1}{2}}) \left[W_{z,y}(x_{i+1})\right]^k\,, \quad \left[W_{z,x}(y_j)\right]^k \hat{W}_{z,x}(y_{j+\frac{1}{2}}) \left[W_{z,x}(y_{j+1})\right]^k\,,
    \end{equation}
and
    \begin{equation}
        \left[W (x_{i-1},x_{i})\right]^k \hat{W}(x_{i-\frac{1}{2}},x_{i+\frac{1}{2}}) \left[W (x_i,x_{i+1})\right]^k\,,\quad \left[W (y_{j-1},y_{j})\right]^k \hat{W}(y_{j-\frac{1}{2}},y_{j+\frac{1}{2}}) \left[W (y_j,y_{j+1})\right]^k\,.
    \end{equation}
To obtain $|\mathbf{0}\rangle_{\hat{\mathcal{L}}_k}$, notice that the Kramers-Wannier transformation in \eqref{Dirichlet-Neumann-relation} will swap $|\mathbf{w}\rangle$ to $|\hat{\mathbf{w}}\rangle$ and thus exchange the role between $W$ with $\hat{W}$. Based on this observation, we can first do a $\left(T^2\right)^k$ transformation to the Dirichlet boundary state $|\mathbf{w}\rangle$ so that the operators given above become the generators of $\mathbb{Z}_N$ subsystem symmetry of $\left(T^2\right)^k|\mathbf{w}\rangle$. Then we consider a Kramers-Wannier transformation to make the resulting states diagonalized by those operators.

To be concrete, notice that the Dirichlet boundary state $|\mathbf{w}\rangle$ can be created by acting $\hat{W}$ operators on the vacuum $|\mathbf{0}\rangle$
    \begin{equation}
        |\mathbf{w}\rangle = \left[\hat{W}(y_{j-\frac{1}{2}},y_{j+\frac{1}{2}})\right]^{w_{z,x;j}}\left[\hat{W}(x_{i-\frac{1}{2}},x_{i+\frac{1}{2}})\right]^{w_{z,y;i}} \left[\hat{W}_{z,x}(y_{j+\frac{1}{2}})\right]^{w_{x;j+\frac{1}{2}}}\left[\hat{W}_{z,y}(x_{i+\frac{1}{2}})\right]^{w_{y;i+\frac{1}{2}}} |\mathbf{0}\rangle\,.
    \end{equation}
Assuming the vacuum $|\mathbf{0}\rangle$ is invariant under the $T^2$-transformation, then the action of $\left(T^2\right)^k$ can be simply achieved by replacing the $\hat{W}$ operators to the transformed version $W\hat{W}W$ as
    \begin{equation}\label{ZN-SSPT-phase-boundary-state}
        \begin{split}
        &(T^2)^k|\mathbf{w}\rangle\\ =& \left\{\left[W (y_{j-1},y_{j})\right]^k \hat{W}(y_{j-\frac{1}{2}},y_{j+\frac{1}{2}}) \left[W (y_j,y_{j+1})\right]^k
        \right\}^{w_{z,x;j}}\\
        &\left\{\left[W (x_{i-1},x_{i})\right]^k \hat{W}(x_{i-\frac{1}{2}},x_{i+\frac{1}{2}}) \left[W (x_i,x_{i+1})\right]^k\right\}^{w_{z,y;i}}\\
        &\left\{\left[W_{z,x}(y_j)\right]^k \hat{W}_{z,x}(y_{j+\frac{1}{2}}) \left[W_{z,x}(y_{j+1})\right]^k\right\}^{w_{x;j+\frac{1}{2}}}\\
        &\left\{\left[W_{z,y}(x_i)\right]^k \hat{W}_{z,y}(x_{i+\frac{1}{2}}) \left[W_{z,y}(x_{i+1})\right]^k\right\}^{w_{y;i+\frac{1}{2}}} |\mathbf{0}\rangle\,.
        \end{split}
    \end{equation}
Move all $W$-operators to the right and use the algebra \eqref{Dirichlet-algebra-1} and \eqref{Dirichlet-algebra-2}, one obtains
    \begin{equation}
        (T^2)^k|\mathbf{w}\rangle = \omega^{k w_{x;j+\frac{1}{2}}(w_{z,x;j}+w_{z,x;j+1}) + k w_{y;i+\frac{1}{2}}(w_{z,y;i}+w_{z,y;i+1})} |\mathbf{w}\rangle\,,
    \end{equation}
where we stack a phase on the state $|\mathbf{w}\rangle$. Consider a further Kramers-Wannier transformation, and we get
    \begin{equation}\label{Boundary-state-k-hat}
    \begin{split}
|\hat{\mathbf{w}}\rangle_{\hat{\mathcal{L}}_{k}} =& \frac{1}{N^{(L_x+L_y-1)}}\sum_{\mathbf{w}\in M_{w}}\omega^{\sum_{i}(\hat{w}_{z,y;i+\frac12}w_{y;i+\frac12}-\hat{w}_{y;i}w_{z,y;i})+\sum_{j}(\hat{w}_{z,x;j+\frac12}w_{x;j+\frac12}-\hat{w}_{x;j}w_{z,x;j})}\nonumber\\
&\times \omega^{k w_{x;j+\frac{1}{2}}(w_{z,x;j}+w_{z,x;j+1}) + k w_{y;i+\frac{1}{2}}(w_{z,y;i}+w_{z,y;i+1})}\ket{\mathbf{w}}\,.
    \end{split}
    \end{equation}
Setting $\hat{\mathbf{w}}=\hat{\mathbf{0}}$ in $|\hat{\mathbf{w}}\rangle_{\hat{\mathcal{L}}_{k}}$, the partition function in \eqref{SSPT-partition-function} reads
    \begin{equation}\label{ZN-SSPT}
        Z_{\textrm{sub,SSPT};k}[\mathbf{w}] = \omega^{k w_{x;j+\frac{1}{2}}(w_{z,x;j}+w_{z,x;j+1}) + k w_{y;i+\frac{1}{2}}(w_{z,y;i}+w_{z,y;i+1})}\,,
    \end{equation}
which is the phase we stack under the $(T^2)^k$-transformation.

\begin{figure}[!h]
    \begin{equation}
            \begin{gathered}
        \begin{tikzpicture}
            \node at (-1,1.5) {$y$};
            \node at (-0.5+1.75,-0.75) {$x$};
            \draw[line,thick] (-0.5,0-2)--(2.5,3-2);    
            \draw[line,thick] (-0.5,0-1)--(2.5,3-1);    
            \draw[line,thick] (-0.5,0)--(2.5,3);
            \draw[line,thick] (-0.5,1)--(2.5,3+1);
            \draw[line,thick] (-0.5,2)--(2.5,3+2);
            \draw[line,thick] (-0.5,3)--(2.5,3+3);
            \draw[line,thick] (-0.5,4)--(2.5,3+4);
            \draw[line,thick] (-0.5,5)--(2.5,3+5);
            \draw[line,thick] (-0.5,-2)--(-0.5,5);
            \draw[line,thick] (-0.5+0.75,0+0.75-2)--(-0.5+0.75,5+0.75);
            \draw[line,thick] (-0.5+1.5,0+1.5-2)--(-0.5+1.5,5+1.5);
            \draw[line,thick] (-0.5+2.25,0+2.25-2)--(-0.5+2.25,5+2.25);
            \draw[line,thick] (-0.5+3,0+3-2)--(-0.5+3,5+3);
            \begin{scope}[line,thick,decoration={markings,mark=at position 0.75 with {\arrow{<}}}] 
            \draw [blue,postaction={decorate}] (0.25+0.375,3.625+1)--(0.25+0.375+4,3.625+1);
            \draw[red,postaction={decorate}] (0.25+0.375+0.375,3.625+1+0.875)--(0.25+0.375+0.375+4,3.625+1+0.875);      
            \draw[red,postaction={decorate}] (0.25,3.625+1-0.875)--(0.25+4,3.625+1-0.875);
        \end{scope}

            \begin{scope}[line,thick,decoration={markings,mark=at position 0.75 with {\arrow{>}}}] 
            \draw[blue,postaction={decorate}] (0.25+0.375,3.625+1-4)--(0.25+0.375+4,3.625+1-4);
            \draw[red,postaction={decorate}] (0.25+0.375+0.375,3.625+1+0.875-4)--(0.25+0.375+0.375+4,3.625+1+0.875-4);
            \draw[red,postaction={decorate}] (0.25,3.625+1-0.875-4)--(0.25+4,3.625+1-0.875-4);
            \end{scope}




            \filldraw[red] (0.25,3.625+1-0.875-4) circle (2.5pt);  
            \filldraw[red] (0.25,3.625+1-0.875) circle (2.5pt); 
            \filldraw[red] (0.25+0.375+0.375,3.625+1+0.875) circle (2.5pt);  
            \filldraw[red] (0.25+0.375+0.375,3.625+1+0.875-4) circle (2.5pt); 
            \node (v1) at (0.25+0.375,3.625+1) {};
            \node (v2) at (0.25+0.375,3.625+1-4) {};
            \node (v3) at (0.25+2.25,3.625+2.875) {};
            \node (v4) at (0.25+2.25,3.625+2.875-4) {};        \draw[opacity=0.25,fill=blue] (v1.center)--(v2.center)--(v4.center)--(v3.center)--(v1.center);          
            \node at (0.25+0.375+5.5,3.625+1) {$\hat{W}_{\tau}(x_{i+\frac{1}{2}},y_{j+\frac{1}{2}})$};
            \node at (0.25+0.375+0.375+5.5,3.625+1+0.875) {$W^k_{\tau}(x_{i+1},y_{j+1})$};
            \node at (0.25+5,3.625+1-0.875) {$W^k_{\tau}(x_{i},y_{j})$};
            \node at (0.25+0.375+5.5,3.625+1-4) {$\hat{W}_{\tau}(x_{i+\frac{1}{2}},y_{j-\frac{7}{2}})$};
            \node at (0.25+0.375+0.375+5.5,3.625+1+0.875-4) {$W^k_{\tau}(x_{i+1},y_{j-3})$};
            \node at (0.25+5.25,3.625+1-0.875-4) {$W^k_{\tau}(x_{i},y_{j-4})$};
        \end{tikzpicture}
    \end{gathered}\quad 
    \begin{gathered}
        \begin{tikzpicture}
        \draw[line,thick] (0,0)--(4,0);  
        \draw[line,thick] (0,1)--(4,1);  
        \draw[line,thick] (0,2)--(4,2); 
        \draw[line,thick] (0,3)--(4,3);  
        \draw[line,thick] (0,4)--(4,4);  
        \draw[line,thick] (0,5)--(4,5);  
        \draw[line,thick] (0,6)--(4,6);  
        \draw[line,thick] (0,0)--(0,6);
        \draw[line,thick] (1,0)--(1,6);
        \draw[line,thick] (2,0)--(2,6);
        \draw[line,thick] (3,0)--(3,6);
        \draw[line,thick] (4,0)--(4,6);
        \node (v1) at (1.5,4.5) {};
        \node (v2) at (1.5,1.5) {};
        \node (v3) at (4,1.5) {};
        \node (v4) at (4,4.5) {};
        \draw[opacity=0.25,fill=blue] (v1.center)--(v2.center)--(v3.center)--(v4.center)--(v1.center);    
        \filldraw[red] (1,1) circle (2.5pt); 
        \filldraw[red] (1,4) circle (2.5pt); 
        \filldraw[red] (2,2) circle (2.5pt); 
        \filldraw[red] (2,5) circle (2.5pt); 
        \node at (1.7,5.3) {$k$};
        \node at (0.7,4.3) {$k$};
        \node at (1.6,2.3) {$-k$};
        \node at (0.6,1.3) {$-k$};
        \node at (2,-0.5) {$x$};
        \node at (-0.5,3) {$y$};
        \end{tikzpicture}
    \end{gathered}\nonumber
    \end{equation}
    \caption{The SymTFT description of $\mathbb{Z}_N$ SSPT phase.}
    \label{Fig-SSPT-ZN-1}
\end{figure}
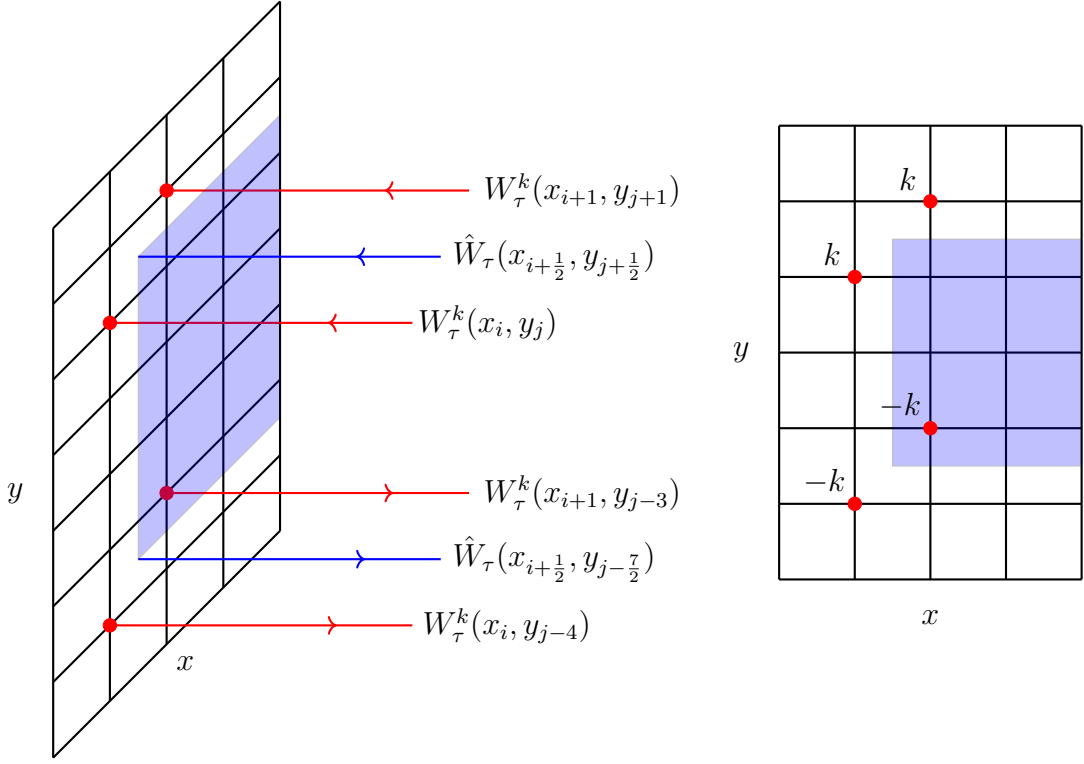

To connect the results in \cite{Devakul2018classifcation} where the $\bZ_N$ SSPT phases are classified by $\mathbb{Z}_N$, suppose we have a pair of $W_{\tau}\hat{W}_{\tau}W_{\tau}$ line operators, with different orientations, starting from the physical boundary as shown in the Figure~\ref{Fig-SSPT-ZN-1}, where their $x$-coordinates are the same. The $W_{\tau}(x,y)$ line can be absorbed by the topological boundary and it ends at $(x,y)$, while the two $\hat{W}_{\tau}$ operators can transit to a strip operator along the $x$-direction.

After we shrink the $\tau$-interval, we obtain a truncated subsystem symmetry generator along the $(x,y)$ plane at some fixed $z$, as depicted on the right of Figure~\ref{Fig-SSPT-ZN-1}. Moreover, we have a pair of fracton operators at each corner, and they carry charge $k$ at the left-top corner and $-k$ at the left-bottom corner under the $\mathbb{Z}_N$ subsystem symmetry, depending on the orientations of $W_{\tau}$. We will denote the pair of fracton operators at the top-left (TL) corner as $V^{TL}_{x,y}(x_{i+\frac{1}{2}},y_{j+\frac{1}{2}})$ and the pair at the bottom-left (BL) as $V^{BL}_{x,y}(x_{i+\frac{1}{2}},y_{j-\frac{7}{2}})$.

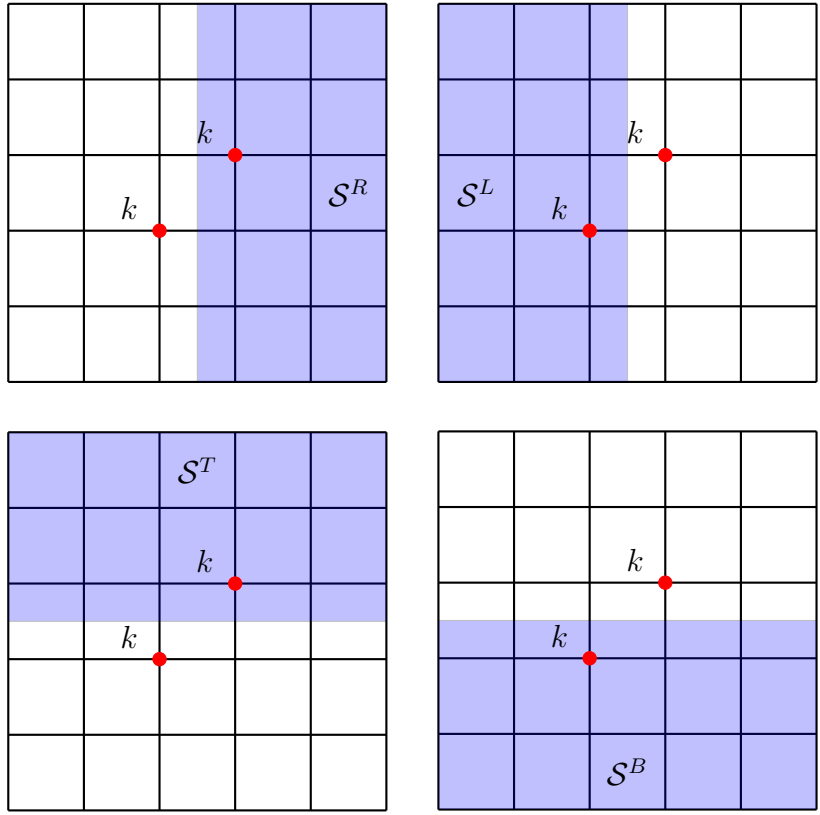
\begin{figure}[!h]
\begin{equation}
    \begin{gathered}
    \begin{gathered}
        \begin{tikzpicture}
        \draw[line,thick] (0,0)--(5,0);  
        \draw[line,thick] (0,1)--(5,1);  
        \draw[line,thick] (0,2)--(5,2); 
        \draw[line,thick] (0,3)--(5,3);  
        \draw[line,thick] (0,4)--(5,4);  
        \draw[line,thick] (0,5)--(5,5);  
        \draw[line,thick] (0,0)--(0,5);
        \draw[line,thick] (1,0)--(1,5);
        \draw[line,thick] (2,0)--(2,5);
        \draw[line,thick] (3,0)--(3,5);
        \draw[line,thick] (4,0)--(4,5);
        \draw[line,thick] (5,0)--(5,5);
        \node (v1) at (2.5,5) {};
        \node (v2) at (2.5,0) {};
        \node (v3) at (5,0) {};
        \node (v4) at (5,5) {};
        \draw[opacity=0.25,fill=blue] (v1.center)--(v2.center)--(v3.center)--(v4.center)--(v1.center);    
        \filldraw[red] (2,2) circle (2.5pt); 
        \filldraw[red] (3,3) circle (2.5pt); 
        \node at (1.6,2.3) {$k$};
        \node at (2.6,3.3) {$k$};
        \node at (4.5,2.5) {$\Scal^R$};
        \end{tikzpicture}
    \end{gathered} \quad
    \begin{gathered}
        \begin{tikzpicture}
        \draw[line,thick] (0,0)--(5,0);  
        \draw[line,thick] (0,1)--(5,1);  
        \draw[line,thick] (0,2)--(5,2); 
        \draw[line,thick] (0,3)--(5,3);  
        \draw[line,thick] (0,4)--(5,4);  
        \draw[line,thick] (0,5)--(5,5);  
        \draw[line,thick] (0,0)--(0,5);
        \draw[line,thick] (1,0)--(1,5);
        \draw[line,thick] (2,0)--(2,5);
        \draw[line,thick] (3,0)--(3,5);
        \draw[line,thick] (4,0)--(4,5);
        \draw[line,thick] (5,0)--(5,5);
        \node (v1) at (2.5,5) {};
        \node (v2) at (2.5,0) {};
        \node (v3) at (0,0) {};
        \node (v4) at (0,5) {};
        \draw[opacity=0.25,fill=blue] (v1.center)--(v2.center)--(v3.center)--(v4.center)--(v1.center);    
        \filldraw[red] (2,2) circle (2.5pt); 
        \filldraw[red] (3,3) circle (2.5pt); 
        \node at (1.6,2.3) {$k$};
        \node at (2.6,3.3) {$k$};
        \node at (0.5,2.5) {$\Scal^L$};
        \end{tikzpicture}
    \end{gathered} \\
        \begin{gathered}
        \begin{tikzpicture}
        \draw[line,thick] (0,0)--(5,0);  
        \draw[line,thick] (0,1)--(5,1);  
        \draw[line,thick] (0,2)--(5,2); 
        \draw[line,thick] (0,3)--(5,3);  
        \draw[line,thick] (0,4)--(5,4);  
        \draw[line,thick] (0,5)--(5,5);  
        \draw[line,thick] (0,0)--(0,5);
        \draw[line,thick] (1,0)--(1,5);
        \draw[line,thick] (2,0)--(2,5);
        \draw[line,thick] (3,0)--(3,5);
        \draw[line,thick] (4,0)--(4,5);
        \draw[line,thick] (5,0)--(5,5);
        \node (v1) at (0,2.5) {};
        \node (v2) at (5,2.5) {};
        \node (v3) at (5,5) {};
        \node (v4) at (0,5) {};
        \draw[opacity=0.25,fill=blue] (v1.center)--(v2.center)--(v3.center)--(v4.center)--(v1.center);    
        \filldraw[red] (2,2) circle (2.5pt); 
        \filldraw[red] (3,3) circle (2.5pt); 
        \node at (1.6,2.3) {$k$};
        \node at (2.6,3.3) {$k$};
        \node at (2.5,4.5) {$\Scal^T$};
        \end{tikzpicture}
    \end{gathered} \quad
    \begin{gathered}
        \begin{tikzpicture}
        \draw[line,thick] (0,0)--(5,0);  
        \draw[line,thick] (0,1)--(5,1);  
        \draw[line,thick] (0,2)--(5,2); 
        \draw[line,thick] (0,3)--(5,3);  
        \draw[line,thick] (0,4)--(5,4);  
        \draw[line,thick] (0,5)--(5,5);  
        \draw[line,thick] (0,0)--(0,5);
        \draw[line,thick] (1,0)--(1,5);
        \draw[line,thick] (2,0)--(2,5);
        \draw[line,thick] (3,0)--(3,5);
        \draw[line,thick] (4,0)--(4,5);
        \draw[line,thick] (5,0)--(5,5);
        \node at (0,5.1) {};
        \node (v1) at (0,2.5) {};
        \node (v2) at (5,2.5) {};
        \node (v3) at (5,0) {};
        \node (v4) at (0,0) {};
        \draw[opacity=0.25,fill=blue] (v1.center)--(v2.center)--(v3.center)--(v4.center)--(v1.center);    
        \filldraw[red] (2,2) circle (2.5pt); 
        \filldraw[red] (3,3) circle (2.5pt); 
        \node at (1.6,2.3) {$k$};
        \node at (2.6,3.3) {$k$};
        \node at (2.5,0.5) {$\Scal^B$};
        \end{tikzpicture}
    \end{gathered}
    \end{gathered}\nonumber
\end{equation}
\caption{The half-plane symmetry generators $\Scal^R(x_{i+\frac{1}{2}}), \Scal^L(x_{i+\frac{1}{2}}), \Scal^T(y_{j+\frac{1}{2}}), \Scal^B(y_{j+\frac{1}{2}})$ acting on the operator $V^{TL}_{x,y}(x_{i+\frac{1}{2}},y_{j+\frac{1}{2}})$.}
\label{Fig-SSPT-ZN-2}
\end{figure}

Let us focus on the $V^{TL}_{x,y}(x_{i+\frac{1}{2}},y_{j+\frac{1}{2}})$ centered at $(x_{i+\frac{1}{2}},y_{j+\frac{1}{2}})$ and define the half-plane symmetry operators following \cite{Devakul2018classifcation}
    \begin{equation}\label{half-plane-symmetry-operators-W}
        \begin{gathered}
        \Scal^R(x_{i+\frac{1}{2}}) = \prod_{i' \geq i} \hat{W}(x_{i'+\frac{1}{2}},x_{i'+\frac{3}{2}})\,, \quad     \Scal^L(x_{i+\frac{1}{2}}) = \prod_{i' < i} \hat{W}(x_{i'+\frac{1}{2}},x_{i'+\frac{3}{2}}) \,,\\
        \Scal^T(y_{j+\frac{1}{2}}) = \prod_{j'\geq j} \hat{W}(y_{j'+\frac{1}{2}},y_{j'+\frac{3}{2}})\,,\quad \Scal^B(y_{j+\frac{1}{2}}) = \prod_{j'< j} \hat{W}(y_{j'+\frac{1}{2}},y_{j'+\frac{3}{2}})\,,
        \end{gathered}
    \end{equation}
where the superscript denotes that we are acting the symmetry to all sites to the right, left, top, and bottom of the coordinate $(x_{i+\frac{1}{2}},y_{j+\frac{1}{2}})$. They are illustrated as the shaded region in Figure~\ref{Fig-SSPT-ZN-2}. The SSPT is characterized by the phase factor $\beta^{\Lambda}_{x,y}(x_{i+\frac{1}{2}},y_{j+\frac{1}{2}})$ defined as
    \begin{equation}\label{beta-factor-W}
        \Scal^{\Lambda}(x_{i+\frac{1}{2}},y_{j+\frac{1}{2}}) V^{TL}_{x,y}(x_{i+\frac{1}{2}},y_{j+\frac{1}{2}}) = \beta^{\Lambda}_{x,y}(x_{i+\frac{1}{2}},y_{j+\frac{1}{2}}) V^{TL}_{x,y}(x_{i+\frac{1}{2}},y_{j+\frac{1}{2}}) \Scal^{\Lambda}(x_{i+\frac{1}{2}},y_{j+\frac{1}{2}})\,,
    \end{equation}
with $\Lambda=R,L,T,B$. Since each shaded region only covers a single fracton with charge $k$, it is easy to deduce that the phase $\beta^{\Lambda}_{x,y}(x_{i+\frac{1}{2}},y_{j+\frac{1}{2}})$ is independent of $\Lambda$ and we have
    \begin{equation}
        \beta^{\Lambda}_{x,y}(x_{i+\frac{1}{2}},y_{j+\frac{1}{2}}) = \omega^{k}\,,\quad (\Lambda=R,L,T,B)
    \end{equation}
which is read from the algebra \eqref{Foliated-Theory-Operators-Albetra-Discrete-1} and \eqref{Foliated-Theory-Operators-Albetra-Discrete-2} by switching $z\leftrightarrow \tau$ therein. Similarly, the bottom-left operator $V^{BL}_{x,y}(x_{i+\frac{1}{2}},y_{j-\frac{7}{2}})$ satisfies
    \begin{equation}
        \Scal^{\Lambda}(x_{i+\frac{1}{2}},y_{j-\frac{7}{2}}) V^{BL}_{x,y}(x_{i+\frac{1}{2}},y_{j-\frac{7}{2}}) = \beta^{*\Lambda}_{x,y}(x_{i+\frac{1}{2}},y_{j-\frac{7}{2}}) V^{BL}_{x,y}(x_{i+\frac{1}{2}},y_{j-\frac{7}{2}}) \Scal^{\Lambda}(x_{i+\frac{1}{2}},y_{j-\frac{7}{2}})\,,
    \end{equation}
with the complex conjugate phase factor $\beta^{*\Lambda}_{x,y}(x_{i+\frac{1}{2}},y_{j-\frac{7}{2}})$ for any $\Lambda=R,L,T,B$. Therefore, we recover the classification of the SSPT phase in \cite{Devakul2018classifcation} for $\mathbb{Z}_N$ case.

The $N-1$ non-trivial SSPT phases, together with the trivial phase, are in one-to-one correspondence to the group elements of $\mathbb{Z}_N$ group, and the group law is implemented by stacking phases. From the SymTFT picture, the $\mathbb{Z}_N$ group is generated by $T^2$-transformation regularized on the lattice.

\subsubsection*{Kramers-Wannier transformation of SSPT}
We can also consider the physical boundary to be $\mathcal{L}_{\textrm{phys}}=\mathcal{L}_{\textrm{sub},k}$ in \eqref{Subsystem-Lagrangian-Algebra-2} where $W_{\tau}$ with $k$-pair of $\hat{W}_{\tau}$ operators can end at the physical boundary. The topological boundary state $|\mathbf{w}\rangle_{\mathcal{L}_{\textrm{sub},k}}$ is similarly obtained as
    \begin{equation}\label{Boundary-state-k}
    \begin{split}
|\mathbf{w}\rangle_{\mathcal{L}_{\textrm{sub},k}} =& \frac{1}{N^{(L_x+L_y-1)}}\sum_{\hat{\mathbf{w}}\in M_{\hat{w}}}\omega^{\sum_{i}(\hat{w}_{y;i}w_{z,y;i}-\hat{w}_{z,y;i+\frac12}w_{y;i+\frac12})+\sum_{j}(\hat{w}_{x;j}w_{z,x;j}-\hat{w}_{z,x;j+\frac12}w_{x;j+\frac12})}\nonumber\\
&\times \omega^{k \hat{w}_{x;j}(\hat{w}_{z,x;j-\frac{1}{2}}+\hat{w}_{z,x;j+\frac{1}{2}}) + k \hat{w}_{y;i}(\hat{w}_{z,y;i-\frac{1}{2}}+\hat{w}_{z,y;i+\frac{1}{2}})}\ket{\hat{\mathbf{w}}}\,,
    \end{split}
    \end{equation}
where we simply exchange the role of $w$ and $\hat{w}$ in \eqref{Boundary-state-k-hat}. Then the corresponding partition function is
    \begin{equation}
        Z_{\textrm{sub,KW;k}}[\mathbf{w}] = \langle\mathbf{w}|\textrm{SSPTKW},k\rangle\,, \quad |\textrm{SSPTKW},k\rangle=N^{L_x+L_y-1}|\mathbf{0}\rangle_{\mathcal{L}_{\textrm{sub},k}}\,,
    \end{equation}
and one has
    \begin{equation}
        \begin{split}
        Z_{\textrm{sub,SSPTKW;k}}[\mathbf{w}] =& \sum_{\hat{\mathbf{w}}\in M_{\hat{w}}} \langle \mathbf{w}|\hat{\mathbf{w}}\rangle\omega^{k \hat{w}_{x;j}(\hat{w}_{z,x;j-\frac{1}{2}}+\hat{w}_{z,x;j+\frac{1}{2}}) + k \hat{w}_{y;i}(\hat{w}_{z,y;i-\frac{1}{2}}+\hat{w}_{z,y;i+\frac{1}{2}})}\\
        =&\frac{1}{N^{(L_x+L_y-1)}} \sum_{\hat{\mathbf{w}}\in M_{\hat{w}}} \omega^{\sum_{i}(\hat{w}_{y;i}w_{z,y;i}-\hat{w}_{z,y;i+\frac12}w_{y;i+\frac12})+\sum_{j}(\hat{w}_{x;j}w_{z,x;j}-\hat{w}_{z,x;j+\frac12}w_{x;j+\frac12})} Z_{\textrm{sub,SSPT;k}}[\hat{\mathbf{w}}]\,,
        \end{split}
    \end{equation}
which is just the partition function of the Kramers-Wannier transformation of the SSPT phase introduced in the last section.

\subsection{$G=\mathbb{Z}_N \times \mathbb{Z}_M$ subsystem symmetry}
\label{sec:ZNZMSymTFT}
Let us move on to $G=\mathbb{Z}_N \times \mathbb{Z}_M$ where we have two pairs of abelian subsystem symmetries and the SymTFT is simply the product of two copies of exotic tensor theories in \eqref{eq:exotic} with level $N$ and level $M$
\begin{equation}
    \begin{aligned}
    S =& \frac{N}{2\pi} \int \left[A^{\tau} (\partial_z \hat{A}^{xy} - \partial_x \partial_y \hat{A}^z) - A^z (\partial_{\tau} \hat{A}^{xy} - \partial_x \partial_y \hat{A}^{\tau})- A^{xy} (\partial_{\tau} \hat{A}^z - \partial_z \hat{A}^{\tau})  \right] \\
    +&\frac{M}{2\pi} \int \left[A'^{\tau} (\partial_z \hat{A}'^{xy} - \partial_x \partial_y \hat{A}'^z) - A'^z (\partial_{\tau} \hat{A}'^{xy} - \partial_x \partial_y \hat{A}'^{\tau})- A'^{xy} (\partial_{\tau} \hat{A}'^z - \partial_z \hat{A}'^{\tau})  \right]\,,
    \end{aligned}
\end{equation}
where we introduce another copy of exotic tensor fields $(A'^{\tau},A'^z,A'^{xy})$ and $(\hat{A}'^{\tau},\hat{A}'^{z},\hat{A}'^{xy})$. In addition to the $SL(2,\mathbb{Z}_N)$ and $SL(2,\mathbb{Z}_M)$ symmetries for each copy, there exists a $\widetilde{T}$-transformation which shifts $\hat{A}$ and $\hat{A}'$ fields according to
    \begin{equation}
        \hat{A} \rightarrow \hat{A} + \frac{M}{\gcd(N,M)} A'\,, \quad \hat{A} '\rightarrow \hat{A}'+\frac{N}{\gcd(N,M)} A\,,
    \end{equation}
and the new $\hat{A}$ and $\hat{A}'$ still take value in $\mathbb{Z}_N$ and $\mathbb{Z}_M$ separately. Moreover, one has $\widetilde{T}^{\gcd(N,M)} =1$ since $MA'$ and $NA$ are trivial as $\mathbb{Z}_N$- and $\mathbb{Z}_M$-valued gauge fields. We also need to formulate the transformation correctly on the lattice.

There are two sets of line/strip operators.
One set contains $A$ and $\hat{A}$ operators and they are still given by $W$ and $\hat{W}$ in \eqref{Foliated-Theory-Line-Operators} and \eqref{Foliated-Theory-Strip-Operators}. The other set of line/strip operators is similarly defined by replacing $A$ and $\hat{A}$ with $A'$ and $\hat{A}'$, and we will label them as $W'$ and $\hat{W}'$. For example, after canonical quantization, the first set satisfies the algebra
\begin{align}
    \begin{split}
    W(x_{i},x_{i+1}) \hat{W}_{z,y}(x_{i+\frac{1}{2}}) &= e^{\frac{2\pi i}{N}} \hat{W}_{z,y}(x_{i+\frac{1}{2}}) W(x_{i},x_{i+1})\,,\\
    W(y_{i},y_{i+1}) \hat{W}_{z,x}(y_{j+\frac{1}{2}}) &= e^{\frac{2\pi i}{N}} \hat{W}_{z,x}(y_{j+\frac{1}{2}}) W(y_{i},y_{i+1})\,,
    \end{split}
\end{align}
and
\begin{align}
    \begin{split}
    \hat{W}(x_{i-\frac{1}{2}},x_{i+\frac{1}{2}}) W_{z,y}(x_i) &= e^{-\frac{2\pi i}{N}} W_{z,y}(x_i) \hat{W}(x_{i-\frac{1}{2}},x_{i+\frac{1}{2}})\,,\\
    \hat{W}(y_{j-\frac{1}{2}},y_{j+\frac{1}{2}}) W_{z,x}(y_j) &= e^{-\frac{2\pi i}{N}} W_{z,x}(y_j) \hat{W}(y_{j-\frac{1}{2}},y_{j+\frac{1}{2}})\,.
    \end{split}
\end{align}
The other set satisfies the algebra
\begin{align}
    \begin{split}
    W'(x_{i},x_{i+1}) \hat{W}'_{z,y}(x_{i+\frac{1}{2}}) &= e^{\frac{2\pi i}{M}} \hat{W}'_{z,y}(x_{i+\frac{1}{2}}) W'(x_{i},x_{i+1})\,,\\
    W'(y_{i},y_{i+1}) \hat{W}'_{z,x}(y_{j+\frac{1}{2}}) &= e^{\frac{2\pi i}{M}} \hat{W}'_{z,x}(y_{j+\frac{1}{2}}) W'(y_{i},y_{i+1})\,,
    \end{split}
\end{align}
and
\begin{align}
    \begin{split}
    \hat{W}'(x_{i-\frac{1}{2}},x_{i+\frac{1}{2}}) W'_{z,y}(x_i) &= e^{-\frac{2\pi i}{M}} W'_{z,y}(x_i) \hat{W}'(x_{i-\frac{1}{2}},x_{i+\frac{1}{2}})\,,\\
    \hat{W}'(y_{j-\frac{1}{2}},y_{j+\frac{1}{2}}) W'_{z,x}(y_j) &= e^{-\frac{2\pi i}{M}} W'_{z,x}(y_j) \hat{W}'(y_{j-\frac{1}{2}},y_{j+\frac{1}{2}})\,.    
    \end{split}
\end{align}

The admissible topological boundaries are richer in this model since we can choose the boundary condition for each set of operators separately, which gives four different types of operators
    \begin{equation}
    (\mathcal{L}_{\textrm{sub},k},\mathcal{L}'_{\textrm{sub},k'})\,,\quad (\hat{\mathcal{L}}_{\textrm{sub},k},\mathcal{L}'_{\textrm{sub},k'})\,,\quad (\mathcal{L}_{\textrm{sub},k},\hat{\mathcal{L}}'_{\textrm{sub},k'})\,,\quad (\hat{\mathcal{L}}_{\textrm{sub},k},\hat{\mathcal{L}}'_{\textrm{sub},k'})\,,
    \end{equation}
where we use $\mathcal{L}'_{\textrm{sub},k'}$ and $\hat{\mathcal{L}}'_{\textrm{sub},k'}$ with $k'=0,\cdots,M-1$ to label the algebras for the second copy. In particular, we will choose the topological boundary to be $(\mathcal{L}_{\textrm{sub},0},\mathcal{L}'_{\textrm{sub},0}) = (\mathcal{L}_{\textrm{Dir}},\mathcal{L}'_{\textrm{Dir}})$ type such that $W_{\tau}$ and $W'_{\tau}$ operators can end on the boundary and the $\mathbb{Z}_N\times \mathbb{Z}_M$ subsystem symmetry is generated by $\hat{W}$ and $\hat{W}'$ operators on the topological boundary. If we consider the physical boundary characterized by one of the four types, we will simply obtain a gapped phase that is the product of the $\mathbb{Z}_N$ and $\mathbb{Z}_M$ subsystem gapped phase.

We can obtain more topological boundaries by performing $\widetilde{T}$-transformation. The $\widetilde{T}$-transformation is expected to dress a 
$W'^{\frac{M}{\gcd(N,M)}}$ operator to $\hat{W}$ and a $W^{\frac{N}{\gcd(N,M)}}$ operator to $\hat{W}'$. On the lattice, let us consider the $\widetilde{T}$ transformation defined by
    \begin{equation}\label{T-tilde-plus-1}
    \widetilde{T}(\hat{W}):\quad\left\{
        \begin{array}{l}
            \hat{W}_{z,x}(y_{j+\frac{1}{2}})\rightarrow\hat{W}_{z,x}(y_{j+\frac{1}{2}}) \left[W'_{z,x}(y_{j+1})\right]^{\frac{M}{\gcd(N,M)}}\\
            \hat{W}_{z,y}(x_{i+\frac{1}{2}})\rightarrow\hat{W}_{z,y}(x_{i+\frac{1}{2}}) \left[W'_{z,y}(x_{i+1})\right]^{\frac{M}{\gcd(N,M)}}\\
            \hat{W}(y_{j-\frac{1}{2}},y_{j+\frac{1}{2}})\rightarrow\hat{W}(y_{j-\frac{1}{2}},y_{j+\frac{1}{2}})\left[W'(y_{j},y_{j+1})\right]^{\frac{M}{\gcd(N,M)}}\\
            \hat{W}(x_{i-\frac{1}{2}},x_{i+\frac{1}{2}})\rightarrow\hat{W}(x_{i-\frac{1}{2}},x_{i+\frac{1}{2}})\left[W'(x_{i},x_{i+1})\right]^{\frac{M}{\gcd(N,M)}}
        \end{array}\right.
    \end{equation}
and also
    \begin{equation}\label{T-tilde-plus-2}
    \widetilde{T}(\hat{W}'):\quad\left\{
        \begin{array}{l}
            \hat{W}'_{z,x}(y_{j+\frac{1}{2}})\rightarrow \left[W_{z,x}(y_{j})\right]^{\frac{N}{\gcd(N,M)}} \hat{W}'_{z,x}(y_{j+\frac{1}{2}})\\
            \hat{W}'_{z,y}(x_{i+\frac{1}{2}})\rightarrow\left[W_{z,y}(x_{i})\right]^{\frac{N}{\gcd(N,M)}} \hat{W}'_{z,y}(x_{i+\frac{1}{2}})\\
            \hat{W}'(y_{j-\frac{1}{2}},y_{j+\frac{1}{2}})\rightarrow\left[W(y_{j-1},y_{j})\right]^{\frac{N}{\gcd(N,M)}} \hat{W}'(y_{j-\frac{1}{2}},y_{j+\frac{1}{2}})\\
            \hat{W}'(x_{i-\frac{1}{2}},x_{i+\frac{1}{2}})\rightarrow\left[W(x_{i-1},x_{i})\right]^{\frac{N}{\gcd(N,M)}} \hat{W}'(x_{i-\frac{1}{2}},x_{i+\frac{1}{2}})
        \end{array}\right.
    \end{equation}
where we dress a nearby $W'$ operators to the right of $\hat{W}$, and $W$ operators to the left of $\hat{W}'$\footnote{By applying parity transformation $x\rightarrow-x$ and/or $y\rightarrow-y$ on the $(x,y)$-plane, one can generate other three copies of $\widetilde{T}$ transformation. They will give the same gapped phases and we will not distinguish them.}. Notice that, unlike the $T^2$ (or $T'^2$ for $A’$) transformation in \eqref{ZN-T-squre-transformation}, we do not need to dress $W$ and $W'$ on both sides to preserve the quantum algebra. For $\hat{W}$-operators, since the $W'$-operators they dressed belong to another copy of the exotic theory, they do not talk to $\hat{W}$-operators, and the operators on the RHS of \eqref{T-tilde-plus-1} still commute among themselves as before. Similar to $\hat{W}'$-operators. One only needs to check that the operators on the RHS of \eqref{T-tilde-plus-1} commute with those of \eqref{T-tilde-plus-2}. 
As an example, let us begin with
    \begin{align}
        \hat{W}_{z,x}(y_{j+\frac{1}{2}}) \left[W'_{z,x}(y_{j+1})\right]^{\frac{M}{\gcd(N,M)}}\left[W(y_{j},y_{j+1})\right]^{\frac{N}{\gcd(N,M)}} \hat{W}'(y_{j+\frac{1}{2}},y_{j+\frac{3}{2}})\,.
    \end{align}
Exchange the $\hat{W}_{z,x}(y_{j+\frac{1}{2}})$ with $W(y_{j},y_{j+1})$ gives
    \begin{equation}
        \hat{W}_{z,x}(y_{j+\frac{1}{2}}) \left[W(y_{j},y_{j+1})\right]^{\frac{N}{\gcd(N,M)}} = \exp\left[-\frac{2\pi i }{\gcd(N,M)} \right]\left[W(y_{j},y_{j+1})\right]^{\frac{N}{\gcd(N,M)}}\hat{W}_{z,x}(y_{j+\frac{1}{2}})\,,
    \end{equation}
and exchange the $W'_{z,x}(y_{j+1})$ with $\hat{W}'(y_{j+\frac{1}{2}},y_{j+\frac{3}{2}})$ gives
    \begin{equation}
        \left[W'_{z,x}(y_{j+1})\right]^{\frac{M}{\gcd(N,M)}}\hat{W}'(y_{j+\frac{1}{2}},y_{j+\frac{3}{2}}) = \exp \left[\frac{2\pi i}{\gcd(N,M)} \right] \hat{W}'(y_{j+\frac{1}{2}},y_{j+\frac{3}{2}})\left[W'_{z,x}(y_{j+1})\right]^{\frac{M}{\gcd(N,M)}}\,,
    \end{equation}
so the two phases cancel each other.

According to the decomposition $W_z(x,y) = W_{z,y}(y) W_{z,x}(y)$, we conclude that the $\widetilde{T}$ transformation will maps the line operators $\hat{W}_z(x_{i+\frac{1}{2}},y_{j+\frac{1}{2}})$ and $\hat{W}'_z(x_{i+\frac{1}{2}},y_{j+\frac{1}{2}})$ according to
\begin{equation}
    \widetilde{T}:\quad \left\{\begin{array}{l}
\hat{W}_z(x_{i+\frac{1}{2}},y_{j+\frac{1}{2}}) \rightarrow \hat{W}_z(x_{i+\frac{1}{2}},y_{j+\frac{1}{2}}) \left[W'_z(x_{i+1},y_{j+1}) \right]^{\frac{M}{\gcd(N,M)}}\,,\\ \hat{W}'_z(x_{i+\frac{1}{2}},y_{j+\frac{1}{2}}) \rightarrow \left[W_z(x_{i},y_{j}) \right]^{\frac{N}{\gcd(N,M)}}\hat{W}'_z(x_{i+\frac{1}{2}},y_{j+\frac{1}{2}}) \,,
    \end{array}\right.
\end{equation}
and we will introduce the set of operators $\widetilde{\mathcal{L}}_{k}$ defined as
    \begin{equation}\label{Z2Z2-algebra-1}
        \widetilde{\mathcal{L}}_{k} = \left(\bigoplus_{i,j}\hat{W}_{\tau}(x_{i+\frac{1}{2}},y_{j+\frac{1}{2}}) \left[W'_{\tau}(x_{i+1},y_{j+1}) \right]^{\frac{kM}{\gcd(N,M)}}\right)\bigoplus\left(\bigoplus_{i,j} \left[W_{\tau}(x_{i},y_{j}) \right]^{\frac{kN}{\gcd(N,M)}}\hat{W}'_{\tau}(x_{i+\frac{1}{2}},y_{j+\frac{1}{2}})\right)\,,
    \end{equation}
with $k=0,\cdots, \gcd(N,M)-1$. We can also consider other kinds of algebras by exchanging the role between $W$ with $\hat{W}$ or/and $W'$ with $\hat{W}'$, and we will not repeat them here.

\subsubsection*{SSPT phase}

Let us choose the topological boundary to be the $(\mathcal{L}_{\textrm{Dir}},\mathcal{L}'_{\textrm{Dir}})$-type which supports the $\mathbb{Z}_N\times \mathbb{Z}_M$ subsystem symmetry generated by $\hat{W}$ and $\hat{W}'$ operators, and the physical boundary to be the $\widetilde{\mathcal{L}}_{k}$ type in \eqref{Z2Z2-algebra-1}. Among all operators, only the identity operator starting from the physical boundary can end on the topological boundary, and we have a unique ground state. Other $\hat{W}_{\tau}$ and $\hat{W}'_{\tau}$ operators will transit to the $\mathbb{Z}_N \times \mathbb{Z}_M$ symmetry defects along the $z$-direction that creates different twist sectors after we shrink the interval. Since $\hat{W}_{\tau}$ operators are decorated by $[W'_{\tau}]^{\frac{kM}{\gcd(N,M)}}$, the $\mathbb{Z}_N$ twist-operators will carry $\mathbb{Z}_M$ subsystem charge. On the other hand, the $\hat{W}'_{\tau}$ operators are decorated by $W_{\tau}^{\frac{N}{\gcd(N,M)}}$ so that the $\mathbb{Z}_M$ twist-operators will carry $\mathbb{Z}_N$ subsystem charge.

The partition function is given by
    \begin{equation}
        Z_{\widetilde{\textrm{SSPT}},k}[\mathbf{w},\mathbf{w}'] = \langle \mathbf{w},\mathbf{w}'| \widetilde{\textrm{SSPT}},k\rangle\,,\quad | \widetilde{\textrm{SSPT}},k\rangle = \left(NM\right)^{L_x+L_y-1}|\hat{\mathbf{0}},\hat{\mathbf{0}}\rangle_{\widetilde{\mathcal{L}}_{k}}\,,
    \end{equation}
where $|\mathbf{w},\mathbf{w}'\rangle \equiv |\mathbf{w}\rangle \otimes |\mathbf{w}'\rangle$ with $|\mathbf{w}\rangle $ and $|\mathbf{w}'\rangle$ the Dirichlet boundary states for $\mathbb{Z}_N$ and $\mathbb{Z}_M$ separately. $|\hat{\mathbf{0}},\hat{\mathbf{0}}\rangle_{\widetilde{\mathcal{L}}_{k}}$ is the vacuum of the topological boundary state diagonalizing the operators in \eqref{Z2Z2-algebra-1}. We can use the same trick as in \eqref{ZN-SSPT-phase-boundary-state} to obtain the boundary state. One can find that the $\widetilde{T}^k$-transformation will stack the phase
    \begin{equation}
        \widetilde{T}^k |\mathbf{w},\mathbf{w}'\rangle = \widetilde{\omega}^{ \sum_i k \left(w_{y;i+\frac{1}{2}}w'_{z,y;i+1} + w'_{y;i+\frac{1}{2}}w_{z,y;i}\right) + \sum_j k \left(w_{x;j+\frac{1}{2}}w'_{z,x;j+1} + w'_{x;j+\frac{1}{2}}w_{z,x;j}\right)} |\mathbf{w},\mathbf{w}'\rangle\,,
    \end{equation}
where $\widetilde{\omega}= \exp\left(2\pi i/\gcd(N,M) \right)$ is the $\gcd(N,M)$-root of unity. And the topological boundary states $|\hat{\mathbf{w}},\hat{\mathbf{w}}'\rangle_{\widetilde{\mathcal{L}}_{k}}$ diagonalizing the operators in \eqref{Z2Z2-algebra-1} is the Kramers-Wannier transformation of the above for both $\mathbb{Z}_N$ and $\mathbb{Z}_M$ factors. Then we have 
    \begin{equation}
        Z_{\widetilde{\textrm{SSPT}},k}[\mathbf{w},\mathbf{w}'] = \widetilde{\omega}^{ \sum_i k \left(w_{y;i+\frac{1}{2}}w'_{z,y;i+1} + w'_{y;i+\frac{1}{2}}w_{z,y;i}\right) + \sum_j k \left(w_{x;j+\frac{1}{2}}w'_{z,x;j+1} + w'_{x;j+\frac{1}{2}}w_{z,x;j}\right)}\,.
    \end{equation}

\begin{figure}[!h]
    \begin{equation}
            \begin{gathered}
        \begin{tikzpicture}
            \node at (-1,1.5) {$y$};
            \node at (-0.5+1.75,-0.75) {$x$};
            \draw[line,thick] (-0.5,0-2)--(2.5,3-2);    
            \draw[line,thick] (-0.5,0-1)--(2.5,3-1);    
            \draw[line,thick] (-0.5,0)--(2.5,3);
            \draw[line,thick] (-0.5,1)--(2.5,3+1);
            \draw[line,thick] (-0.5,2)--(2.5,3+2);
            \draw[line,thick] (-0.5,3)--(2.5,3+3);
            \draw[line,thick] (-0.5,4)--(2.5,3+4);
            \draw[line,thick] (-0.5,5)--(2.5,3+5);
            \draw[line,thick] (-0.5,-2)--(-0.5,5);
            \draw[line,thick] (-0.5+0.75,0+0.75-2)--(-0.5+0.75,5+0.75);
            \draw[line,thick] (-0.5+1.5,0+1.5-2)--(-0.5+1.5,5+1.5);
            \draw[line,thick] (-0.5+2.25,0+2.25-2)--(-0.5+2.25,5+2.25);
            \draw[line,thick] (-0.5+3,0+3-2)--(-0.5+3,5+3);
            \begin{scope}[line,thick,decoration={markings,mark=at position 0.75 with {\arrow{<}}}] 
            \draw [blue,postaction={decorate}] (0.25+0.375,3.625+1)--(0.25+0.375+4,3.625+1);
            \draw [blue,dashed] (0.25+0.375,3.625+1-0.1)--(0.25+0.375+4,3.625+1-0.1);
            \draw[red,dashed,postaction={decorate}] (0.25+0.375+0.375,3.625+1+0.875)--(0.25+0.375+0.375+4,3.625+1+0.875);      
            \draw[red,postaction={decorate}] (0.25,3.625+1-0.875)--(0.25+4,3.625+1-0.875);
        \end{scope}
            \begin{scope}[line,thick,decoration={markings,mark=at position 0.75 with {\arrow{>}}}] 
            \draw[blue,postaction={decorate}] (0.25+0.375,3.625+1-4)--(0.25+0.375+4,3.625+1-4);
            \draw[blue,dashed] (0.25+0.375,3.625+1-4+0.1)--(0.25+0.375+4,3.625+1-4+0.1);
            \draw[red,dashed,postaction={decorate}] (0.25+0.375+0.375,3.625+1+0.875-4)--(0.25+0.375+0.375+4,3.625+1+0.875-4);
            \draw[red,postaction={decorate}] (0.25,3.625+1-0.875-4)--(0.25+4,3.625+1-0.875-4);
            \end{scope}
            \filldraw[red] (0.25,3.625+1-0.875-4) circle (2.5pt);  
            \filldraw[red] (0.25,3.625+1-0.875) circle (2.5pt); 
            \filldraw[red] (0.25+0.375+0.375-0.075,3.625+1+0.875-0.075) rectangle ++(0.15,0.15);  
            \filldraw[red] (0.25+0.375+0.3750-0.075,3.625+1+0.875-4-0.075) rectangle ++(0.15,0.15);  
            \node (v1) at (0.25+0.375,3.625+1) {};
            \node (v2) at (0.25+0.375,3.625+1-4) {};
            \node (v3) at (0.25+2.25,3.625+2.875) {};
            \node (v4) at (0.25+2.25,3.625+2.875-4) {};        \draw[opacity=0.25,fill=blue] (v1.center)--(v2.center)--(v4.center)--(v3.center)--(v1.center);          
            \node at (0.25+0.375+7,3.625+1) {$\hat{W}_{\tau}(x_{i+\frac{1}{2}},y_{j+\frac{1}{2}})\hat{W}'_{\tau}(x_{i+\frac{1}{2}},y_{j+\frac{1}{2}})$};
            \node at (0.25+0.375+0.375+6.25,3.625+1+0.875) {$\left[W'_{\tau}(x_{i+1},y_{j+1})\right]^{\frac{kM}{\gcd(N,M)}}$};
            \node at (0.25+6,3.625+1-0.875) {$\left[W_{\tau}(x_{i},y_{j})\right]^{\frac{kN}{\gcd(N,M)}}$};
            \node at (0.25+0.375+7,3.625+1-4) {$\hat{W}_{\tau}(x_{i+\frac{1}{2}},y_{j-\frac{7}{2}})\hat{W}'_{\tau}(x_{i+\frac{1}{2}},y_{j-\frac{7}{2}})$};
            \node at (0.25+0.375+0.375+6.25,3.625+1+0.875-4) {$\left[W'_{\tau}(x_{i+1},y_{j-3})\right]^{\frac{k M}{\gcd(N,M)}}$};
            \node at (0.25+6.25,3.625+1-0.875-4) {$\left[W_{\tau}(x_{i},y_{j-4})\right]^{\frac{k N}{\gcd(N,M)}}$};
        \end{tikzpicture}
    \end{gathered}\quad
    \begin{gathered}
        \begin{tikzpicture}
        \draw[line,thick] (0,0)--(4,0);  
        \draw[line,thick] (0,1)--(4,1);  
        \draw[line,thick] (0,2)--(4,2); 
        \draw[line,thick] (0,3)--(4,3);  
        \draw[line,thick] (0,4)--(4,4);  
        \draw[line,thick] (0,5)--(4,5);  
        \draw[line,thick] (0,6)--(4,6);  
        \draw[line,thick] (0,0)--(0,6);
        \draw[line,thick] (1,0)--(1,6);
        \draw[line,thick] (2,0)--(2,6);
        \draw[line,thick] (3,0)--(3,6);
        \draw[line,thick] (4,0)--(4,6);
        \node (v1) at (1.5,4.5) {};
        \node (v2) at (1.5,1.5) {};
        \node (v3) at (4,1.5) {};
        \node (v4) at (4,4.5) {};
        \draw[opacity=0.25,fill=blue] (v1.center)--(v2.center)--(v3.center)--(v4.center)--(v1.center);    
        \filldraw[red] (1,1) circle (2.5pt); 
        \filldraw[red] (1,4) circle (2.5pt); 
        \filldraw[red] (2-0.075,2-0.075) rectangle ++(0.15,0.15); ; 
        \filldraw[red] (2-0.075,5-0.075) rectangle ++(0.15,0.15); 
        \node at (1.7,5.3) {$k$};
        \node at (0.7,4.3) {$k$};
        \node at (1.6,2.3) {$-k$};
        \node at (0.6,1.3) {$-k$};
        \node at (2,-0.5) {$x$};
        \node at (-0.5,3) {$y$};
        \end{tikzpicture}
    \end{gathered}\nonumber
    \end{equation}
\caption{SymTFT description of $\mathbb{Z}_N \times \mathbb{Z}_M$ SSPT phase.}
    \label{Fig-SSPT-ZNZM-1}
\end{figure}
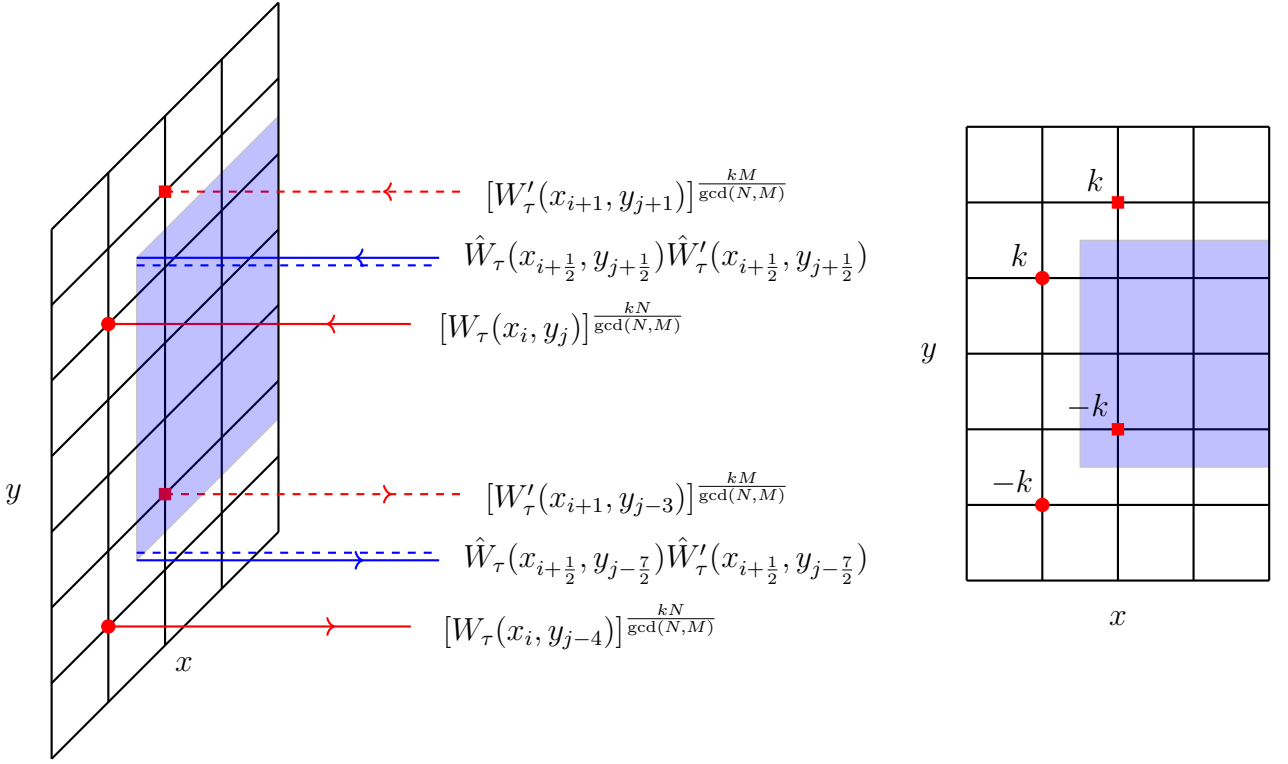

Let us connect the results in \cite{Devakul2018classifcation}. Suppose we have a pair of $\hat{W}_{\tau}(x,y) \hat{W}'_{\tau}(x,y)$ with different orientations, starting from the physical boundary with the same $x$-coordinates and decorated by $W_{\tau}$ and $W'_{\tau}$ operators as shown in Figure~\ref{Fig-SSPT-ZNZM-1}. At the topological boundary, the $\hat{W}_{\tau} \hat{W}'_{\tau}$ pair transit into the strip operator which implements a $\mathbb{Z}_N \times \mathbb{Z}_M$ symmetry transformation labeled by $(g_o,g_e)$. Here we choose the generators of $\mathbb{Z}_N$ and $\mathbb{Z}_M$ separately as $g_o$ and $g_e$ satisfying $g_o^N=g_e^M=1$. On the other hand, the $W_{\tau}$ and $W'_{\tau}$ operators will simply end at the topological boundary.
After we shrink the $\tau$-interval, we obtain a truncated subsystem symmetry generator labeled by $(g_o,g_e)$ along the $(x,y)$ plane at some fixed $z$, as depicted on the right of Figure~\ref{Fig-SSPT-ZNZM-1}. We will denote the pair of fracton operators at the top-left corner as $\widetilde{V}^{TL}_{x,y}(x_{i+\frac{1}{2}},y_{j+\frac{1}{2}})$ and the pair at the bottom-left as $\widetilde{V}^{BL}_{x,y}(x_{i+\frac{1}{2}},y_{j-\frac{7}{2}})$.

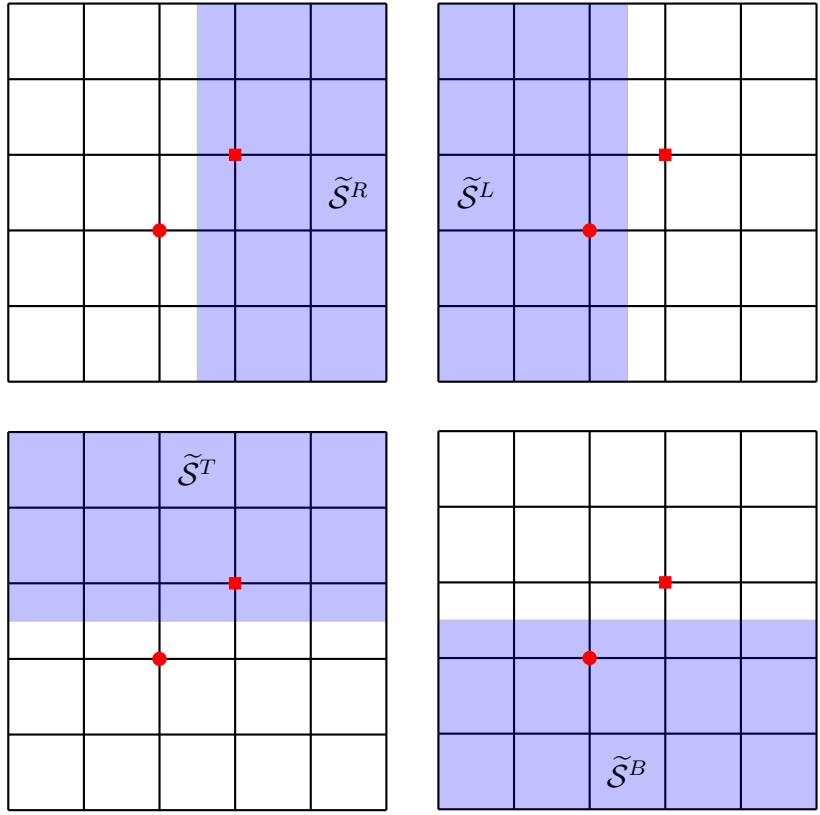
\begin{figure}[!h]
\begin{equation}
    \begin{gathered}
    \begin{gathered}
        \begin{tikzpicture}
        \draw[line,thick] (0,0)--(5,0);  
        \draw[line,thick] (0,1)--(5,1);  
        \draw[line,thick] (0,2)--(5,2); 
        \draw[line,thick] (0,3)--(5,3);  
        \draw[line,thick] (0,4)--(5,4);  
        \draw[line,thick] (0,5)--(5,5);  
        \draw[line,thick] (0,0)--(0,5);
        \draw[line,thick] (1,0)--(1,5);
        \draw[line,thick] (2,0)--(2,5);
        \draw[line,thick] (3,0)--(3,5);
        \draw[line,thick] (4,0)--(4,5);
        \draw[line,thick] (5,0)--(5,5);
        \node (v1) at (2.5,5) {};
        \node (v2) at (2.5,0) {};
        \node (v3) at (5,0) {};
        \node (v4) at (5,5) {};
        \draw[opacity=0.25,fill=blue] (v1.center)--(v2.center)--(v3.center)--(v4.center)--(v1.center);    
        \filldraw[red] (2,2) circle (2.5pt); 
        \filldraw[red] (3-0.075,3-0.075) rectangle ++(0.15,0.15); 
        \node at (4.5,2.5) {$\widetilde{\Scal}^R$};
        \end{tikzpicture}
    \end{gathered} \quad
    \begin{gathered}
        \begin{tikzpicture}
        \draw[line,thick] (0,0)--(5,0);  
        \draw[line,thick] (0,1)--(5,1);  
        \draw[line,thick] (0,2)--(5,2); 
        \draw[line,thick] (0,3)--(5,3);  
        \draw[line,thick] (0,4)--(5,4);  
        \draw[line,thick] (0,5)--(5,5);  
        \draw[line,thick] (0,0)--(0,5);
        \draw[line,thick] (1,0)--(1,5);
        \draw[line,thick] (2,0)--(2,5);
        \draw[line,thick] (3,0)--(3,5);
        \draw[line,thick] (4,0)--(4,5);
        \draw[line,thick] (5,0)--(5,5);
        \node (v1) at (2.5,5) {};
        \node (v2) at (2.5,0) {};
        \node (v3) at (0,0) {};
        \node (v4) at (0,5) {};
        \draw[opacity=0.25,fill=blue] (v1.center)--(v2.center)--(v3.center)--(v4.center)--(v1.center);    
        \filldraw[red] (2,2) circle (2.5pt); 
        \filldraw[red] (3-0.075,3-0.075) rectangle ++(0.15,0.15); 
        \node at (0.5,2.5) {$\widetilde{\Scal}^L$};
        \end{tikzpicture}
    \end{gathered} \\
        \begin{gathered}
        \begin{tikzpicture}
        \draw[line,thick] (0,0)--(5,0);  
        \draw[line,thick] (0,1)--(5,1);  
        \draw[line,thick] (0,2)--(5,2); 
        \draw[line,thick] (0,3)--(5,3);  
        \draw[line,thick] (0,4)--(5,4);  
        \draw[line,thick] (0,5)--(5,5);  
        \draw[line,thick] (0,0)--(0,5);
        \draw[line,thick] (1,0)--(1,5);
        \draw[line,thick] (2,0)--(2,5);
        \draw[line,thick] (3,0)--(3,5);
        \draw[line,thick] (4,0)--(4,5);
        \draw[line,thick] (5,0)--(5,5);
        \node (v1) at (0,2.5) {};
        \node (v2) at (5,2.5) {};
        \node (v3) at (5,5) {};
        \node (v4) at (0,5) {};
        \draw[opacity=0.25,fill=blue] (v1.center)--(v2.center)--(v3.center)--(v4.center)--(v1.center);    
        \filldraw[red] (2,2) circle (2.5pt); 
        \filldraw[red] (3-0.075,3-0.075) rectangle ++(0.15,0.15); 
        \node at (2.5,4.5) {$\widetilde{\Scal}^T$};
        \end{tikzpicture}
    \end{gathered} \quad
    \begin{gathered}
        \begin{tikzpicture}
        \draw[line,thick] (0,0)--(5,0);  
        \draw[line,thick] (0,1)--(5,1);  
        \draw[line,thick] (0,2)--(5,2); 
        \draw[line,thick] (0,3)--(5,3);  
        \draw[line,thick] (0,4)--(5,4);  
        \draw[line,thick] (0,5)--(5,5);  
        \draw[line,thick] (0,0)--(0,5);
        \draw[line,thick] (1,0)--(1,5);
        \draw[line,thick] (2,0)--(2,5);
        \draw[line,thick] (3,0)--(3,5);
        \draw[line,thick] (4,0)--(4,5);
        \draw[line,thick] (5,0)--(5,5);
        \node at (0,5.1) {};
        \node (v1) at (0,2.5) {};
        \node (v2) at (5,2.5) {};
        \node (v3) at (5,0) {};
        \node (v4) at (0,0) {};
        \draw[opacity=0.25,fill=blue] (v1.center)--(v2.center)--(v3.center)--(v4.center)--(v1.center);    
        \filldraw[red] (2,2) circle (2.5pt); 
        \filldraw[red] (3-0.075,3-0.075) rectangle ++(0.15,0.15); 
        \node at (2.5,0.5) {$\widetilde{\Scal}^B$};
        \end{tikzpicture}
    \end{gathered}
    \end{gathered}\nonumber
\end{equation}
\caption{The half-plane symmetry generators $\widetilde{\Scal}^R(x_{i+\frac{1}{2}}),\widetilde{\Scal}^L(x_{i+\frac{1}{2}}),\widetilde{\Scal}^T(y_{j+\frac{1}{2}}),\widetilde{\Scal}^B(y_{j+\frac{1}{2}})$ acting on the operators $\widetilde{V}^{TL}_{x,y}(x_{i+\frac{1}{2}},y_{j+\frac{1}{2}})$.}
\label{Fig-SSPT-ZNZM-2}
\end{figure}

Let us focus on $\widetilde{V}^{TL}_{x,y}(x_{i+\frac{1}{2}},y_{j+\frac{1}{2}})$ centered at $(x_{i+\frac{1}{2}},y_{j+\frac{1}{2}})$ and define the half-plane symmetry operators of $(g_o,g_e)$ as
    \begin{equation}
        \begin{gathered}
        \widetilde{\Scal}^R(x_{i+\frac{1}{2}}) = \prod_{i' \geq i} \hat{W}(x_{i'+\frac{1}{2}},x_{i'+\frac{3}{2}})\hat{W}'(x_{i'+\frac{1}{2}},x_{i'+\frac{3}{2}})\,, \quad     
        \widetilde{\Scal}^L(x_{i+\frac{1}{2}}) = \prod_{i' < i} \hat{W}(x_{i'+\frac{1}{2}},x_{i'+\frac{3}{2}})\hat{W}'(x_{i'+\frac{1}{2}},x_{i'+\frac{3}{2}}) \,,\\
        \widetilde{\Scal}^T(y_{j+\frac{1}{2}}) = \prod_{j'\geq j} \hat{W}(y_{j'+\frac{1}{2}},y_{j'+\frac{3}{2}})\hat{W}'(y_{j'+\frac{1}{2}},y_{j'+\frac{3}{2}})\,,\quad \widetilde{\Scal}^B(y_{j+\frac{1}{2}}) = \prod_{j'< j} \hat{W}(y_{j'+\frac{1}{2}},y_{j'+\frac{3}{2}})\hat{W}'(y_{j'+\frac{1}{2}},y_{j'+\frac{3}{2}})\,,
        \end{gathered}
    \end{equation}
where the superscript denotes that we are acting the symmetry to all sites to the right, left, top, and bottom of the coordinate $(x_{i+\frac{1}{2}},y_{j+\frac{1}{2}})$. They are represented by the shaded region in Figure~\ref{Fig-SSPT-ZNZM-2}.
The SSPT is characterized by the phase factor $\widetilde{\beta}^{\Lambda}_{xy}(x_{i+\frac{1}{2}},y_{j+\frac{1}{2}})$ defined as
    \begin{equation}
        \widetilde{\Scal}^{\Lambda}(x_{i+\frac{1}{2}},y_{j+\frac{1}{2}}) \widetilde{V}^{TL}_{x,y}(x_{i+\frac{1}{2}},y_{j+\frac{1}{2}}) = \widetilde{\beta}_{x,y}^{\Lambda}(x_{i+\frac{1}{2}},y_{j+\frac{1}{2}}) \widetilde{V}^{TL}_{x,y}(x_{i+\frac{1}{2}},y_{j+\frac{1}{2}}) \widetilde{\Scal}^{\Lambda}(x_{i+\frac{1}{2}},y_{j+\frac{1}{2}})\,,
    \end{equation}
with $\Lambda=R,L,T,B$. For $\Lambda=R$ and $T$, since the fracton at $(x_{i+1},y_{j+1})$ carries charge $\frac{kM}{\gcd(N,M)}$ of $\mathbb{Z}_M$ only, then the phase factor $\widetilde{\beta}^{R/T}_{x,y}(x_{i+\frac{1}{2}},y_{j+\frac{1}{2}})$ is
    \begin{equation}
        \widetilde{\beta}^{R}_{x,y}(x_{i+\frac{1}{2}},y_{j+\frac{1}{2}})=\widetilde{\beta}^{T}_{x,y}(x_{i+\frac{1}{2}},y_{j+\frac{1}{2}})= \exp\left(\frac{2\pi i}{M} \times \frac{kM}{\gcd(N,M)} \right) = \widetilde{\omega}^{k}\,.
    \end{equation}
On the other hand, for $\Lambda=L$ and $B$, the fracton at $(x_i,y_j)$ carries charge $\frac{kN}{\gcd(N,M)}$ of $\mathbb{Z}_N$ only, then the phase factor $\widetilde{\beta}^{L/B}_{x,y}(x_{i+\frac{1}{2}},y_{j+\frac{1}{2}})$ is
    \begin{equation}
        \widetilde{\beta}^{L}_{x,y}(x_{i+\frac{1}{2}},y_{j+\frac{1}{2}})=\widetilde{\beta}^{B}_{x,y}(x_{i+\frac{1}{2}},y_{j+\frac{1}{2}})= \exp\left(\frac{2\pi i}{N} \times \frac{kN}{\gcd(N,M)} \right) = \widetilde{\omega}^{k}\,,
    \end{equation}
and we have $\widetilde{\beta}_{x,y}^{\Lambda}(x_{i+\frac{1}{2}},y_{j+\frac{1}{2}}) = \widetilde{\omega}^k$ for all $\Lambda=R,L,T,B$. Similarly, the bottom-left operator $\widetilde{V}^{BL}_{x,y}(x_{i+\frac{1}{2}},y_{j-\frac{7}{2}})$ satisfies
    \begin{equation}
        \widetilde{\Scal}^{\Lambda}(x_{i+\frac{1}{2}},y_{j-\frac{7}{2}}) \widetilde{V}^{BL}_{x,y}(x_{i+\frac{1}{2}},y_{j-\frac{7}{2}}) = \widetilde{\beta}_{x,y}^{* \Lambda}(x_{i+\frac{1}{2}},y_{j-\frac{7}{2}}) \widetilde{V}^{BL}_{x,y}(x_{i+\frac{1}{2}},y_{j-\frac{7}{2}}) \widetilde{\Scal}^{\Lambda}(x_{i+\frac{1}{2}},y_{j-\frac{7}{2}})\,,
    \end{equation}
with the complex conjugate factor $\widetilde{\beta}_{x,y}^{* \Lambda}(x_{i+\frac{1}{2}},y_{j-\frac{7}{2}})$ for any $\Lambda = R,L,T,B$.

We can still consider the truncated symmetry generator for $\mathbb{Z}_N$ or $\mathbb{Z}_M$ only, labeled by $(g_o,1)$ or $(1,g_e)$. In the SymTFT picture, they are created by a pair of $\hat{W}_{\tau}$ operators or a pair of $\hat{W}'_{\tau}$ operators. Compared to Figure~\ref{Fig-SSPT-ZNZM-1}, we can remove either the dashed lines or the solid lines to achieve that. In the former case, we can use the half-plane symmetry operators given in \eqref{half-plane-symmetry-operators-W} for symmetry $\mathbb{Z}_N$ only to measure the phase $\beta^{\Lambda}_{xy}$ defined in \eqref{beta-factor-W}. However, since the dressing operator $W'_{\tau}$ do not carry charges of $\mathbb{Z}_N$, we will simply get $\beta^{\Lambda}_{xy}=1$. In the latter case, we can similarly introduce another set of half-plane symmetry operators for symmetry $\mathbb{Z}_M$ to measure the phase $\beta'^{\Lambda}_{xy}$. Since the dressing operator $W_{\tau}$ do not carry charges of $\mathbb{Z}_M$, we will also get $\beta'^{\Lambda}_{xy}=1$. They agree with the results in \cite{Devakul2018classifcation}.

As summary, for (2+1)D system with the $\mathbb{Z}_N\times\mathbb{Z}_M$ subsystem symmetry, the SSPT phase is classified by
    \begin{equation} \label{eq:ZmZNClass}
        \mathcal{C}\left[\mathbb{Z}_N \times \mathbb{Z}_M \right] = \mathbb{Z}_N \times \mathbb{Z}_M \times \mathbb{Z}_{\gcd(N,M)}\,,
    \end{equation}
where the first two factors classify the individual SSPT phases for the $\mathbb{Z}_N$ and $\mathbb{Z}_M$ symmetry, and they have been discussed in Section~\ref{sec-ZN-SSPT}. The last factor classifies the mixed SSPT phases between $\mathbb{Z}_N$ and $\mathbb{Z}_M$ symmetry discussed above.

The results for $\mathbb{Z}_N\times \mathbb{Z}_M$ can be easily generalized to a general non-anomalous abelian group $\mathbb{Z}_{N_1} \times \mathbb{Z}_{N_2} \times \cdots \times \mathbb{Z}_{N_k}$. We need to introduce $k$ copies of exotic tensor theories in Eq.~\eqref{eq:exotic} with level $N_1, \cdots, N_k$.
In this case the classification is given by
    \begin{equation}\label{eq:classMult}
        \mathcal{C}\left[\mathbb{Z}_{N_1} \times \mathbb{Z}_{N_2} \times \cdots \times \mathbb{Z}_{N_k}\right] = \prod_{i=1}^k \mathbb{Z}_{N_i} \times \prod_{i<j}^k \mathbb{Z}_{\gcd(N_i,N_j)}\,,
    \end{equation}    
where we have the individual SSPT phases for each $\mathbb{Z}_{N_i}$ factor, and also the SSPT phases for each $(N_i,N_j)$ pair. 
Our foliated SymTFT provides an intuitive illustration of the classification in Eq.~\eqref{eq:exotic}. In comparison, the lattice-model-based discussion mainly addresses the classification given in an equivalent expression (Eq.~\eqref{eq:ClassGgeneral}), which may otherwise appear rather mysterious.

\section{Example: $(2+1)$D Cluster State Model}
\label{sec:cluster}

We have used the subsystem SymTFT, which is based on the 2-foliated exotic gauge field theory, to classify SSPT phases. The invariant of each phase is computed using half-space and corner operators. In this section, we take the cluster state model as an illustrative example to meticulously analyze how the invariant associated with it relates to our SymTFT perspective.

A systematic method for constructing lattice Hamiltonian realizations of subsystem SymTFT remains an open problem and is currently under active investigation. Nonetheless, certain special classes of lattice models are known, with a prototypical example being the cluster state model.
The qubit cluster state model corresponds to a lattice Hamiltonian that exhibits $\mathbb{Z}_2 \times \mathbb{Z}_2$ strong SSPT order~\cite{You2018SSPT,Devakul2018classifcation}. Originally introduced in Ref.~\cite{Raussendorf2001} as a resource state for measurement-based quantum computation, it has since become a key example in the study of SSPTs.
In Ref.~\cite{Devakul2018classifcation}, it was argued, based on considerations of the linearly symmetric local unitary circuit, that strong SSPT phases protected by subsystem $G$ symmetry are classified by
\begin{equation}\label{eq:classSSPT}
    \mathcal{C}[G] = H^{2}(G^{\times 2}, U(1)) \big/ \left(H^2(G, U(1))\right)^3\,.
\end{equation}
For the case $G = \mathbb{Z}_N \times \mathbb{Z}_M$, this expression reduces to Eq.~\eqref{eq:ZmZNClass}. In the following, we will revisit the $\mathbb{Z}_2 \times \mathbb{Z}_2$ case as discussed in Ref.~\cite{Devakul2018classifcation}, and then extend the discussion to the $\mathbb{Z}_N \times \mathbb{Z}_M$ setting.
From this analysis, we will see that the preceding discussion based on the 2-foliated exotic gauge field theory aligns well with the results obtained from the cluster state model.

\subsection{$\mathbb{Z}_2\times \mathbb{Z}_2$ cluster state model}

\begin{figure}
    \centering
    \includegraphics[width=0.8\linewidth]{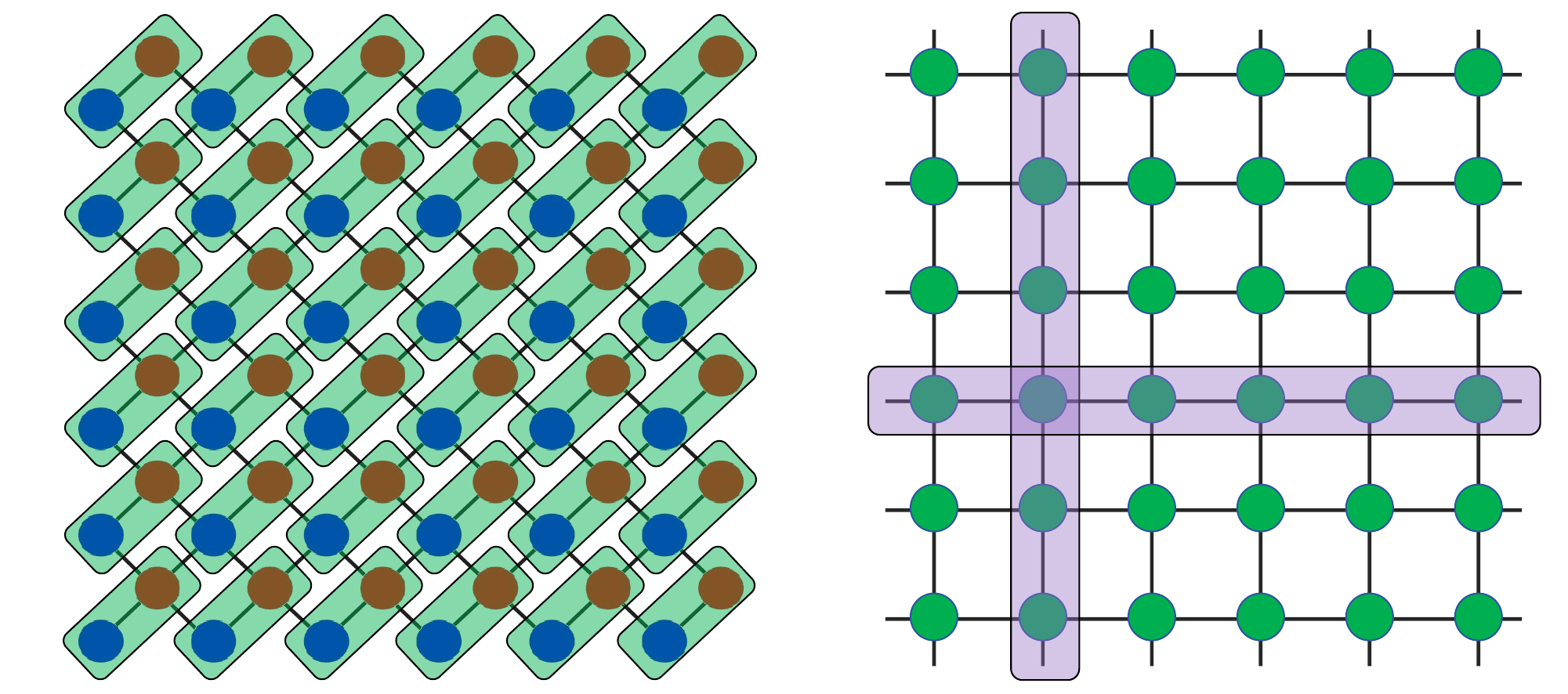}
    \caption{Illustration of the cluster state lattice. On the left is the cluster square lattice, with odd and even vertices colored red and blue, respectively. As the lattice is bipartite, we can pair one odd vertex with one even vertex to form a basis set. The resulting Bravais lattice is also square, but rotated by an angle of $\pi/4$ relative to the original lattice.}
    \label{fig:ClusterLattice}
\end{figure}

Consider a square lattice (shown on the left of Figure~\ref{fig:ClusterLattice}), where each vertex is assigned a qubit and classified as either odd or even (represented by red and blue dots, respectively, in Figure~\ref{fig:ClusterLattice}). Odd vertices are connected only to even vertices, and even vertices are connected only to odd vertices, making the lattice a bipartite graph.
We then pair each odd vertex with a neighboring even vertex to form a basis set. This results in a Bravais square lattice (shown on the right of Figure~\ref{fig:ClusterLattice}), where each vertex now hosts two qubits.

To distinguish the two qubits on each vertex, we denote the Pauli operators for the odd ($v_o$) and even ($v_e$) qubits as $X_{v_o},Y_{v_o},Z_{v_o}=\sigma^{x,y,z}$ and $X_{v_e},Y_{v_e},Z_{v_e}=\tau^{x,y,z}$, respectively. To define the cluster state Hamiltonian, it is more convenient to work with the original cluster lattice. The local stabilizers are defined as
\begin{equation}
    K_{v_o}=\sigma_{v_o}^x \otimes (\bigotimes_{v_e\in N(v_o)} \tau^z_{v_e})=
\begin{aligned}
    \begin{tikzpicture}
\draw[black] (0,0) -- (0.5, 0.5);
\draw[black] (0,0) -- (0.5, -0.5);
\draw[black] (0,0) -- (-0.5, 0.5);
\draw[black] (0,0) -- (-0.5, -0.5);
\fill[red] (0,0) circle (0.15);
\fill[blue] (0.5, 0.5) circle (0.15);
\fill[blue] (0.5, -0.5) circle (0.15);
\fill[blue] (-0.5, 0.5) circle (0.15);
\fill[blue] (-0.5, -0.5) circle (0.15);
\node at (0, 0.4) {$\sigma^x$};
\node at (0.7, 0.9) {$\tau^z$};
\node at (0.7, -0.9) {$\tau^z$};
\node at (-0.7, 0.9) {$\tau^z$};
\node at (-0.7, -0.9) {$\tau^z$};
\end{tikzpicture}
\end{aligned}\,,
\quad
  K_{v_e}=\tau^x_{v_e} \otimes (\bigotimes_{v_o \in N(v_e)} \sigma^z_{v_o})
=
\begin{aligned}
    \begin{tikzpicture}
\draw[black] (0,0) -- (0.5, 0.5);
\draw[black] (0,0) -- (0.5, -0.5);
\draw[black] (0,0) -- (-0.5, 0.5);
\draw[black] (0,0) -- (-0.5, -0.5);
        \fill[blue] (0,0) circle (0.15);
\fill[red] (0.5, 0.5) circle (0.15);
\fill[red] (0.5, -0.5) circle (0.15);
\fill[red] (-0.5, 0.5) circle (0.15);
\fill[red] (-0.5, -0.5) circle (0.15);
\node at (0, 0.4) {$\tau^x$};
\node at (0.7, 0.9) {$\sigma^z$};
\node at (0.7, -0.9) {$\sigma^z$};
\node at (-0.7, 0.9) {$\sigma^z$};
\node at (-0.7, -0.9) {$\sigma^z$};
    \end{tikzpicture}
\end{aligned}.
\end{equation}
The Hamiltonian is given by
\begin{equation}
    H = -\sum_{v_o \in \textrm{odd}} K_{v_o} - \sum_{v_e \in \textrm{even}} K_{v_e}\,,
\end{equation}
which is a local commutative Hamiltonian, meaning that all stabilizers are local and commute with each other.
The ground state of the model is of the form~\cite{Raussendorf2001} (called cluster state or graph state)
\begin{equation}\label{eq:clusterstateZ2}
    |\Psi_{\rm GS}\rangle = \prod_{e} CZ_{e} \left( \bigotimes_{v} |+\rangle_v \right)\,,
\end{equation}
where $CZ_e = (1 + Z_{v_1} + Z_{v_2} - Z_{v_1} Z_{v_2}) /2$ is the control-$Z$ operation that acts on the edge $e = \langle v_1 v_2 \rangle$, and $|+\rangle = \frac{|0\rangle + |1\rangle}{\sqrt{2}}$ is $+1$ eigenstate of $\sigma_x$.
It is easy to verify that $K_{v_o}|\Psi_{\rm GS}\rangle= K_{v_e} |\Psi_{\rm GS}\rangle =|\Psi_{\rm GS}\rangle$ for all $v_o,v_e$, thus $|\Psi_{\rm GS}\rangle$ is the ground state of $H$.

We will focus on the Bravais square lattice (shown on the right of Figure~\ref{fig:ClusterLattice}) from here on, unless otherwise specified. Consider a 2-foliation of the lattice, consisting of a horizontal decomposition into a set of codimension-1 sublattices and a vertical decomposition into another set of codimension-1 sublattices. On each site $(x,y) \in \mathbb{Z} \times \mathbb{Z}$ of lattice, we assign odd and even qubits, thus $\mathcal{H}_{x,y} = \mathcal{H}_{x,y}^{\rm odd} \otimes \mathcal{H}_{x,y}^{\rm even}$, with $\mathcal{H}_{x,y}^{\rm odd} = \mathcal{H}_{x,y}^{\rm even} = \mathbb{C}[\bZ_2]$. The total Hilbert space is $\mathcal{H}_{\rm tot}=\bigotimes_{x,y\in \bZ} \mathcal{H}_{x,y}$. Notice that each codimension-1 leaf of the foliation now contains two subleaves, corresponding to those that contain only odd or even vertices, respectively.

The on-site symmetry group is $G = \mathbb{Z}_2 \times \mathbb{Z}_2 = \{1, g_o, g_e, g_o g_e\}$, where $g_o$ and $g_e$ are the two generators of the group. The local representation of the group is given by
\begin{equation}
    u_{x,y}(g_o) = \sigma_{x,y}^x, \quad u_{x,y}(g_e) = \tau_{x,y}^x\,.
\end{equation}
On an $x$-leaf, the subsystem symmetry is
\begin{equation}
    S_x(g_o) = \prod_{y=-\infty}^{+\infty} \sigma_{x,y}^x, \quad S_x(g_e) = \prod_{y=-\infty}^{+\infty} \tau_{x,y}^x\,.
\end{equation}
Similarly, on a $y$-leaf, the subsystem symmetry is
\begin{equation}
    S_y(g_o) = \prod_{x=-\infty}^{+\infty} \sigma_{x,y}^x, \quad S_y(g_e) = \prod_{x=-\infty}^{+\infty} \tau_{x,y}^x.
\end{equation}
It is straightforward to verify that the cluster state Hamiltonian commutes with these subsystem symmetries.

The truncated symmetry for a square region $[x_0,x_1] \times [y_0,y_1]$ is given by  
\begin{equation}
    U_{[x_0,x_1] \times [y_0,y_1]} (g) = \prod_{(x,y) \in [x_0,x_1] \times [y_0,y_1]} u_{x,y}(g).
\end{equation}  
When acting on the ground state $|\Psi_{\rm GS}\rangle$, it creates four local excitations at the four corners of the square region, commonly denoted as bottom-left(BL), bottom-right(BR), top-right(TR), and top-left(TL). There exist four local operators, $V_{x,y}(g)$, satisfying the relation  
\begin{equation}
    V_{x,y}(g) = V^{BL}_{x,y}(g) = V_{x,y}^{TL}(g^{-1}) = V_{x,y}^{TR}(g) =V^{BR}_{x,y}(g^{-1}),
\end{equation}  
which can annihilate these excitations:  
\begin{equation}
    V^{BL}_{x_0,y_0}(g) V^{TL}_{x_0,y_1}(g^{-1}) V^{TR}_{x_1,y_1}(g) V^{BR}_{x_1,y_0}(g^{-1}) U_{[x_0,x_1] \times [y_0,y_1]} |\Psi_{\rm GS}\rangle = |\Psi_{\rm GS}\rangle.
\end{equation}  
These local operators form a projective representation of the symmetry group. For the cluster state model, they can be chosen as
\begin{equation}
  V_{x,y}(g)=   V_{x,y}(g_o^ag_e^b) = (\sigma^z_{x-1,y-1})^{b}\otimes (\tau_{x,y}^z)^a=
   \begin{aligned}
    \begin{tikzpicture}
\definecolor{emerald}{rgb}{0.31, 0.78, 0.47}
    \draw[fill=emerald, draw=black, rounded corners=0.1cm, rotate=45] 
        (0.4,-0.3) rectangle ++ (1.3,0.6);
   \draw[fill=emerald, draw=black, rounded corners=0.1cm, rotate=45] 
        (0.4,-0.3)++(-0.05,-0.05) rectangle ++ (-1.3,0.6); 
\draw[black] (0,0) -- (1, 1);
\draw[black] (0,0) -- (-0.5, -0.5);     
\fill[red] (0,0) circle (0.15);
\fill[blue] (0.5, 0.5) circle (0.15);
\fill[red] (1, 1) circle (0.15);
\fill[blue] (-0.5, -0.5) circle (0.15);
\node at (1.4, -0.6) {\color{red}$(\sigma_{x-1,y-1}^z)^b$};
\node at (1.4, 1.3) {$I$};
\node at (-.4, 0.9) {\color{blue}$(\tau^z_{x,y})^a$};
\node at (-1, -0.7) {$I$};
\end{tikzpicture}
\end{aligned}, \quad a,b=0,1\,.
\end{equation}

\begin{figure}
    \centering
    \includegraphics[width=0.7\linewidth]{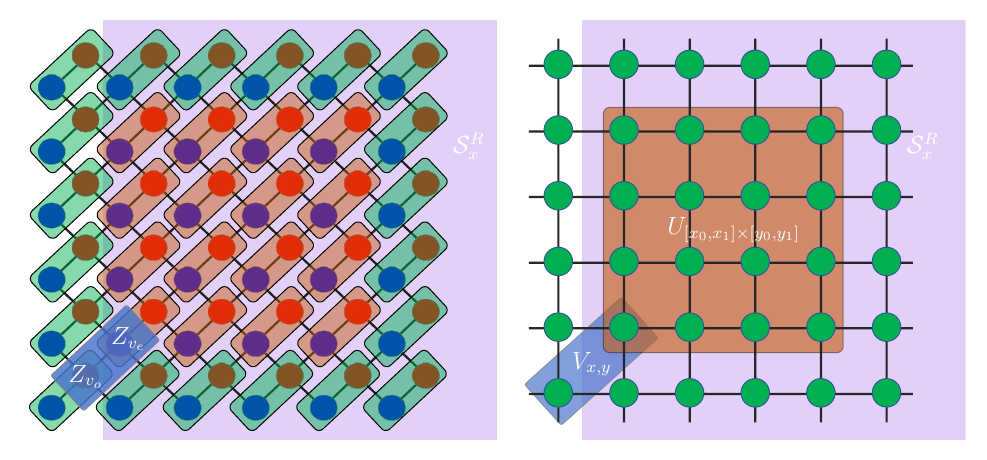}
  \caption{Illustration of the truncated symmetry operator, corner operators, and half-space operators in both the original cluster lattice and the Bravais square lattice.}
    \label{fig:ClusterLatticeOPE}
\end{figure}

Consider the left (L), right (R), top (T), and bottom (B) half-spaces of the $\mathbb{Z} \times \mathbb{Z}$ lattice. We introduce the corresponding half-space symmetry operators: 
\begin{equation}
\begin{aligned}
    \mathcal{S}_{x}^L (g) &= \prod^{x' = x}_{-\infty} S_{x'}(g),\quad  
    \mathcal{S}_{x}^R (g) = \prod_{x' = x}^{+\infty} S_{x'}(g),\\  
    \mathcal{S}_{y}^B (g) &= \prod^{y' = y}_{-\infty} S_{y'}(g),\quad  
    \mathcal{S}_{y}^T (g) = \prod_{y' = y}^{+\infty} S_{y'}(g).
\end{aligned}
\end{equation}  
It was proven in \cite{Devakul2018classifcation} that the invariant of the quantum phase is given by  
\begin{equation}
    \beta_{x,y}^R(g) = \langle \Psi| (\mathcal{S}_x^R(g))^{\dagger} (V_{x,y}(g))^{\dagger} \mathcal{S}_x^R(g) V_{x,y}(g) |\Psi\rangle.
\end{equation}  
Notably, this invariant quantity is independent of the coordinates of the corner as well as the choice of the half-plane symmetry operator. Thus, it can be simply denoted as $\beta(g)$. For cluster state we have
\begin{equation}
    \beta(g_o)=\beta(g_e)=1, \beta(g_og_e)=-1.
\end{equation}
Notice that for symmetry group $G = \mathbb{Z}_2 \times \mathbb{Z}_2$, Eq.~\eqref{eq:classSSPT} yields $\mathcal{C}[\mathbb{Z}_2 \times \mathbb{Z}_2] = \mathbb{Z}_2 \times \mathbb{Z}_2 \times \mathbb{Z}_2$, whose elements correspond to $(\beta(g_o), \beta(g_e), \beta(g_o g_e))$. Different evaluations correspond to distinct strong SSPT phases.

\subsection{$\mathbb{Z}_N\times \mathbb{Z}_M$ cluster state model}

The $\mathbb{Z}_N \times \mathbb{Z}_M$ cluster state model can be constructed in a manner similar to the $\mathbb{Z}_2 \times \mathbb{Z}_2$ cluster state model. Consider the left lattice in Figure~\ref{fig:ClusterLattice}, where odd vertices are assigned a $\mathcal{H}_{x,y}^{\rm odd}=\mathbb{C}[\mathbb{Z}_N]$ qudit\footnote{By qubit we mean the $d>2$ dimensional generalization of qubit.} (red dot), and even vertices are assigned a $\mathcal{H}_{x,y}^{\rm even}=\mathbb{C}[\mathbb{Z}_M]$ (blue dot). We can pair each odd vertex with a nearby even vertex $\mathcal{H}_{x,y}=\mathcal{H}_{x,y}^{\rm odd}\otimes \mathcal{H}_{x,y}^{\rm even}$ (green dot), forming a Bravais square lattice, as shown in Figure~\ref{fig:ClusterLattice}.

The Weyl-Heisenberg operators for the cyclic group $\mathbb{Z}_q$ are defined as
\begin{equation}
    X_q = \sum_{h \in \mathbb{Z}_q} |h+1\rangle \langle h|\,, \quad Z_q = \sum_{h \in \mathbb{Z}_q} \omega_q^h |h\rangle \langle h|\,,
\end{equation}  
where $\omega_q = e^{2\pi i / q}$ and $h=0,\cdots,q-1$. Notice that $X_q^a, a=0,\cdots, q-1$ form regular representation of $\bZ_q$ and $Z^b_q, b=0,\cdots, q-1$ form representation of dual group. These operators satisfy the commutation relation  
\begin{equation}
    Z_q X_q = \omega_q X_q Z_q.
\end{equation}
The eigenstates of $Z_q$ are denoted by $|h\rangle$, with corresponding eigenvalues $\omega_q^h$. The eigenstates of $X_q$ are given by
\begin{equation}\label{eq:Xeigen}
    |\hat{g}\rangle := \frac{1}{\sqrt{q}} \sum_{h \in \mathbb{Z}_q} \omega_q^{-g h} |h\rangle,
\end{equation}
with eigenvalues $\omega_q^{g}$. The $|\hat{g}\rangle$ basis and the $|h\rangle$ basis are thus related by a discrete Fourier transform. The corresponding basis change unitary can be regarded as a generalized Hadamard matrix $H_q=\sum_g |\hat{g}\rangle \langle g|$.

Using the fundamental theorem of abelian groups, any abelian group can be written as  
$G = \mathbb{Z}_{N_1} \times \cdots \times \mathbb{Z}_{N_k}$.  
The corresponding qudit space is given by  
$\mathbb{C}[G] = \mathbb{C}[\mathbb{Z}_{N_1}] \otimes \cdots \otimes \mathbb{C}[\mathbb{Z}_{N_k}]$.  
The Weyl-Heisenberg operators can be generalized accordingly as 
$X_G = \bigotimes_{i=1}^{k} X_{N_i}$, $Z_G = \bigotimes_{i=1}^{k} Z_{N_i}$.  
The model we present below can be extended to any abelian group, but the generalization is more complicated and tedious. Here, we will focus on the cyclic group.

The local stabilizers for the generalized cluster state model are defined for odd ($v_o$) and even ($v_e$) vertices, respectively, as
\begin{equation}
 \begin{aligned}
             K^s_{v_o}&=X_{N,v_o} \otimes (Z_{M,v_{e,1}} \otimes Z_{M,v_{e,2}}^{\dagger} \otimes Z_{M,v_{e,3}} \otimes Z_{M,v_{e,4}}^{\dagger})^{sM/\operatorname{gcd}(M,N)}\\
&=
\begin{aligned}
    \begin{tikzpicture}
\draw[black] (0,0) -- (0.5, 0.5);
\draw[black] (0,0) -- (0.5, -0.5);
\draw[black] (0,0) -- (-0.5, 0.5);
\draw[black] (0,0) -- (-0.5, -0.5);
\fill[red] (0,0) circle (0.15);
\fill[blue] (0.5, 0.5) circle (0.15);
\fill[blue] (0.5, -0.5) circle (0.15);
\fill[blue] (-0.5, 0.5) circle (0.15);
\fill[blue] (-0.5, -0.5) circle (0.15);
\node at (0.5, 0) {$X_N$};
\node at (1.7, 0.9) {$Z_{M,v_{e,3}}^{sM/\operatorname{gcd}(N,M)}$};
\node at (1.7, -1) {$(Z_{M,v_{e,2}}^{\dagger})^{sM/\operatorname{gcd}(N,M)}$};
\node at (-1.5, 0.9) {$(Z_{M,v_{e,4}}^{\dagger})^{sM/\operatorname{gcd}(N,M)}$};
\node at (-1.5, -1) {$Z_{M,v_{e,1}}^{sM/\operatorname{gcd}(N,M)}$};
\end{tikzpicture}
\end{aligned},
 \end{aligned}
\end{equation}

\begin{equation}
 \begin{aligned}
             K^s_{v_e}&=X_{M,v_e} \otimes (Z_{N,v_{o,1}}^{\dagger} \otimes Z_{N,v_{o,2}} \otimes Z_{N,v_{o,3}}^{\dagger} \otimes Z_{N,v_{o,4}})^{sN/\operatorname{gcd}(N,M)}\\
&=
\begin{aligned}
    \begin{tikzpicture}
\draw[black] (0,0) -- (0.5, 0.5);
\draw[black] (0,0) -- (0.5, -0.5);
\draw[black] (0,0) -- (-0.5, 0.5);
\draw[black] (0,0) -- (-0.5, -0.5);
\fill[blue] (0,0) circle (0.15);
\fill[red] (0.5, 0.5) circle (0.15);
\fill[red] (0.5, -0.5) circle (0.15);
\fill[red] (-0.5, 0.5) circle (0.15);
\fill[red] (-0.5, -0.5) circle (0.15);
\node at (0.5,0) {$X_M$};
\node at (1.7, 0.9) {$(Z_{N,v_{o,3}}^{\dagger})^{sN/\operatorname{gcd}(N,M)}$};
\node at (1.7, -1) {$Z_{N,v_{o,2}}^{sN/\operatorname{gcd}(N,M)}$};
\node at (-1.5, 0.9) {$Z_{N,v_{o,4}}^{sN/\operatorname{gcd}(N,M)}$};
\node at (-1.5, -1) {$(Z_{N,v_{o,1}}^{\dagger})^{sN/\operatorname{gcd}(N,M)}$};
\end{tikzpicture}
\end{aligned}.
 \end{aligned}
\end{equation}
It is straightforward to verify that all local terms commute with each other.
Since Weyl-Heisenberg operators are unitary but not Hermitian for $\mathbb{Z}_q$ with $q > 2$, we can Hermitianize the local stabilizers by adding their Hermitian conjugates to the Hamiltonian \cite{You2018SSPT}: 
\begin{equation}
    H= -\sum_{v_o: \text{ odd}} (K_{v_o}^s + (K_{v_o}^s)^{\dagger}) - \sum_{v_e: \text{ even}} (K^s_{v_e} + (K_{v_e}^s)^{\dagger}).
\end{equation}
The  $\bZ_M\times \bZ_N$ strong SSPT phases are classified by    $ \mathcal{C}[\bZ_M\times \bZ_N] = \bZ_M\times \bZ_N \times \bZ_{\operatorname{gcd}(N,M)}$ \cite{Devakul2018classifcation}.
Different choices of $s \in \mathbb{Z}_{\operatorname{gcd}(N,M)}$ in the above Hamiltonian correspond to different strong SSPT lattice models that lies in $\bZ_{\operatorname{gcd}(M,N)}$.

Note that the eigenvalues of $K_{v_o}^s$ are $\exp\left(2\pi i \frac{l + s k t_N}{N}\right)$, where $t_N = N / \gcd(N, M)$, $l = 0, \ldots, N - 1$, and $k = 0, \ldots, \gcd(N, M) - 1$. The ground state corresponds to $l = k = 0$, and the projector onto the $K_{v_o}^s = 1$ subspace can thus be constructed as
\begin{equation}
    P^s_{v_o} = \frac{1}{N} \sum_{r \in \mathbb{Z}_N} \left(K_{v_o}^s\right)^r.
\end{equation}
Similarly, the eigenvalues of $K_{v_e}^s$ are $\exp\left(2\pi i \frac{l + s k t_M}{M}\right)$, where $t_M = M / \gcd(N, M)$, $l = 0, \ldots, M - 1$, and $k = 0, \ldots, \gcd(N, M) - 1$. The ground state corresponds again to $l = k = 0$, and the projector onto the $K_{v_e}^s = 1$ subspace is
\begin{equation}
    P^s_{v_e} = \frac{1}{M} \sum_{r \in \mathbb{Z}_M} \left(K_{v_e}^s\right)^r.
\end{equation}
The ground state of the model is therefore given by
\begin{equation}
  |\Psi_{\rm GS}\rangle =  \prod_{v_o} P^s_{v_o} \prod_{v_e} P^s_{v_e} 
    \left( \bigotimes_{v_o} |\hat{0}\rangle_{v_o} \otimes \bigotimes_{v_e} |\hat{0}\rangle_{v_e} \right),
\end{equation}
where $|\hat{0}\rangle_{v_o}$ and $|\hat{0}\rangle_{v_e}$ are the $+1$ eigenstates of $X_N$ and $X_M$, respectively, as defined in Eq.~\eqref{eq:Xeigen}. For $ N = M = 2 $ and $s=1$, the state reduces to the cluster state in Eq.~\eqref{eq:clusterstateZ2}.

The on-site symmetry group is $\mathbb{Z}_N \times \mathbb{Z}_M = \{ g=g_o^a g_e^b \mid a = 0, \dots, N-1, \; b = 0, \dots, M-1 \}$, where $g_o$ and $g_e$ are the generators (on a local site (green dot), $g_o,g_e$ corresponds to odd and even vertices respectively). On each lattice site, we assign a unitary representation  
\begin{equation}
  u_{x,y}(g)=   u_{x,y}(g_o^a g_e^b) = X_N^a \otimes X_M^b.
\end{equation}
On an $x$-leaf, the subsystem symmetry is given by  
\begin{equation}
  S_x(g) =  S_x(g_o^a g_e^b) = \prod_{y=-\infty}^{+\infty} u_{x,y}(g_o^a g_e^b).
\end{equation}
Similarly, on a $y$-leaf, the subsystem symmetry takes the form  
\begin{equation}
 S_y(g)=    S_y(g_o^a g_e^b) = \prod_{x=-\infty}^{+\infty} u_{x,y}(g_o^a g_e^b).
\end{equation}
Since each leaf subsystem symmetry operator $S_x(g^a_o g^b_e)$ and $S_y(g^a_o g^b_e)$ overlaps with one $Z$ and one $Z^{\dagger}$ factor in local stabilizers $K_{v_o}^s, K_{v_e}^s$, it is straightforward to verify that the generalized cluster state Hamiltonian commutes with these subsystem symmetries:  
\begin{equation}
    [S_x(g^a_o g^b_e), H] = [S_y(g^a_o g^b_e), H] = 0\,, \quad (\forall a, b)
\end{equation}
The generalized cluster state model has $\mathbb{Z}_N \times \mathbb{Z}_M$ subsystem symmetries.

The truncated symmetry operator on a square region $[x_0,x_1]\times [y_0,y_1]$ is defined as:
\begin{equation}
  U_{[x_0,x_1]\times [y_0,y_1]}(g)=   U_{[x_0,x_1]\times [y_0,y_1]}(g_o^a g_e^b)=\bigotimes_{(x,y)\in [x_0,x_1]\times [y_0,y_1]} u_{x,y}(g_o^a g_e^b).
\end{equation}
The corner operator can be chosen as
\begin{equation}
V_{x,y}(g) =     V_{x,y}(g_o^ag_e^b)= Z_N^{sbN/\operatorname{gcd}(N,M)} \otimes Z_M^{saM/\operatorname{gcd}(N,M)}
=
    \begin{aligned}
    \begin{tikzpicture}
\definecolor{emerald}{rgb}{0.31, 0.78, 0.47}
    \draw[fill=emerald, draw=black, rounded corners=0.1cm, rotate=45] 
        (0.4,-0.3) rectangle ++ (1.3,0.6);
   \draw[fill=emerald, draw=black, rounded corners=0.1cm, rotate=45] 
        (0.4,-0.3)++(-0.05,-0.05) rectangle ++ (-1.3,0.6); 
\draw[black] (0,0) -- (1, 1);
\draw[black] (0,0) -- (-0.5, -0.5);     
\fill[red] (0,0) circle (0.15);
\fill[blue] (0.5, 0.5) circle (0.15);
\fill[red] (1, 1) circle (0.15);
\fill[blue] (-0.5, -0.5) circle (0.15);
\node at (1.4, -0.6) {\color{red}$Z_N^{sbN/\operatorname{gcd}(N,M)}$};
\node at (1.4, 1.3) {$I$};
\node at (-1.0, 0.9) {\color{blue}$Z_M^{saM/\operatorname{gcd}(N,M)}$};
\node at (-0.9, -0.8) {$I$};
\end{tikzpicture}
\end{aligned}
\end{equation}
where $s=0,\cdots, \operatorname{gcd}(N,M)-1$; $a=0,\cdots, N-1$, and $b=0,\cdots, M-1$.
A direct calculation shows that for all $v_o, v_e$, we have  
\begin{equation}
\begin{aligned}
   & [K^s_{v_o}, V^{BL}_{x_0,y_0}(g) V^{TL}_{x_0,y_1}(g^{-1}) V^{TR}_{x_1,y_1}(g) V^{BR}_{x_1,y_0}(g^{-1}) U_{[x_0,x_1] \times [y_0,y_1]} ] \\
   =& [K^s_{v_e}, V^{BL}_{x_0,y_0}(g) V^{TL}_{x_0,y_1}(g^{-1}) V^{TR}_{x_1,y_1}(g) V^{BR}_{x_1,y_0}(g^{-1}) U_{[x_0,x_1] \times [y_0,y_1]} ] = 0.
\end{aligned}
\end{equation}
This further implies
\begin{equation}
\begin{aligned}
   & K_{v_o} V^{BL}_{x_0,y_0}(g) V^{TL}_{x_0,y_1}(g^{-1}) V^{TR}_{x_1,y_1}(g) V^{BR}_{x_1,y_0}(g^{-1}) U_{[x_0,x_1] \times [y_0,y_1]} |\Psi_{\rm GS} \rangle \\
   = & V^{BL}_{x_0,y_0}(g) V^{TL}_{x_0,y_1}(g^{-1}) V^{TR}_{x_1,y_1}(g) V^{BR}_{x_1,y_0}(g^{-1}) U_{[x_0,x_1] \times [y_0,y_1]} |\Psi_{\rm GS} \rangle \\
   =& K_{v_e} V^{BL}_{x_0,y_0}(g) V^{TL}_{x_0,y_1}(g^{-1}) V^{TR}_{x_1,y_1}(g) V^{BR}_{x_1,y_0}(g^{-1}) U_{[x_0,x_1] \times [y_0,y_1]} |\Psi_{\rm GS} \rangle.
\end{aligned}
\end{equation}
Since the ground state $|\Psi_{\rm GS} \rangle$ is invariant under local stabilizers and is unique, we obtain  
\begin{equation}
    V^{BL}_{x_0,y_0}(g) V^{TL}_{x_0,y_1}(g^{-1}) V^{TR}_{x_1,y_1}(g) V^{BR}_{x_1,y_0}(g^{-1}) U_{[x_0,x_1] \times [y_0,y_1]} |\Psi_{\rm GS} \rangle = |\Psi_{\rm GS} \rangle.
\end{equation}
Thus, the corner operators annihilate the excitations created by the truncated symmetry operator.

The half-space symmetry operators are defined as
\begin{equation}
\begin{aligned}
    \mathcal{S}_{x}^L (g) &= \prod_{x' \leq x} S_{x'}(g_o^a g_e^b),\quad  
    \mathcal{S}_{x}^R (g) = \prod_{x' \geq x} S_{x'}(g_o^a g_e^b),\\  
    \mathcal{S}_{y}^B (g) &= \prod_{y' \leq y} S_{y'}(g_o^a g_e^b),\quad  
    \mathcal{S}_{y}^T (g) = \prod_{y' \geq y} S_{y'}(g_o^a g_e^b).
\end{aligned}
\end{equation}  
A direct calculation of the topological invariant of the model gives, for $g = g_o^a g_e^b$,  
\begin{equation}
    \beta_{x,y}^R(g) = \langle \Psi| \mathcal{S}_x^R(g)^\dagger V_{x,y}(g)^\dagger \mathcal{S}_x^R(g) V_{x,y}(g) |\Psi\rangle = e^{2\pi i  \frac{sab}{\operatorname{gcd}(N,M)}}.
\end{equation}  
This expression is independent of the specific choices of the four corner operators and the four half-space symmetry operators, provided that the corner operator has a nontrivial overlap with the half-space operator. The result matches well with result we obtain in Section~\ref{sec:ZNZMSymTFT} via subsystem SymTFT.

\section{Discussion and future directions}

In this work, we provide a SymTFT realization of the classification of SSPT phases, which shows good agreement with results obtained from the lattice Hamiltonian formalism. 
This demonstrates that subsystem symmetries also satisfy the correspondence between symmetries and one-dimensional higher field theories. 
By carefully comparing the topological invariants calculated from lattice models with those derived from the foliated tensor gauge theory (or equivalently foliated BF theory), we confirm that they coincide, give an intuitive interpretation of classification from the SymTFT perspective, thereby validating the effectiveness of the subsystem SymTFT framework in capturing the essential features of SSPT phases.

We mainly employ the foliated tensor gauge theory, and the methodology developed here can be naturally extended to other types of foliated field theories, potentially providing a systematic SymTFT framework for understanding subsystem symmetries and their associated quantum phases.

Despite the progress made, there are several interesting questions to be investigated for the subsystem SymTFT:

(1) \emph{Higher Dimensional Subsystem SymTFT.} In this work, we have focused on 2d SSPT phases. While higher-dimensional foliated BF theories can be naturally constructed, the classification of their topological boundary conditions—and the corresponding implications for classifying higher-dimensional SSPT phases—remains largely open. 
Ref.~\cite{Devakul2020planar} provides some discussion of 3d SSPT phases from the lattice formalism. We believe that their results may be consistent with those obtainable via the subsystem SymTFT framework, and may also shed light on the structure of SSPT phases on general $n$-dimensional spatial manifolds.

(2) \emph{Subsystem SymTFT for General Group and Fusion Categories.}  
A comprehensive understanding of subsystem SymTFTs for general symmetry groups remains an open problem. Moreover, subsystem symmetries can be generalized to non-invertible symmetries, a direction that, to the best of our knowledge, remains largely unexplored.
A thorough understanding of the topological boundary conditions in SymTFT requires the language of higher fusion categories \cite{Kong2020algebraic,johnson2022classification}. However, the mathematical theory of higher fusion categories is still under development. Similarly, within the context of subsystem SymTFT, gaining a comprehensive understanding of topological boundary conditions remains a crucial topic that requires further investigation. It is also possible to generalize subsystem symmetries to non-invertible ones; however, the study of such generalizations remains largely open and under active investigation \cite{Cao2023subsystem}.

(3) \emph{Gapless SSPT and Its Topological Holography.}
While we now have a relatively complete understanding of gapped quantum phases and their classification, the classification of gapless phases remains an outstanding open problem. The generalization of SPT phases and SymTFTs to gapless systems has recently attracted considerable attention~\cite{kong2020mathematical,Chatterjee2024gapless,bhardwaj2024clubsandwich,huang2023topologicalholo,wen2023classification11dgaplesssymmetry,wen2025stringconden}. In particular, the club sandwich construction offers a promising framework for extending subsystem SymTFTs to gapless settings, opening up a rich and intriguing direction for future investigation.
Very recently, several related developments have appeared in this direction (see, e.g.,~\cite{ohmori2025gaplessfoliatedexoticduality,apruzzi2025symtftfoliated}).

(4) \emph{Systematic Lattice Realization of Subsystem SymTFTs.}
From the perspective of lattice models, the systematic construction of SymTFTs has become an increasingly active area of research; see, for instance, \cite{fechisin2023noninvertible, bhardwaj2024lattice, Bhardwaj:2023fca, Seiberg2024noninvertible, Seifnashri2024cluster, inamura202411dsptphasesfusion, jia2024generalized, jia2024weakhopfnoninvertible,meng2024noninvertiblespt,Pace2025gauging,Pace2025lattice}. 
In the context of subsystem symmetries, a natural direction is to investigate how the existing lattice constructions for global symmetries can be generalized to accommodate the richer and more intricate structure of subsystem symmetries. 
Current lattice models of SSPT phases have largely been developed on a case-by-case basis \cite{You2018SSPT,Devakul2018classifcation,Devakul2020planar}, and understanding their associated subsystem SymTFTs remains an important challenge. 
Conversely, constructing lattice models that systematically realize a given subsystem SymTFT is also a crucial open problem deserving further exploration.

(5) \emph{(Para-)Fermionic SSPT Phase.} In \cite{Cao2022boson}, the authors generalize the Jordan-Wigner transformation in 2D to (2+1)-dimensional $\mathbb{Z}_2$ subsystem symmetry, and the corresponding fermionic topological boundaries have been discussed in a following paper \cite{Cao2024SymTFT} and are also reviewed in Section~\ref{sec:subSymTFT}. It is also interesting to explore the fermionic SSPT phase from the SymTFT picture, and consider the generalization to para-fermionic cases~\cite{Duan:2023ykn}.

\subsection*{Acknowledgements}

Z. J. acknowledges Dagomir Kaszlikowski for his support, and Hiromi Ebisu and Masahito Yamazaki for insightful discussions on SSPT, and he is supported by the National Research Foundation in Singapore, the A*STAR under its CQT Bridging Grant and CQT-Return of PIs EOM YR1-10 Funding and  CQT Young Researcher Career Development Grant. Q. J. is supported by National Research Foundation of Korea (NRF) Grant No. RS-2024-00405629 and Jang Young-Sil Fellow Program at the Korea Advanced Institute of Science and Technology.
We are also grateful to Nathanan Tantivasadakarn for his valuable comments.




%

\end{document}